\documentclass[prd,aps,showpacs,superscriptaddress,nofootinbib,amsmath,amssymb]{revtex4}
\usepackage{graphicx}% Include figure files
\usepackage{dcolumn}% Align table columns on decimal point
\usepackage{bm}% bold math
\usepackage{multirow}
\usepackage{amssymb}
\usepackage{adjustbox}
\usepackage{textcomp}
\usepackage{hyperref}
\usepackage{xcolor}
\graphicspath{{./eps/}}

\begin{document}
 \title{\Large Weak quasielastic hyperon production leading to pions in the antineutrino-nucleus reactions}
\author{A. \surname{Fatima}}
% \email{atikafatima1706@gmail.com}
\affiliation{Department of Physics, Aligarh Muslim University, Aligarh-202002, India}
\author{M. Sajjad \surname{Athar}}
\email{sajathar@gmail.com}
\affiliation{Department of Physics, Aligarh Muslim University, Aligarh-202002, India}
\author{S. K. \surname{Singh}}
% \email{sksingh.amu@gmail.com}
\affiliation{Department of Physics, Aligarh Muslim University, Aligarh-202002, India}

\begin{abstract}
 In this review, we have studied the quasielastic production cross sections and polarization components of $\Lambda$, 
 $\Sigma^0$ and $\Sigma^-$ hyperons induced by the weak charged currents in the antineutrino reactions on the nucleon 
 and the nuclear targets like $^{12}$C, $^{16}$O, $^{40}$Ar and $^{208}$Pb. It is shown that the energy and the $Q^2$ 
 dependence of the cross sections and the various polarization components can be effectively used to determine the 
 axial vector transition form factors in the strangeness sector and test the validity of various symmetry properties 
 of the weak hadronic currents like G-invariance, T-invariance and SU(3) symmetry. In particular, the energy and the 
 $Q^2$ dependence of the polarization components of the hyperons is found to be sensitive enough to determine the 
 presence of the second class current with or without T-invariance.

These hyperons decay dominantly into pions giving an additional contribution to the weak pion production induced by the 
antineutrinos. This contribution is shown to be quantitatively significant as compared to the pion production by the 
$\Delta$ excitation in the nuclear targets in the sub-GeV energy region relevant for the $\bar{\nu}_\mu$ cross section 
measurements in the oscillation experiments. We have also included a few new results, based on our earlier works, which 
are in the kinematic region of the present and future (anti)neutrino experiments being done with the accelerator 
(anti)neutrinos at T2K, MicroBooNE, MiniBooNE, NO$\nu$A, MINER$\nu$A and DUNE, as well as for the atmospheric 
(anti)neutrino experiments in this energy region.
\end{abstract}
\maketitle

\section{Introduction}
A simultaneous knowledge of the neutrino and antineutrino cross sections in the same energy region for the nuclear 
targets is highly desirable in order to understand the systematics relevant for the analyses of various neutrino 
oscillation experiments being done in search of CP violation in the leptonic sector and in the determination of 
neutrino mass hierarchy~\cite{Abe:2014tzr, Paley:2013sta, NOvA:2018gge, Abe:2015zbg, Acciarri:2016ooe, Galymov:2016nwz, 
Goodman:2015gmv}. Experimentally there are many results available in the cross section measurements for the various 
weak processes induced by the neutrinos in nuclei in the sub-GeV and few-GeV energy region~\cite{Abe:2017ufe, 
Tsai:2017pta, AguilarArevalo:2010bm, Mariani:2011xm}. There are very few measurements reported for the processes 
induced by the antineutrinos in the same energy region specially around $E_{\bar{\nu}_\mu} \approx 1$ 
GeV~\cite{Abe:2017ufe, Aguilar-Arevalo:2013nkf}. Theoretically, however, there exists quite a few calculations for the 
antineutrino-nucleus cross sections and some of them have been incorporated in most of the neutrino event generators 
like GENIE~\cite{Andreopoulos:2009rq}, NEUT~\cite{Hayato:2009zz}, NuWro~\cite{Golan:2012wx} and 
GiBUU~\cite{Buss:2011mx}. In this energy region of antineutrinos, $E_{\bar{\nu}_\mu} \approx 0.5-1.2$ GeV, the most 
important processes contributing to the nuclear cross sections are the quasielastic (QE) scattering and the inelastic 
scattering where the excitation of $\Delta$ resonance is the dominant process contributing to the single pion 
production~(CC1$\pi$). There is some contribution from the excitation of higher resonances and very little contribution 
from the deep inelastic scattering~(DIS)~\cite{Alvarez-Ruso:2017oui, Katori:2016yel, Alvarez-Ruso:2014bla, 
Formaggio:2013kya, Morfin:2012kn}. 

It is well known that the cross sections for the various weak processes induced by the neutrinos and antineutrinos 
differ by the sign of the interference terms between the vector and the axial vector currents making the antineutrino 
cross sections smaller and fall faster with $Q^2$ as compared to the neutrino cross sections~\cite{Alam:2015gaa, 
Akbar:2015yda, prd1, a6, Athar:2007wd, Hernandez:2007qq}. There is another difference between the neutrino and 
antineutrino induced processes on the nucleon and the nuclear targets which has not been adequately emphasized in 
the context of the discussion of the systematics in the neutrino oscillation experiments. This difference arises due 
to the phenomenological $\Delta S =\Delta Q$ rule implicit in the standard model (SM) in the charged current sector 
which allows the quasielastic production of hyperons on nucleons induced by the antineutrinos, i.e. $\bar{\nu}_l + N 
\rightarrow l^+ + Y;~ N=n \text{ or }p,~ Y = \Lambda,~ \Sigma^0 \text{ or } \Sigma^-$, but not with the neutrinos i.e. 
$\nu_l + N \not\rightarrow l^- + Y$. The hyperon production process is Cabibbo suppressed and its cross section is 
generally small as compared to the quasielastic or $\Delta$ production in the $\Delta S=0$ sector. However, in the 
lower energy region of the antineutrinos i.e. $E_{\bar{\nu}_\mu} << 1$~GeV where the production of $\Delta$ resonance 
is kinematically inhibited due to a higher threshold for the $\Delta$ production as compared to $\Lambda$ production, 
the hyperon production cross section may not be too small. These hyperons dominantly decay into $\pi^-$ and $\pi^o$ and 
give additional contribution to the pion production induced by the antineutrinos from the nucleon and the nuclear 
targets.

Since $\pi^-$ and $\pi^o$ are the largest misidentified background for the $\bar{\nu}_\mu$ disappearance and 
$\bar{\nu}_e$ appearance channels in the present neutrino oscillation experiments with the antineutrino beams, the 
hyperon production becomes an important process to be considered in the accelerator experiments specially at 
T2K~\cite{Abe:2017ufe}, MicroBooNE~\cite{Tsai:2017pta}, MiniBooNE~\cite{AguilarArevalo:2010bm, Aguilar-Arevalo:2013nkf} 
and NO$\nu$A~\cite{Adamson:2016xxw}, where the antineutrino energies are in the sub-GeV energy region. Moreover, these 
experiments are being done using nuclear targets like $^{12}$C, $^{16}$O, $^{40}$Ar, etc., where the pion production 
cross sections through the $\Delta$ excitations are considerably suppressed due to the nuclear medium effect~(NME) and 
the final state interaction~(FSI) effect~\cite{Alvarez-Ruso:2017oui, Katori:2016yel}. On the other hand, the pions 
arising from the hyperons are expected to be less affected by these effects due to the fact that the hyperon decay 
widths are highly suppressed in the nuclear medium making them live longer and travel through most of the nuclear 
medium before they decay~\cite{Holstein,Oset:1989ey}. Therefore, the two effects discussed above i.e. the lower 
threshold energy of the hyperon production and near absence of the FSI for the pions coming from the hyperon decay 
compensate for the $\tan^2 \theta_c$ suppression as compared to the pions coming from the $\Delta$ production. This 
makes these processes important in the context of oscillation experiments with antineutrino beams in the sub-GeV energy 
region.

Notwithstanding the importance of the hyperon production in the context of present day oscillation experiments with 
the accelerator antineutrino beams at lower energies, the study of these processes is important in their own right as 
these processes give us an opportunity to understand the weak interactions at higher energies in the $\Delta S = 1$ 
sector through the study of the nucleon-hyperon transition form factors at higher energy and $Q^2$. The information 
about these form factors is obtained through the analysis of semileptonic hyperon decays which is limited to very low 
$Q^2$~\cite{Cabibbo:2003cu, Gaillard:1984ny, Gazia}. It is for this reason that the work in the quasielastic production 
of hyperons induced by the antineutrino was started more than 50 years back and many theoretical papers have reported 
results for the cross section and the polarization of the hyperons in the literature~\cite{Henley:1969uz, 
Cannata:1970br, DeRujula:1970ek, Adler:1963, Berman:1964zza, Fujii2, Fujii1, Cabibbo:1964zza, Glashow:1965zz, 
Okamura:1971pn, Ketley, Egardt, Block:1965zol, Block:NAL, Cabibbo:1965zza, Singh:2006xp, Alam:2014bya, Cabibbo:1963yz, 
Block:1964gj, Chilton:1964zza, sirlin, Finjord:1975zy} which have been summarized in the early works of Marshak et 
al.~\cite{Marshak}, Llewellyn Smith~\cite{LlewellynSmith:1971uhs} and Pais~\cite{Pais:1971er}. Experimentally, however, 
there are very few attempts made where the quasielastic production of $\Lambda$, $\Sigma^0$, $\Sigma^-$ have been 
studied, like at CERN~\cite{Erriquez:1977tr, Eichten:1972bb, Erriquez:1978pg}, BNL~\cite{Fanourakis:1980si}, 
FNAL~\cite{Ammosov:1986jn, Ammosov:1986xv} and SKAT~\cite{Brunner:1989kw}. A summary of all the experimental results 
on the energy dependence of the total cross sections on the hyperon production and its comparison with the theoretical 
calculations has been given by Kuzmin and Naumov~\cite{Kuzmin:2003ji} and Rafi Alam et al.~\cite{Alam:2013cra}. With 
the availability of the high intensity antineutrino beams at JPARC~\cite{JPARC} and FNAL~\cite{fermi} and the advances 
made in the detector technology, the feasibility of studying the quasielastic production of hyperons and their 
polarizations have been explored in many theoretical calculations~\cite{Kuzmin:2008zz, Bilenky:2013fra, 
Bilenky:2013iua, Akbar:2017qsf, Graczyk:2017rti, Kuzmin:2003ji, Kuzmin:2004ke, Graczyk:2004vg, Hagiwara:2004gs, 
Graczyk:2004uy, Valverde:2006yi, Fatima:2018tzs, Fatima:2018gjy, Akbar:2016awk, Mintz:2004eu, Mintz:2002cj, 
Mintz:2001jc, Wu:2013kla}. Experimentally, while the MINER$\nu$A~\cite{Drakoulakos:2004gn} collaboration has included 
the study of quasielastic production of hyperons in its future plans, some other collaborations are also considering 
the feasibility of making such measurements~\cite{DUNE}. 

In this review, we have attempted to give an overview of the present and the earlier works done in the study of the 
quasielastic production of hyperons induced by the antineutrinos from the nucleon~\cite{Fatima:2018tzs} and the nuclear 
targets~\cite{Akbar:2016awk} and its implications for the pion production~\cite{Alam:2013cra, Alam:2014bya} relevant 
for the analysis of the oscillation experiments being done with the antineutrino beams in the sub-GeV energy region. 
Specifically, we describe the energy and the $Q^2$ dependence of the production cross section and polarizations of 
$\Lambda$, $\Sigma^0$ and $\Sigma^-$ hyperons in the quasielastic reactions on the nucleon and the nuclear targets like 
$^{12}$C, $^{16}$O, $^{40}$Ar and $^{208}$Pb. In view of the future experiments to be done with the antineutrino beams 
in the medium energy region of few GeV, it is useful to review the current status of the theoretical and experimental 
work on this subject.

We also take into account the nuclear medium effects on the production cross section of hyperons in a local density 
approximation~\cite{Akbar:2016awk, Alam:2013cra}. The effect of the final state interaction of the hyperons on the 
production cross section and its $Q^2$ dependence arising due to the strong interactions in the presence of the 
nucleons in the nuclear medium leading to elastic and charge exchange reactions like $\Sigma N \rightarrow \Lambda N$ 
and $\Lambda N \rightarrow \Sigma N$ is also taken into account in a simple model~\cite{Singh:2006xp}. The effect of 
the second class current with or without the presence of T-invariance~\cite{Fatima:2018tzs, Fatima:2018gjy} on the 
total and the differential cross sections, and the $Q^2$ dependence of the polarization components of the hyperons have 
also been presented. 

These hyperons decay into pions through the different $Y \longrightarrow N \pi$ decay modes and contribute to the pion 
production cross sections induced by the antineutrinos which is in addition to the pion production cross section 
through the excitation of $\Delta^{0}$ and $\Delta^{-}$ resonances. Keeping in mind the present and future 
(anti)neutrino experiments being done with the accelerator (anti)neutrinos at T2K, MicroBooNE, MiniBooNE, NO$\nu$A, 
MINER$\nu$A and DUNE, as well as the atmospheric (anti)neutrino experiments being planned in this energy region, we 
have also presented some new results on the pion production  in the kinematic region of these experiments based on the 
formalism discussed here in brief in Sections~\ref{formalism} and \ref{NME+FSI}.

In Section~\ref{formalism}, we describe in brief the formalism for calculating the cross sections and the polarization 
components of the $\Lambda$, $\Sigma^0$ and $\Sigma^-$ hyperons produced in the quasielastic reactions of the 
antineutrinos from the nucleons in the presence of the second class currents. We also reproduce the essential formalism 
for the excitation of $\Delta$ in this section and describe the process of pion production from the hyperon (Y) and 
$\Delta$ decay. We describe in Section~\ref{NME+FSI} the effect of the nuclear medium on the $\Delta$ and the hyperon 
productions, and in Section~\ref{NME+FSI+Pion} final state interactions of the hyperons in the nuclear medium and the 
final state interactions on the production of pions as a result of the $\Delta$ excitations. In Section~\ref{results}, 
we present our results and finally in Section~\ref{conclusions} conclude the findings.

\section{Formalism}\label{formalism}
\subsection{Hyperon production off the free nucleon}
\subsubsection{Matrix element and form factors}

\begin{figure}
 \begin{center}
    \includegraphics[height=3cm,width=6cm]{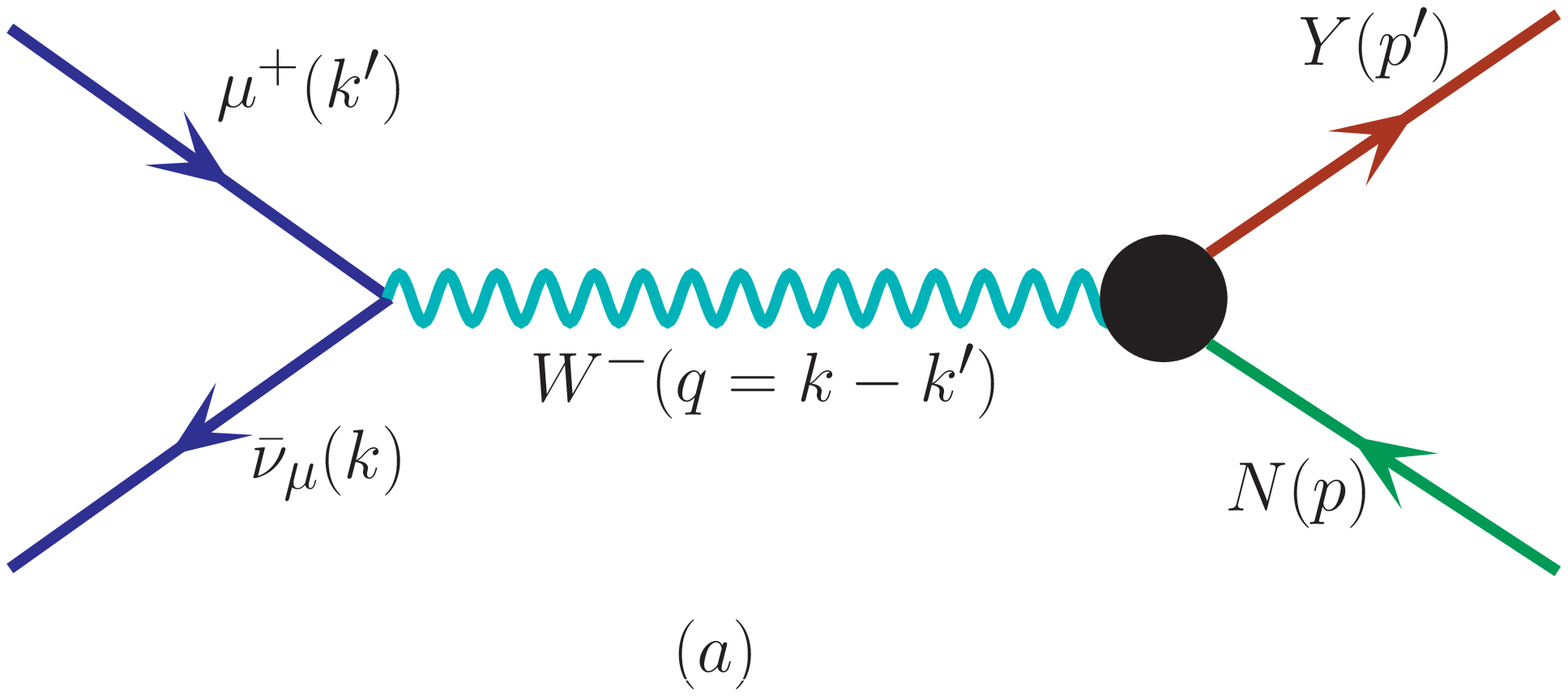}
%  \end{subfigure}
  \hspace{5mm}
% \begin{subfigure}
  \includegraphics[height=3cm,width=6cm]{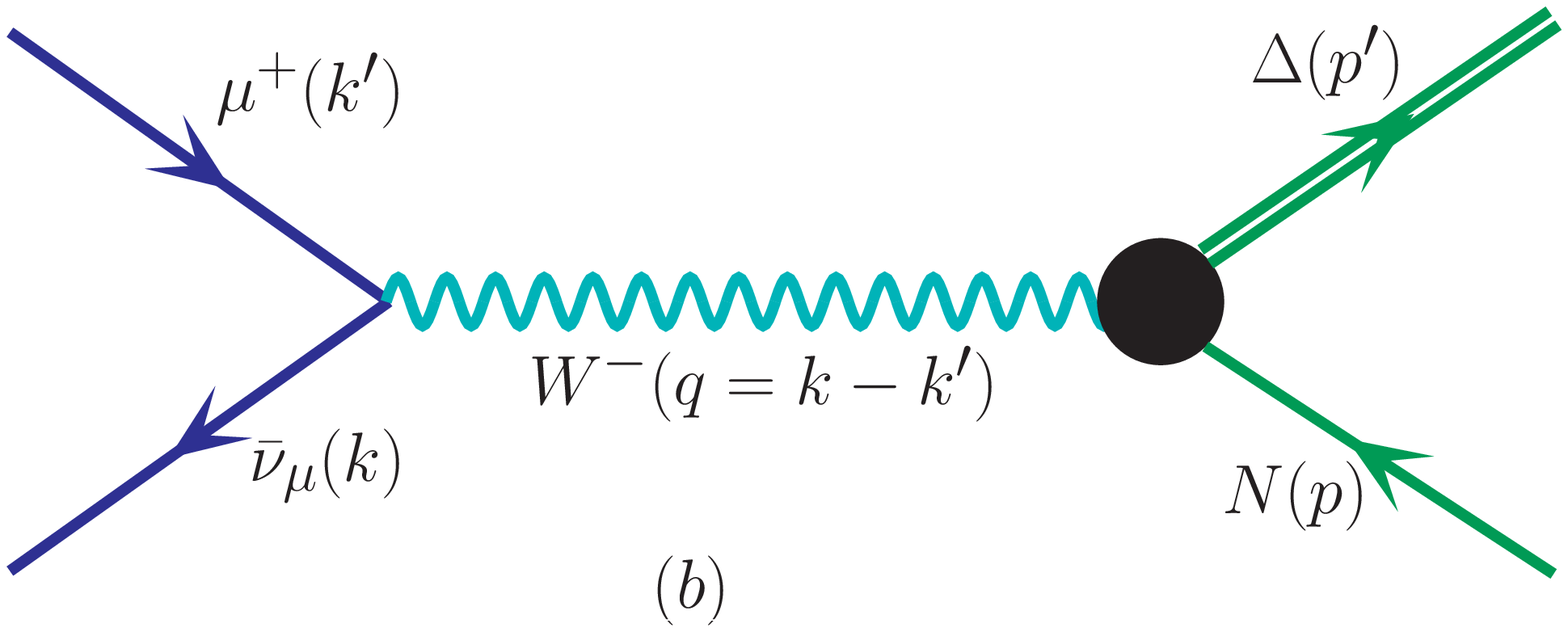}  
%   \end{subfigure}
  \caption{Feynman diagram for the processes (a) $\bar{\nu}_\mu(k) + N(p) \longrightarrow \mu^+ (k^\prime) + 
  Y(p^\prime)$ (left panel) and (b) $\bar{\nu}_\mu(k) + N(p) \longrightarrow \mu^+ (k^\prime) + \Delta(p^\prime)$~(right
  panel). The quantities in the bracket represent four momenta of the corresponding particles. $N$ stands for a $n$ or 
  $p$, $Y$ may be a $\Lambda$ or $\Sigma^0$ or $\Sigma^-$ and the $\Delta$ may be a $\Delta^0$ or $\Delta^-$ depending 
  upon the initial nucleon state.}
  \label{fyn_hyp}
   \end{center}
 \end{figure}

The transition matrix element for the processes,
\begin{eqnarray}\label{process3}
 \bar{\nu}_\mu (k) + p (p) &\rightarrow& \mu^+ (k^\prime) + \Lambda (p^\prime), \\
 \label{process4}
 \bar{\nu}_\mu (k) + p (p) &\rightarrow& \mu^+ (k^\prime) + \Sigma^0 (p^\prime), \\
 \label{process5}
 \bar{\nu}_\mu (k) + n (p) &\rightarrow& \mu^+ (k^\prime) + \Sigma^- (p^\prime), 
\end{eqnarray}
shown in Fig.~\ref{fyn_hyp}(a), may be written as
 \begin{eqnarray}\label{matrix}
  {\cal{M}} = \frac{G_F}{\sqrt{2}} sin \theta_c ~ l^\mu {{J}}_\mu,
 \end{eqnarray}
where the quantities in the brackets represent the four momenta of the corresponding particles, $G_F$ is the Fermi 
coupling constant and $\theta_c$ is the Cabibbo mixing angle. 

The leptonic current $l^\mu$ is given by 
  \begin{equation}\label{l}
 l^\mu = \bar{u} (k^\prime) \gamma^\mu (1 + \gamma_5) u (k),
\end{equation}
 and the hadronic current ${J}_\mu$ is expressed as:
\begin{equation}\label{j}
 {{J}}_\mu =  \bar{u} (p^\prime) {\Gamma_\mu} u (p)
\end{equation}
with
\begin{equation}\label{gamma}
 {\Gamma_\mu} = V_\mu - A_\mu.
\end{equation}

The vector ($V_\mu$) and the axial vector ($A_\mu$) currents are given by~\cite{Fatima:2018tzs}:
\begin{eqnarray}\label{vx}
 \langle Y(p^\prime) | V_\mu| N(p) \rangle &=& \bar{u}(p^\prime) \left[ \gamma_\mu f_1^{NY}(Q^2)+i\sigma_{\mu \nu} 
 \frac{q^\nu}{M+M^\prime} f_2^{NY}(Q^2) + \frac{2 ~q_\mu}{M+M^\prime} f_3^{NY}(Q^2) \right] u(p),
  \end{eqnarray}
and 
\begin{eqnarray}\label{vy}
  \langle Y(p^\prime) | A_\mu| N(p) \rangle &=& \bar{u} (p^\prime) \left[ \gamma_\mu \gamma_5 g_1^{NY}(Q^2) + 
  i\sigma_{\mu \nu} \frac{q^\nu}{M+M^\prime} \gamma_5 g_2^{NY}(Q^2) + \frac{2 ~q_\mu} {M+M^\prime} g_3^{NY}(Q^2) 
  \gamma_5 \right] u(p), 
\end{eqnarray}
which may be rewritten as
\begin{eqnarray}\label{vy2}
  \langle Y(p^\prime) | A_\mu| N(p) \rangle &=& \bar{u} (p^\prime) \Bigg[ \gamma_\mu \gamma_5 g_1^{NY}(Q^2) + 
  \left(\frac{\Delta_M}{M + M^\prime} \gamma_\mu \gamma_5 - \frac{p_\mu + p_\mu^\prime}{M + M^\prime} \gamma_5 \right) 
  g_2^{NY}(Q^2) \nonumber \\
  &+& \frac{2 ~q_\mu} {M+M^\prime} g_3^{NY}(Q^2) \gamma_5 \Bigg] u(p), ~~~~~
\end{eqnarray}
where $Y(=\Lambda, \Sigma^0 \text{ and } \Sigma^-)$ represents a hyperon, $\Delta_M = M^\prime - M$ with $M$ and 
 $M^\prime$ being the masses of the initial nucleon and the final hyperon. $q_\mu (= k_\mu - k_\mu^\prime = 
 p_\mu^\prime -p_\mu)$ is the four momentum transfer with $Q^2 = - q^2,~Q^2 \ge 0$. $f_1^{NY}(Q^2)$, $f_2^{NY}(Q^2)$ 
 and $f_3^{NY}(Q^2)$ are the vector, weak magnetic and induced scalar form factors and $g_1^{NY}(Q^2)$,~$g_2^{NY}(Q^2)$ 
 and $g_3^{NY}(Q^2)$ are the axial vector, induced tensor (or weak electric) and induced pseudoscalar form factors, 
respectively.
\vspace{1mm}

 The six form factors $f_i^{NY} (Q^2)$ and $g_i^{NY} (Q^2) ~ (i=1-3)$ are determined using the following assumptions 
 about the symmetry properties of the weak vector (V) and axial vector (A) currents in the SM. Under 
 the assumption of the SU(3) symmetry, the initial and the final baryons as well as the weak V and A currents belong to 
 the octet (8) representation, each form factor $f^{NY}_i(Q^2)$ and $g^{NY}_i(Q^2)$ occurring in the definition of the 
 transition form factors defined in Eqs.~(\ref{vx}) and (\ref{vy2}) can be written in terms of the two functions 
 $D(Q^2)$ and $F(Q^2)$ corresponding to the symmetric octet~($8^{S}$) and antisymmetric octet~($8^{A}$) couplings of 
 the octets of the vector and the axial vector currents to the octets of the initial and the final baryons. 
 Specifically, we write
\begin{eqnarray}
 f^{NY}_i(Q^2) &=& a F_{i}^{V}(Q^2) + b D_{i}^{V}(Q^2)\label{coef1},\\
 g^{NY}_i(Q^2) &=& a F_{i}^{A}(Q^2) + b D_{i}^{A}(Q^2)\label{coef2},
\end{eqnarray}
where $a$ and $b$ are the SU(3) Clebsch-Gordan coefficients given in Table-\ref{tabI} for the reactions shown in 
Eqs.~(\ref{process3})--(\ref{process5}). $F_i^{V}(Q^2) (D_i^{V}(Q^2))$ and $F_i^{A}(Q^2) (D_i^{A}(Q^2));~(i=1,2)$, 
are the couplings corresponding to the antisymmetric~(symmetric) couplings of the vector and the axial vector currents.

\begin{table*}[h]
\begin{tabular}{|c|c|c|}\hline
~~ Transitions ~~ & ~~ $a$ ~~ & ~~ $b$ ~~ \\ \hline
 ~~ $n\rightarrow p$ ~~ & ~~1~~ & ~~1~~ \\
 ~~ $p\rightarrow \Lambda$ ~~ & ~~ $-\sqrt{\frac{3}{2}}$ ~~ & ~~ $-\frac{1}{\sqrt{6}}$ ~~\\
 ~~ $p\rightarrow \Sigma^0$ ~~ & ~~ $-\frac{1}{\sqrt{2}}$ ~~ & ~~ $\frac{1}{\sqrt{2}}$ ~~\\
~~ $n\rightarrow \Sigma^-$ ~~ & ~~ $-1$ ~~ & ~~ 1 ~~ \\ \hline
% \hline
\end{tabular}
\caption{Values of the coefficients $a$ and $b$ given in Eqs.~(\ref{coef1})$-$(\ref{coef2}).}
\label{tabI}
\end{table*}

For the determination of the $N-Y$ transition form factors, we take the following considerations into account: 
\begin{enumerate} 
\item[a)] The assumption of the conserved vector current (CVC) hypothesis leads to $f_3^{NY} (Q^2) = 0$ and the two 
vector form factors \textit{viz.} $f^{NY}_1(Q^2)$ and $f^{NY}_2(Q^2)$ are determined in terms of the electromagnetic 
form factors of the nucleon, \textit{i.e.} $f_{1}^{N}(Q^{2})$ and $f_{2}^{N}(Q^{2}),~ N=(p,n)$ as
 \begin{eqnarray}
 \label{fplambda}
 f_{1,2}^{p \Lambda}(Q^2)&=& -\sqrt{\frac{3}{2}}~f_{1,2}^p(Q^2), \\
 \label{fnsigma}
 f_{1,2}^{n \Sigma^-}(Q^2)&=& -\left[f_{1,2}^p(Q^2) + 2 f_{1,2}^n(Q^2) \right],  \\
 \label{fpsigma}
 f_{1,2}^{p \Sigma^0}(Q^2)&=& -\frac{1}{\sqrt2}\left[f_{1,2}^p(Q^2) + 2 f_{1,2}^n(Q^2) \right].
 \end{eqnarray}
 The electromagnetic form factors $f_{1,2}^p (Q^2)$ and $f_{1,2}^n (Q^2)$ are expressed in terms of 
 the Sachs electric and magnetic form factors $G_E^{p,n} (Q^2)$ and $G_M^{p,n} (Q^2)$ of the nucleons as
\begin{eqnarray}\label{f1pn}
f_1^{p,n}(Q^2)&=&\left(1+\frac{Q^2}{4M^2}\right)^{-1}~\left[G_E^{p,n}(Q^2)+\frac{Q^2}{4M^2}~G_M^{p,n}(Q^2)\right],\\
\label{f2pn}
f_2^{p,n}(Q^2)&=&\left(1+\frac{Q^2}{4M^2}\right)^{-1}~\left[G_M^{p,n}(Q^2)-G_E^{p,n}(Q^2)\right].
\end{eqnarray}
For $G_E^{p,n}(Q^2)$ and $G_M^{p,n}(Q^2)$ various parameterizations are available in the literature and in our 
numerical calculations, we have used the parameterization given by Bradford et al.~\cite{Bradford:2006yz}.
 
\item[b)]  The axial vector form factors $g^{NY}_{1}(Q^{2})$ and $g^{NY}_{2}(Q^{2})$ are determined from 
Eq.~(\ref{coef2}) in terms of the two functions $F_{1,2}^A(Q^2)$ and $D_{1,2}^A(Q^2)$. $g_{1,2}^{p\Lambda}(Q^2)$, 
$g_{1,2}^{p \Sigma^0}(Q^2)$ and $g_{1,2}^{n \Sigma^-}(Q^2)$ are rewritten in terms of $g_{1,2}^{pn}(Q^2)$ and 
$x_{1,2}(Q^2)=\frac{F^A_{1,2}(Q^2)}{F^A_{1,2}(Q^2)+D^A_{1,2}(Q^2)}$ as
\begin{eqnarray} \label{gplam}
 g_{1,2}^{p \Lambda}(Q^2)&=& -\frac{1}{\sqrt{6}}(1+2x_{1,2}) g_{1,2}^{np} (Q^2), \\
 \label{gnsig}
 g_{1,2}^{n \Sigma^-}(Q^2)&=& (1-2x_{1,2})g_{1,2}^{np}(Q^2),  \\
 \label{gpsig}
 g_{1,2}^{p \Sigma^0}(Q^2)&=& \frac{1}{\sqrt2}(1-2x_{1,2})g_{1,2}^{np}(Q^2).
\end{eqnarray}
We further assume that $F^A_{1,2}(Q^2)$ and $D^A_{1,2}(Q^2)$ have the same $Q^2$ dependence, such that $x_{1,2}(Q^2)$ 
become constant given by $x_{1,2}(Q^2)=x_{1,2}=\frac{F^A_{1,2}(0)}{F^A_{1,2}(0)+D_{1,2}^A(0)}$. 

\item [c)] For the axial vector form factor $g_{1}^{pn}(Q^2)$, a dipole parameterization has been used:
\begin{eqnarray}\label{g1}
 g_{1}^{pn}(Q^2)=g_{A}(0)\left(1+\frac{Q^2}{M_{A}^2}\right)^{-2},
\end{eqnarray}
where $M_A$ is the axial dipole mass and $g_A(0)$ is the axial charge. For the numerical calculations, we have used the 
world average value of $M_A=1.026$ GeV. $g_A(0)$ and $x_1$ are taken to be 1.2723 and 0.364, respectively, as 
determined from the experimental data on the $\beta-$decay of neutron and the semileptonic decay of hyperons. 

\item [d)] The induced tensor form factor $g_2^{pn} (Q^2)$ is taken to be of the dipole form, i.e., 
\begin{eqnarray}\label{g2} 
 g_{2}^{pn}(Q^2)=g_{2}^{pn}(0)\left(1+\frac{Q^2}{M_{2}^2}\right)^{-2}.
 \end{eqnarray}
There is limited experimental information about $g_2^{pn} (Q^2)$ which is obtained from the analysis of the weak 
processes made for the search of G-noninvariance assuming T-invariance which implies $g_2^{pn}(0)$ to be real. A 
purely imaginary value of $g_2^{pn}(0)$ implies T-violation~\cite{Weinberg:1958ut}. In the numerical calculations we 
have taken real as well as imaginary values, with $|g_2 (0)|$ varying in the range $0-3$~\cite{Fatima:2018tzs}.

\item [e)] The pseudoscalar form factor $g_3^{NY} (Q^2)$ is proportional to the lepton mass and the contribution is 
small in the case of antineutrino scattering with muon antineutrinos. However, in the numerical calculations, we 
have taken the following expression given by Nambu~\cite{Nambu:1960xd} using the generalized GT relation. 
 \begin{eqnarray}\label{Nambu}
  g_3^{NY} (Q^2) = \frac{(M + M^\prime)^2}{2~(m_K^2 + Q^2)} g_1^{NY} (Q^2),
 \end{eqnarray}
where $m_K$ is the mass of the kaon.
\end{enumerate}

\subsubsection{Cross section}
The general expression of the differential cross section corresponding to the processes (\ref{process3}), 
(\ref{process4}) and (\ref{process5}), in the rest frame of the initial nucleon, is written as:
 \begin{eqnarray}
 \label{crosv.eq}
 d\sigma&=&\frac{1}{(2\pi)^2}\frac{1}{4E_{\bar{\nu}_\mu} M}\delta^4(k+p-k^\prime-p^\prime) \frac{d^3k^\prime}
 {2E_{k^\prime}}  \frac{d^3p^\prime}{2E_{p^\prime}} \overline{\sum} \sum |{\cal{M}}|^2,
 \end{eqnarray}
 where the transition matrix element squared is expressed as:
\begin{equation}\label{matrix}
  \overline{\sum} \sum |{\cal{M}}|^2 = \frac{G_F^2 sin^2 \theta_c}{2} \cal{L}_{\mu \nu} \cal{J}^{\mu \nu}.
\end{equation}
The leptonic ($\cal{L}_{\mu \nu}$) and the hadronic ($\cal{J}^{\mu \nu}$) tensors are given by
% {\small
\begin{eqnarray}\label{L}
\cal{L}^{\mu \nu} &=& ~\mathrm{Tr}\left[\gamma^{\mu}(1 + \gamma_{5}) \Lambda(k) \gamma^{\nu}
(1 + \gamma_{5}) \Lambda(k')\right], \\ 
\label{J}
\cal{J}_{\mu \nu} &=& \frac{1}{2} \mathrm{Tr}\left[\Lambda({p^\prime}) J_{\mu}
  \Lambda({p}) \tilde{J}_{\nu} \right], 
\end{eqnarray} 
with $\Lambda(p)=(p\!\!\!/+M)$, $\Lambda(p^{\prime})=(p\!\!\!/^{\prime}+M^{\prime})$, $\Lambda(k)=k\!\!\!/$, 
$\Lambda(k^{\prime})=(k\!\!\!/^{\prime}+m_\mu)$, $\tilde{J}_{\nu} =\gamma^0 J^{\dagger}_{\nu} \gamma^0$ and $J_\mu$ is 
defined in Eq.~(\ref{j}). 
 
 Using the above definitions, the $Q^2$ distribution is written as
\begin{equation}\label{dsig}
 \frac{d\sigma}{dQ^2}=\frac{G_F^2 ~sin^2 \theta_c}{16 \pi M^2 {E_{\bar{\nu}_\mu}}^2} N(Q^2),
\end{equation}
where the expression of $N(Q^2)$ is given in the Appendix-I of Ref.~\cite{Fatima:2018tzs}.
 
 \subsubsection{Polarization of the hyperon} 
 Using the covariant density matrix formalism, the polarization 4-vector~($\xi^\tau$) of the hyperon produced in 
 the reactions given in Eqs.~(\ref{process3}), (\ref{process4}) and (\ref{process5}) is written as~\cite{Bilenky}:
 \begin{eqnarray}\label{polar4}
\xi^{\tau}&=&\left( g^{\tau\sigma}-\frac{p'^{\tau}p'^{\sigma}}{{M^\prime}^2}\right) \frac{  {\cal L}^{\alpha \beta}  
\mathrm{Tr}\left[\gamma_{\sigma}\gamma_{5}\Lambda(p')J_{\alpha} \Lambda(p)\tilde{J}_{\beta} \right]}
{ {\cal L}^{\alpha \beta} \mathrm{Tr}\left[\Lambda(p')J_{\alpha} \Lambda(p)\tilde{J}_{\beta} \right]}.
\end{eqnarray}
 \begin{figure}
 \begin{center}  
        \includegraphics[height=5cm,width=9cm]{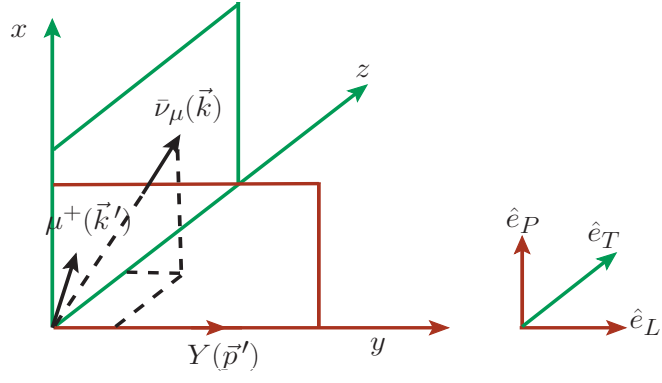}
  \caption{Polarization observables of the hyperon. $\hat{e}_{L}$, $\hat{e}_{P}$ and $\hat{e}_{T}$ represent the 
  orthogonal unit vectors corresponding to the longitudinal, perpendicular and transverse directions with respect to 
  the momentum of the final hadron.}\label{TRI}
   \end{center}
 \end{figure}
One may write the polarization vector $\vec{\xi}$ in terms of the three orthogonal vectors $\hat{e}_{i}~(i=L,P,T)$, 
i.e.
 \begin{equation}\label{polarLab}
\vec{\xi}=\xi_{L} \hat{e}_{L} + \xi_{P} \hat{e}_{P}+\xi_{T} \hat{e}_{T} ,
\end{equation}
where $\hat{e}_{L}$, $\hat{e}_{P}$ and $\hat{e}_{T}$ are chosen to be the set of orthogonal unit vectors corresponding 
to the longitudinal, perpendicular and transverse directions with respect to the momentum of the hyperon, depicted in 
Fig.~\ref{TRI}, and are written as
\begin{equation}\label{vectors}
\hat{ e}_{L}=\frac{\vec{ p}^{\, \prime}}{|\vec{ p}^{\, \prime}|},~~~~~
\hat{ e}_{P}=\hat{ e}_{L}\times \hat{ e}_T, ~~~~ 
\hat{e}_T=\frac{\vec{ p}^{\, \prime}\times \vec{ k}}{|\vec{ p}^{\, \prime}\times \vec{ k}|}.
 \end{equation}
The longitudinal, perpendicular and transverse components of the polarization vector $\vec{\xi}_{L,P,T} (Q^2)$ using 
Eqs.~(\ref{polarLab}) and (\ref{vectors}) may be written as:
% {\small
\begin{equation}\label{PL}
 \xi_{L,P,T}(Q^2)=\vec{\xi} \cdot \hat{e}_{L,P,T}~.
\end{equation}
In the rest frame of the initial nucleon, the polarization vector $\vec{\xi}$ is expressed as
\begin{equation}\label{pol2}
 \vec{\xi} = A(Q^2)~ \vec{k} + B(Q^2)~ \vec{p}^{\, \prime} + C(Q^2)~  M (\vec{k} \times \vec{p}^{\,\prime})
\end{equation}
and is explicitly calculated using Eq.~(\ref{polar4}). The expressions for the coefficients $A(Q^2)$, $B(Q^2)$ and 
$C(Q^2)$ are given in the Appendix-I of Ref.~\cite{Fatima:2018tzs}.

The longitudinal ($P_L(Q^2)$), perpendicular ($P_P(Q^2)$) and transverse ($P_T(Q^2)$) components of the polarization 
vector in the rest frame of the final hyperon are obtained by performing a Lorentz boost and are written 
as~\cite{Fatima:2018tzs}:
\begin{equation}\label{PlPp}
 P_L (Q^2) = \frac{M^\prime}{E^\prime} \xi_L (Q^2), ~~~~~~~ P_P (Q^2) = \xi_P (Q^2), ~~~~~~~ P_T (Q^2) = \xi_T (Q^2).
\end{equation}
The expressions for $P_L (Q^2)$, $P_P (Q^2)$ and $P_T (Q^2)$ are then obtained using Eqs.~(\ref{vectors}), (\ref{PL}) 
and (\ref{pol2}) in Eq.~(\ref{PlPp}) and are expressed as
\begin{eqnarray}
P_L (Q^2) &=& \frac{M^\prime}{E^\prime} \frac{A(Q^2) \vec{k} \cdot \hat{p}^{\prime} + B (Q^2) |\vec{p}^{\,\prime}|}
  {N(Q^2)},  \label{Pl} \\
P_P (Q^2) &=& \frac{A(Q^2) [(\vec{k}.\hat{p}^{\prime})^2 - |\vec{k}|^2]}{N(Q^2) ~|\hat{p}^{\prime} \times \vec{k}|},
 \label{Pp} 
 \end{eqnarray}
 \begin{eqnarray}
 P_T (Q^2) &=& \frac{C(Q^2) M |\vec{p}^{\,\prime}|[(\vec{k}.\hat{p}^{\prime})^2 - |\vec{k}|^2]}{N(Q^2)~
 |\hat{p}^{\prime}  \times \vec{k}|}. \label{Pt}
\end{eqnarray}
If the T-invariance is assumed then all the vector and the axial vector form factors are real and the expression for 
$C(Q^2)$ vanishes which implies that the transverse component of polarization, $ P_T (Q^2)$ perpendicular to the 
production plane, vanishes.

\subsection{$\Delta$ production off the free nucleon}\label{deltafree}
 In the intermediate energy region of about $0.5 - 1$~GeV, the antineutrino induced reactions on a nucleon is 
 dominated by the $\Delta$ excitation, presented in Fig.~\ref{fyn_hyp}(b) and is given by:
\begin{eqnarray}\label{eq1}
\bar\nu_\mu(k)+ n(p)&\rightarrow& \mu^{+}(k^\prime)+\Delta^{-}(p^\prime), \\
\label{eq2}
\bar\nu_\mu(k)+ p(p)&\rightarrow& \mu^{+}(k^\prime)+ \Delta^{0}(p^\prime),
\end{eqnarray}
 and the matrix element for the antineutrino induced charged current process on the free neutron is written 
 as~\cite{Athar:2007wd}:
\begin{equation}\label{eq3}
T = \sqrt{3}\frac{G_F}{\sqrt{2}}\cos{\theta_c} ~l_{\mu} ~J^{\mu},
\end{equation}
where the leptonic current $l_\mu$ is defined in Eq.~(\ref{l}) and the hadronic current $J^\mu$ is given by
\begin{eqnarray}\label{eq4}
J^\mu&=&\overline{\psi}_\alpha(p^\prime)O^{\alpha\mu} u(p).
\end{eqnarray}
In the above expression, ${\psi_\alpha}(p^\prime)$ is the Rarita Schwinger spinor for the $\Delta$ of momentum 
$p^\prime$ and $u(p)$ is the Dirac spinor for the nucleon of momentum $p$. $O^{\alpha\mu}$ is the $N-\Delta$ transition
operator which is the sum of the vector~($O^{\alpha\mu}_V$) and the axial vector~($O^{\alpha\mu}_A$) pieces, and the 
operators $O^{\alpha\mu}_V$ and $O^{\alpha\mu}_A$ are given by:
\begin{eqnarray}\label{eq5}
O^{\alpha\mu}_V&=&\left(\frac{C^V_{3}(q^2)}{M}(g^{\alpha\mu}{\not q}-q^\alpha{\gamma^\mu})+\frac{C^V_{4}(q^2)}{M^2}
(g^{\alpha\mu}q\cdot{p^\prime}-q^\alpha{p^{\prime\mu}})+\frac{C^V_5(q^2)}{M^2}(g^{\alpha\mu}q\cdot p-q^\alpha{p^\mu})
\right. \nonumber \\
&+& \left.\frac{C^V_6(q^2)}{M^2}q^\alpha q^\mu\right)\gamma_5
\end{eqnarray}
and
\begin{eqnarray}\label{eq6}
O^{\alpha\mu}_A=\left(\frac{C^A_{3}(q^2)}{M}(g^{\alpha\mu}{\not q}-q^\alpha{\gamma^\mu})+\frac{C^A_{4}(q^2)}{M^2}
(g^{\alpha\mu}q\cdot{p^\prime}-q^\alpha{p^{\prime\mu}})+C^A_{5}(q^2)g^{\alpha\mu}+\frac{C^A_6(q^2)}{M^2}q^\alpha 
q^\mu\right).
\end{eqnarray}
A similar expression for $J^\mu$ is used for the $\Delta^0$ excitation from the proton target without a factor of 
$\sqrt{3}$ in Eq.~(\ref{eq3}). Here $q(=p^\prime-p=k-k^\prime)$ is the momentum transfer, $Q^2$($=-q^2$) is the 
momentum transfer square and M is the mass of the nucleon. $C^V_i~(i =3-6)$ are the vector and $C^A_i(i=3-6)$ are the 
axial vector transition form factors which have been taken from Ref.~\cite{paschos2}.

The differential scattering cross section for the reactions given in Eqs.~(\ref{eq1}) and (\ref{eq2}) is given 
by~\cite{Akbar:2017qsf, AlvarezRuso:1998hi, ruso}:
\begin{equation}\label{eq9}
\frac{d^2\sigma}{dE_{k^\prime}d\Omega_{k^\prime}}=\frac{1}{64\pi^3}\frac{1}{MM_\Delta}\frac{|{\vec{k}^\prime|}}
{E_k}\frac{\frac{\Gamma(W)}{2}}{(W-M_\Delta)^2+\frac{\Gamma^2(W)}{4}}{|{T}|^2},
\end{equation}
where $M_\Delta$ is the mass of $\Delta$ resonance, $\Gamma$ is the Delta decay width, $W$ is the center of mass energy 
i.e. $W=\sqrt{(p+q)^2}$ and
\begin{eqnarray}
 {|{T}|^2} &=& \frac{{G_F}^2\cos^2{\theta_c}}{2}L_{\mu\nu} J^{\mu\nu} \nonumber\\
 J^{\mu\nu} &=& \overline{\Sigma}\Sigma J^{\mu\dagger} J^\nu =\frac{1}{2} Tr\left[\frac{(\not p+ M)}{2 M}
 {\tilde{\mathcal O}}^{\alpha\mu } {\it P}_{\alpha \beta}{\mathcal O}^{\beta\nu} \right].
  \end{eqnarray}
In the above expression $L_{\mu\nu}$ is given by Eq.~(\ref{L}), 
 ${\tilde{\mathcal O}}^{\alpha\mu } = \gamma^0 {{\mathcal O}^{\alpha\mu }}^\dagger \gamma^0$, 
 ${\mathcal O}^{\alpha\mu }= O^{\alpha\mu}_V + O^{\alpha\mu}_A$ and $P^{\mu\nu}$ is the 
 spin $\frac{3}{2}$ projection operator defined as \[P^{\mu\nu}=
\sum_{spins}\psi^\mu {\overline{\psi}^\nu}\] and is given by:
\begin{equation}\label{prop}
P^{\mu\nu}=-\frac{\not{p^\prime}+M_\Delta}{2M_\Delta}\left(g^{\mu\nu}-\frac{2}{3}\frac{p^{\prime\mu} p^{\prime\nu}}
{M_\Delta^2}+\frac{1}{3}\frac{p^{\prime\mu} \gamma^\nu-p^{\prime\nu} \gamma_\mu}{M_\Delta}-\frac{1}{3}\gamma^\mu
\gamma^\nu\right).
\end{equation}
In Eq.~(\ref{eq9}), the Delta decay width $\Gamma$ is taken to be an energy dependent P-wave decay width given 
by~\cite{a4}:
\begin{eqnarray}\label{gamma}
\Gamma(W)=\frac{1}{6\pi}\left(\frac{f_{\pi N\Delta}}{m_\pi}\right)^2\frac{M}{W}|{{\vec{q}}_{cm}|^3}\Theta(W-M-m_\pi),
\end{eqnarray}
where $f_{\pi N \Delta}$ is the $\pi N \Delta$ coupling constant taken as 2.12 for numerical calculations and 
$|{\vec{q}}_{cm}|$ is defined as
\[|{\vec{q}}_{cm}|=\frac{\sqrt{(W^2-m_\pi^2-M^2)^2-4m_\pi^2M^2}}{2W}.\]
The step function $\Theta$ in Eq.~(\ref{gamma}) denotes the fact that the width is zero for the invariant masses below 
the $N\pi$ threshold, ${|\vec{q}_{cm}|}$ is the pion momentum in the rest frame of the resonance.

\subsection{Pion production from the hyperons and $\Delta$}
The basic reactions for the charged current antineutrino induced one pion production off the nucleon $N$, arising from 
a hyperon in the final state are given by,
\begin{eqnarray}\label{channels_numubar_pi-_lam}
\bar \nu_l + p &\rightarrow& l^{+} + \Lambda\nonumber\\
           &&~~~~~~~                     \searrow n + \pi^o~~~\left[35.8\%\right]\nonumber\\
           &&~~~~~~~                        \searrow p + \pi^-~~~\left[63.9\%\right]\nonumber\\
\end{eqnarray}
\begin{eqnarray}\label{channels_numubar_pi-_sig0}
\bar \nu_l + p &\rightarrow& l^{+} + \Sigma^0\nonumber\\
           &&~~~~~~~~                   \searrow \gamma + \Lambda~~~\left[100\%\right]\nonumber\\
           &&~~~~~~~~~~~~~~~~~~                     \searrow n + \pi^o~~~\left[35.8\%\right]\nonumber\\
           &&~~~~~~~~~~~~~~~~~~                     \searrow p + \pi^-~~~\left[63.9\%\right]\nonumber\\
\end{eqnarray}
\begin{eqnarray}\label{channels_numubar_pi-_sig-}
\bar\nu_l + n &\rightarrow& l^{+} + \Sigma^-\nonumber\\
           &&~~~~~~~                    \searrow n + \pi^-,~~~\left[99.85\%\right]
\end{eqnarray}
where the quantities in the square brackets represent the branching ratios of the respective decay modes.

The basic reactions for the charged current neutrino and antineutrino induced one pion production off the nucleon 
$N$~(proton or neutron) through the production of $\Delta$ are:
\begin{eqnarray}\label{chan_numu_pi+}
\nu_l + p &\rightarrow& l^{-} + \Delta^{++}\nonumber\\
           &&~~~~~~~~~      \searrow p + \pi^+~~~\left[1\right],
\end{eqnarray}
\begin{eqnarray}\label{channels_pi}
\nu_l + n &\rightarrow& l^{-} + \Delta^{+},\nonumber\\
           &&~~~~~~~      \searrow p + \pi^o,~~~\left[\sqrt{\frac{2}{3}}\right]\nonumber\\
           &&~~~~~~~     \searrow n  + \pi^+,~~~\left[\sqrt{\frac{1}{3}}\right]
\end{eqnarray}
\begin{eqnarray}\label{channels_numubar_pi0}
\bar \nu_l + p &\rightarrow& l^{+} + \Delta^{0},\nonumber\\
           &&~~~~~~~~            \searrow p + \pi^-,~~~\left[\sqrt{\frac{1}{3}}\right]\nonumber\\
           &&~~~~~~~~              \searrow n  + \pi^o,~~~\left[\sqrt{\frac{2}{3}}\right]
\end{eqnarray}
\begin{eqnarray}\label{channels_numubar_pi-}
\bar \nu_l + n &\rightarrow& l^{+} + \Delta^{-},\nonumber \\
           &&~~~~~~~~          \searrow n + \pi^-,~~~\left[1\right]
\end{eqnarray}
where the quantities in the square brackets represent the respective Clebsch-Gordan coefficients for $\Delta 
\rightarrow N \pi$ channel.

\section{Nuclear medium effects}\label{NME+FSI}
\subsection{Hyperons produced inside the nucleus}\label{hyp-nu}
 When the reactions shown in Eqs.~(\ref{channels_numubar_pi-_lam}), (\ref{channels_numubar_pi-_sig0}), 
 (\ref{channels_numubar_pi-_sig-}) take place on nucleons which are bound in the nucleus, Fermi motion and Pauli 
 blocking effects of initial nucleons are considered. In the present work the Fermi motion effects are calculated in a 
 local Fermi gas model~(LFGM), and the cross section is evaluated as a function of local Fermi momentum $p_F(r)$ and 
 integrated over the whole nucleus. The incoming antineutrino interacts with the nucleon moving inside the nucleus of 
 density $\rho_N(r)$ such that the differential scattering cross section inside the nucleus is expressed in terms of 
 the  differential scattering cross section for an antineutrino scattering from a free nucleon~(Eq.~(\ref{dsig})) as 
 \begin{equation}\label{hyp-nucl}
\frac{d\sigma}{dQ^{2}}=2{\int d^3r \int 
\frac{d^3p}{{(2\pi)}^3}n_N(p,r)\left[\frac{d\sigma}{dQ^{2}}\right]_{\bar\nu N}},
\end{equation}
where $n_N(p,r)$ is the occupation number of the nucleon. $n_N(p,r)=1$ for $p\le p_{F_N} (r)$ and is equal to zero 
for $p>p_{F_N}(r)$, where $p_{F_N}(r)$ is the Fermi momentum of the nucleon and is given as:
$$
{p_F}_p(r) = \left(  3 \pi^2 \rho_p(r) \right)^\frac13;  \qquad \qquad {p_F}_n(r) = \left(  3 \pi^2 \rho_n(r) 
\right)^\frac13 ,
$$
with $\rho_p(r)$ and $\rho_n(r)$ are, respectively, the proton and the neutron densities inside the nucleus and are, 
in turn, expressed in terms of the nuclear density $\rho(r)$ as
\begin{eqnarray}
  \rho_{p}(r) &\rightarrow& \frac{Z}{A} \rho(r);  \qquad \qquad
  \rho_{n}(r) \rightarrow \frac{A-Z}{A} \rho(r). \nonumber 
 \end{eqnarray}
In the above expression, $\rho(r)$ is determined in the electron scattering experiments for the different 
nuclei~\cite{vries}.

The produced hyperons are further affected by the FSI within the nucleus through the hyperon-nucleon elastic processes 
like $\Lambda N \rightarrow  \Lambda N$, $\Sigma N \rightarrow  \Sigma N$, etc. and the charge exchange scattering
processes like $\Lambda + n \rightarrow \Sigma^- + p$, $\Lambda + n \rightarrow \Sigma^0 + n$, $\Sigma^- + p 
\rightarrow \Lambda + n$, $\Sigma^- + p \rightarrow \Sigma^0 + n$, etc. Because of such types of interaction in the 
nucleus, the probability of $\Lambda$ or $\Sigma$ production changes and has been taken into account by using the 
prescription given in Ref.~\cite{Singh:2006xp}.

\subsection{Delta produced inside the nucleus}\label{del-nu}
 When an antineutrino interacts with a nucleon (Eq.\ref{eq1}) inside a nuclear target, nuclear medium effects come 
 into play like Fermi motion, Pauli blocking, etc. The produced $\Delta$s have no such constraints in the production 
 channel but their decay is inhibited by the Pauli blocking of the final nucleons. Also, there are other disappearance 
 channels open for $\Delta$s through particle hole excitations and this leads to the modification in the mass and width 
 of the propagator defined in Eq.(\ref{gamma}).
 
 To take into account the nuclear medium effects, we have evaluated the cross section using the local density 
 approximation, following the same procedure as mentioned in section-\ref{hyp-nu}, and the differential scattering 
 cross section for the reactions given in Eqs.~(\ref{eq1}) and (\ref{eq2})is defined as :
  \begin{equation}\label{eq99}
\left. \frac{d^2\sigma}{dE_{k^\prime}d\Omega_{k^\prime}}\right|_{\bar\nu A}= \int d^3r \frac{1}{64\pi^3}\frac{1}{M
M_\Delta}\frac{|{\vec{k}^\prime|}} {E_k} \frac{\left(\frac{\tilde\Gamma(W)}{2} - Im\Sigma_\Delta\right)}{\left(W- 
M_\Delta - Re\Sigma_\Delta\right)^2 + \left(\frac{\tilde\Gamma(W)}{2} - Im\Sigma_\Delta \right)^2} ~\left(\rho_{n}(r)
~+~\frac{1}{3}\rho_{p}(r)\right) {|{T}|^2}.
 \end{equation}
In the nuclear medium the properties of $\Delta$ like its mass and decay width $\Gamma$ to be used in 
Eq.~(\ref{gamma}) are modified due to the nuclear medium effect which have been discussed in detail in Ref.~\cite{a4} 
and the modifications are given by
\begin{equation}
\frac{\Gamma}{2}\rightarrow\frac{\tilde\Gamma}{2} - Im\Sigma_\Delta~~\text{and}~~
M_\Delta\rightarrow\tilde{M}_\Delta= M_\Delta + Re\Sigma_\Delta.
\end{equation}
The expressions of $Re\Sigma_\Delta$ and $Im\Sigma_\Delta$ are given in Ref.~\cite{a4}.

\section{Final state interaction effect}\label{NME+FSI+Pion}
\subsection{Pions produced inside the nucleus}\label{hyp-nu+Pion}
 \subsubsection{Delta production}\label{deltafsi}
  When the reactions, given in Eqs.~(\ref{chan_numu_pi+})--(\ref{channels_numubar_pi-}) take place inside the nucleus, 
 the pions may be produced in two ways, through the coherent channel and the incoherent channel. If the target nucleus 
 stays in the ground state and does not loose its identity, giving all the transferred energy in the reaction to the 
 outgoing pion, then the pion production process is called coherent pion production otherwise if the nucleus can be 
 excited and/or broken up then it leads to the incoherent production of pions. The contribution of coherent pion 
 production has been found to be less than $2-3\%$ at the antineutrino energies of the present 
 interest~\cite{Katori:2016yel, Alvarez-Ruso:2014bla}, and is not discussed here. We have not considered the 
 contributions from the nonresonant background terms and higher resonances like $P_{11}(1440)$, $S_{11}(1535)$, etc.
  
The transition amplitude for an incoherent pion production process is written as~\cite{Athar:2007wd}:
\begin{equation}\label{matrix_element}
\mathcal M_{fi}=\sqrt{3}\frac{G_F}{\sqrt{2}}
\frac{f_{\pi N \Delta}}{m_{\pi}} cos\theta_{c} \bar u(p^{\prime}) k^{\sigma}_{\pi} P_{\mu\nu} 
\mathcal O^{\lambda \alpha} u(p) l_{\alpha},
\end{equation}
 where the symbols have the same meaning as in section-\ref{deltafree}.

Starting with the general expression for the differential scattering cross section in the lab frame
 \begin{eqnarray}
  d\sigma &=& \frac{1}{2E_{\bar{\nu}_\mu}2E_{N} (2\pi)^{5}} \frac{d{\vec k}^{\prime}}{ (2 E_{l})} 
\frac{d{\vec p_N\,}^{\prime}}{ (2 E_N^{\prime})} \frac{d{\vec k}_{\pi}}{ (2 E_{\pi})}
 \delta^{4}(k+p_N-k^{\prime}-p_N^{\prime}-k_{\pi}){\bar \sum} \sum|\mathcal M_{fi}|^2,\;\;\;\;\; \nonumber
 \end{eqnarray}
 and using the local density approximation, following the same procedure as mentioned in section-\ref{hyp-nu}, we may 
 write
 \begin{eqnarray}
 \left(\frac{d\sigma}{dQ^2 dcos\theta_{\pi}}\right)_{\bar\nu A} &=& \int d\vec r ~\rho_n(r)~  \left(\frac{d\sigma}{dQ^2 
 dcos\theta_{\pi}}\right)_{\bar\nu N}
 \end{eqnarray}
 which gives 
 \begin{eqnarray}\label{dsig_delta}
 \left( \frac{d\sigma}{dQ^2 dcos\theta_{\pi}} \right)_{\bar\nu A} &=& 2 \int d^3 r \sum_{N=n,p}\frac{d^3 p_N}{(2 
 \pi)^3}~\Theta_{1}(E_{F}^{N}(r)-E_N) \Theta_{2}(E_{N} + q_{0} - E_{\pi} - E_{F}^{N^\prime}(r)) \nonumber \\
 &\times& \left( \frac{d\sigma}{dQ^2 dcos\theta_{\pi}} \right)_{\bar\nu N},
 \end{eqnarray}
 where $q_0$ is the energy transferred to the target particle. 
 Using 
 $$d^3  p_N={|\vec p_N|^{2}\;d |\vec p_N|\; dcos\theta_{N}\; d\phi_{N}},$$ 
 $$E_{N}=\sqrt{|{\vec p_N}|^{2} + M^{2}};~~{\vec k}~+~{\vec p_N}~=~\vec{ k}^\prime~+~{\vec{ p}^{\,\prime}_N}~
 +~{\vec p}_\pi, $$
 and 
 $$ E^{\prime}_{N}=\sqrt{|{\vec {p}^{\,\prime}_N}|^{2} + M^{2}} = \sqrt{|\vec q - \vec p_{\pi} + \vec 
 p_{N}|^{2} + M^{2}} .$$
 Eq.~(\ref{dsig_delta}) may also be written as

\begin{eqnarray}\label{sigma_inelastic}
\left( \frac{d\sigma}{dQ^2 dcos\theta_{\pi}} \right)_{\bar\nu A} &=& \frac{1}{(4\pi)^5}\int_{r_{min}}^{r_{max}}\rho_{N}
(r) d\vec r \int^{{k^\prime}_{max}}_{{k^\prime}_{min}} dk^{\prime} \int_{0}^{2\pi}
d\phi_{\pi}~~\frac{\pi|\vec  k^{\prime}||\vec k_{\pi}|}{M E_{\bar\nu}^2 E_{l}} \frac{1}{E_{p}^{\prime}+E_{\pi}\left(1-
\frac{|\vec q|}{|\vec k_{\pi}|}\cos\theta_{\pi }\right)} \nonumber \\
&\times&{\overline \sum} \sum|\mathcal M_{fi}|^2,~~~~~
\end{eqnarray}
where $\rho_{N}(r)$ is the nucleon density defined in terms of nuclear density $\rho(r)$. In a nucleus, the 
contributions to $\pi^-$ and $\pi^o$ productions come from the neutron and proton targets. These are taken into account 
using the Clebsch-Gordan coefficients written in Eqs.~(\ref{chan_numu_pi+})--(\ref{channels_numubar_pi-}). The total 
production cross section for  $\pi^-$ and $\pi^o$  from a nucleus can be written by replacing $\rho_N$ as
\begin{eqnarray}\label{rho}
 \rho_{N}(r)&=&~\rho_{n}(r)~+~\frac{1}{9}\rho_{p}(r) \qquad \text{ for $\pi^-$ production,} \nonumber \\
 \rho_{N}(r)&=&~\frac{2}{9}\left[\rho_{n}(r)~+~\rho_{p}(r)\right] \qquad \text{ for $\pi^o$ production}.
\end{eqnarray}

The pions produced in these processes inside the nucleus may re-scatter or may produce more pions or may get absorbed 
while coming out from the final nucleus. We have taken the results of Vicente Vacas et al.~\cite{a5} for the final 
state interaction of pions which is calculated in an eikonal approximation using probabilities per unit length as the 
basic input. The details are given Ref.~\cite{a5}.

 \subsubsection{Hyperon production}
 The pions produced as a result of hyperon decays are shown in Eqs.~(\ref{channels_numubar_pi-_lam}), 
 (\ref{channels_numubar_pi-_sig0}) and (\ref{channels_numubar_pi-_sig-}). However, when the hyperons are produced in a 
 nuclear medium, some of them disappear through the hyperon-nucleon interaction processes like $YN \rightarrow NN$, 
 though it is suppressed due to nuclear effects~\cite{Holstein, Oset:1989ey}. The pionic modes of hyperons are Pauli 
 blocked as the momentum of the nucleons available in these decays is considerably below the Fermi level of energy for 
 most nuclei leading to a long lifetime for the hyperons in the nuclear medium~\cite{Holstein, Oset:1989ey}. Therefore, 
 the hyperons which survive the $YN \rightarrow NN$ decay in the medium live long enough to travel the nuclear medium 
 and decay outside the nucleus. In view of this we have assumed no final state interaction of the produced pions with 
 the nucleons inside the nuclear medium. In a realistic situation, all the hyperons produced in these reactions will 
 not survive in the nucleus, and the pions coming from the decay of hyperons will undergo FSI~\cite{Singh:2006xp}. A 
 quantitative analysis of the hyperon disappearance through the $YN \rightarrow NN$ interaction and the pions having 
 FSI effect, will require a dynamic nuclear model to estimate the nonmesonic and mesonic decay of the hyperons in a 
 nucleus which is beyond the scope of the present work. Our results in the following section, therefore, represent an 
 upper limit on the production of pions arising due to the production of hyperons.
 
 \section{Results and Discussion}\label{results}
  In this section, we first present a review of the old experimental results on the total cross sections and their 
  $Q^2$ dependence in the case of hyperon production from CERN on Freon~\cite{Eichten:1972bb} and 
  Propane~\cite{Erriquez:1977tr, Erriquez:1978pg}, FNAL on Neon~\cite{Ammosov:1986jn, Ammosov:1986xv}, SKAT on 
  Freon~\cite{Brunner:1989kw} and BNL on H$_2$~\cite{Fanourakis:1980si} and compare them with the theoretical results. 
  We also present the experimental results on the Lambda hyperon polarizations from CERN~\cite{Erriquez:1978pg} and 
  compare them with the most recent theoretical calculations~\cite{Akbar:2016awk}. The theoretical calculations used 
  for making comparisons with the experimental results are done for the nucleon targets assuming the nuclear medium 
  effects in hyperon production to be small at the energies relevant for these experiments~\cite{Erriquez:1978pg, 
  Singh:2006xp}. The results have also been presented for the total cross sections and their $Q^2$ dependence for 
  nuclear targets like $^{12}$C, $^{16}$O, $^{40}$Ar and $^{208}$Pb with and without the nuclear medium~(NME) and 
  final state interaction~(FSI) effects. The results have also been presented for the longitudinal~($P_L(Q^2)$), 
  perpendicular~($P_P(Q^2)$) and transverse~($P_T(Q^2)$) components of the polarization vector of the hyperon in the 
  presence of the second class currents with and without T-invariance by taking the numerical values of $g_2(0)$ to be 
  real and imaginary, respectively.
  \vspace{2mm}
 
 The following points describe the inputs used for the numerical calculations which have been done to obtain these 
 results: 
 \begin{enumerate}
 \item For the hyperon production cross section off the free nucleon target, we have integrated over $Q^2$ in 
 Eq.~(\ref{dsig}) and obtained the results for the total scattering cross section. For the $\Delta$ production cross 
 section off the free nucleon target in the charged current neutrino and antineutrino induced reactions, we have used 
 Eq.~(\ref{eq9}) and integrated over the final lepton kinematical variables.

  \item In the presence of nuclear medium effects the expression of the cross sections given in Eqs.(\ref{hyp-nucl}) 
  and (\ref{eq99}), respectively for the Y production and the $\Delta$ production, have been used. In the case of 
  hyperon production FSI arising due to the quasielastic and charge exchange hyperon nucleon scattering has been 
  taken into account as described in section-\ref{hyp-nu}.
    
  \item For the pion production cross section from the hyperons, we have used the same expression~(Eq.(\ref{hyp-nucl}))
  with the hyperon-nucleon interaction. For the pions arising from the $\Delta$ decay with NME+FSI, we have used 
  Eq.~(\ref{sigma_inelastic}) with the pion FSI effect as described in section-\ref{deltafsi}. Therefore, FSI effect in 
  the case of pion production from the hyperons is different from the FSI effect for the pion production from the 
  $\Delta$ i.e. there is no pion absorption in the case of hyperons giving rise to pions, whereas there is pion 
  absorption inside the nucleus when $\Delta$s give rise to pions. 
  
  \item  We have used $\rho_\text{p}(r)=\frac{Z}{A}\rho(r)$ for the proton density, and $\rho_\text{n}(r)=\frac{A-Z}{A}
  \rho(r)$ for the neutron density, where $\rho(r)$ is nuclear density taken as 3-parameter Fermi density for $^{12}$C, 
  $^{16}$O, $^{40}$Ar and $^{208}$Pb given by:
  \[\rho(r)=\frac{\rho_0\left(1+w\frac{r^2}{c^2}\right)}{\left(1+exp\left(\frac{r-c}{z}\right)\right)},\]
 with the density parameters $c=2.355$ fm, $z=0.5224$ fm and $w=-0.149$ for $^{12}$C, $c=2.608$ fm, $z=0.513$ fm and 
 $w=-0.051$ for $^{16}$O, $c=3.73$ fm, $z=0.62$ fm and $w=-0.19$ for $^{40}$Ar and $c=6.624$ fm, $z=0.549$ fm and 
 $w=0$ for $^{208}$Pb and have been taken from Ref.~\cite{vries}. 
  
 \item The results for the longitudinal~($P_L(Q^2)$), perpendicular~($P_P(Q^2)$) and transverse~($P_T(Q^2)$) components 
 of the polarization of the $\Lambda$ hyperon have been obtained using Eqs.~(\ref{Pl}), (\ref{Pp}) and (\ref{Pt}) 
 respectively in the presence of second class currents with and without T-invariance. For the axial vector form 
 factors, the expressions used in Eqs.~(\ref{gplam}), (\ref{gnsig}) and (\ref{gpsig}) for $g_{1}^{NY} (Q^2)$ and 
 $g_{2}^{NY} (Q^2)$ have been used while the parameterization of BBBA05 for the nucleon form factors as they appear in 
 Eqs.~(\ref{fplambda}), (\ref{fnsigma}) and (\ref{fpsigma}) for $f_{1}^{NY} (Q^2)$ and $f_{2}^{NY} (Q^2)$ have been 
 used. For the pseudoscalar form factor $g_3^{NY} (Q^2)$, Nambu's parameterization given in Eq.~(\ref{Nambu}) has been 
 used. 
 
 \item We have also presented the results for the flux averaged cross section and the polarization observables in order 
 to compare our results with the experimental results. For this, we have integrated the differential cross section 
 ${d\sigma}/{dQ^2}$ and polarization observables $P_L(Q^2)$ and $P_P(Q^2)$ over $E_{\bar{\nu}_{_\mu}}$ and $Q^2$ 
 distributions to obtain the total cross section $\langle \sigma \rangle $ defined as: 
 \begin{equation}\label{flux_sigma}
  \langle \sigma \rangle = \frac{\int^{E_{max}}_{E_{th}}\int^{Q^2_{max}}_{Q^2_{min}} \frac{d\sigma}{dQ^2}~dQ^2 
  \Phi(E_{\bar \nu_{_\mu}}) dE_{\bar \nu_{_\mu}} }{\int^{E_{max}}_{E_{min}}\Phi(E_{\bar \nu_{_\mu}}) 
  dE_{\bar \nu_{_\mu}} }
 \end{equation}
 and components of hyperon polarization  $\langle P_{L,P} \rangle$ defined as:
\begin{equation}\label{flux_pol}
\langle P_{L,P} \rangle = \frac{1}{\langle \sigma \rangle} \int_{E_{th}}^{E_{max}} \int^{Q^2_{max}}_{Q^2_{min}} 
P_{L,P}(Q^2,E_{\bar \nu_{_\mu}})~ \frac{d\sigma}{dQ^2}~dQ^2 \Phi(E_{\bar \nu_{_\mu}}) d E_{\bar \nu_{_\mu}}.  
\end{equation} 

  \end{enumerate}
  
   \subsection{Hyperon and Delta productions from free nucleons}
 \subsubsection{Hyperon production}
\begin{figure}
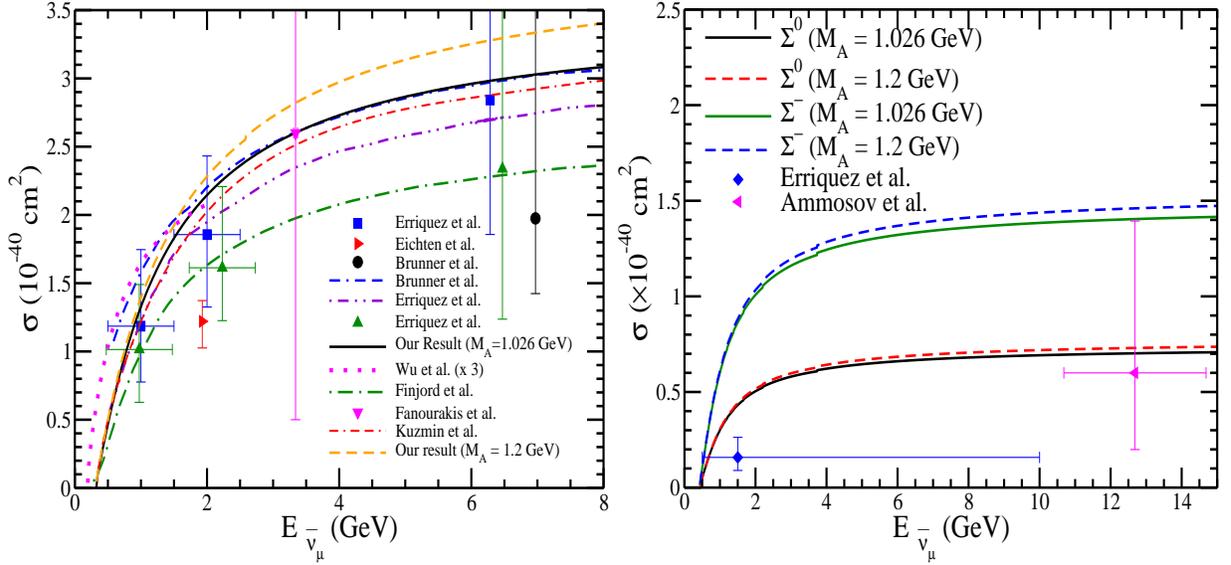

 \begin{center}
    \includegraphics[height=7.5cm,width=8cm]{xsec_comp2.eps} 
    \includegraphics[height=7.5cm,width=8cm]{xsctn_sig0_vs_enu.eps} 
    \end{center}
  \caption{$\sigma$ vs. $E_{\bar{\nu}_\mu}$ for the $\Lambda$ production~(left panel), $\Sigma^{0}$ and $\Sigma^{-}$ 
  production~(right panel) cross sections. Solid~(dashed) line represents the result using $M_A = 1.026~(1.2)$ GeV. 
  Experimental results for the process $\bar \nu_\mu p \to \mu^+ \Lambda$ (triangle right~\cite{Eichten:1972bb}, 
  triangle up~\cite{Erriquez:1977tr}, square~\cite{Erriquez:1978pg}, triangle down($\sigma = 2.6^{+5.9}_{-2.1} \times 
  10^{-40} cm^2$)~\cite{Fanourakis:1980si}, circle~\cite{Brunner:1989kw}) and for the process $\bar \nu_\mu p \to \mu^+ 
  \Sigma^0$ (diamond~\cite{Erriquez:1977tr}) are shown with error bars. Theoretical curves are of Kuzmin and 
  Naumov~\cite{Kuzmin:2008zz}(double dashed-dotted line), Brunner et al.~\cite{Brunner:1989kw}(dashed line), Erriquez 
  et al.~\cite{Erriquez:1978pg}(dashed-double dotted line) obtained using Cabibbo theory with axial vector dipole mass 
  as 0.999GeV, 1.1 GeV and 1 GeV, respectively, while the results of Wu et al.~\cite{Wu:2013kla}(dotted line) and 
  Finjord and Ravndal~\cite{Finjord:1975zy}(dashed dotted line) are obtained using quark model.}
   \label{free_hyperon}
 \end{figure}

  In Fig.~\ref{free_hyperon}, we have presented the results for the hyperon production cross sections from the free 
 nucleons presented in Eqs.~(\ref{process3})--(\ref{process5}) as a function of antineutrino energies. These results 
 are presented for the $\Lambda$, $\Sigma^-$ and $\Sigma^0$ cross sections at the two values of $M_A$ viz. 
 $M_A = 1.026$ GeV and 1.2 GeV. We find that in this region there is very little dependence of $M_A$ on the cross 
 section in the case of $\Sigma^-$ and $\Sigma^0$ production, while in the case of $\Lambda$ production, the cross 
 section increases with energy and the increase is about 5$\%$ at $E_{\bar{\nu}_\mu} = 1$ GeV. In the case of free 
 nucleon, the cross sections for $\bar \nu_\mu + n \rightarrow \mu^+ + \Sigma^- $ and $\bar \nu_\mu + p \rightarrow 
 \mu^+ +  \Sigma^0$ are related by a simple relation i.e. $\sigma(\bar \nu_\mu p \rightarrow \mu^+ \Sigma^0) = \frac12 
 \sigma(\bar \nu_\mu n \rightarrow \mu^+ \Sigma^-)$, while no $\Sigma^+$ is produced off the free nucleon target due to 
 $\Delta S \neq \Delta Q$ rule. A comparison is made with available experimental results from 
 CERN~\cite{Eichten:1972bb, Erriquez:1977tr,Erriquez:1978pg}, BNL~\cite{Fanourakis:1980si}, FNAL~\cite{Ammosov:1986jn, 
 Ammosov:1986xv} and SKAT~\cite{Brunner:1989kw} experiments as well as with the theoretical calculations performed by 
 Wu et al.\cite{Wu:2013kla} and Finjord and Ravndal~\cite{Finjord:1975zy} using quark model and the calculations 
 performed by Erriquez et al.~\cite{Erriquez:1978pg}, Brunner et al.\cite{Brunner:1989kw} and Kuzmin and 
 Naumov~\cite{Kuzmin:2008zz} based on the prediction from Cabibbo theory. A reasonable agreement with the experimental 
 results can be seen.
  
       \begin{figure}
 \begin{center}
    \includegraphics[height=6cm,width=5.5cm]{dsigma_dq2_lambda_sigma_enu_500MeV.eps} 
    \includegraphics[height=6cm,width=5.5cm]{dsigma_dq2_lambda_sigma_enu_750MeV.eps} 
    \includegraphics[height=6cm,width=5.5cm]{dsigma_dq2_lambda_sigma_enu_1GeV.eps}     
    \end{center}
  \caption{$d \sigma/dQ^2$ vs. $Q^2$ for the hyperon production cross sections at different antineutrino energies viz. 
  $E_{\bar{\nu}_\mu} = 0.5$ GeV~(left panel), 0.75 GeV~(middle panel) and 1 GeV~(right panel). Lines and points have 
  the same meaning as in Fig~\ref{free_hyperon}.}
   \label{free_hyperon_dq2}
 \end{figure}
 
 \begin{figure}
  \begin{center}
   \includegraphics[height=6.5cm,width=10cm]{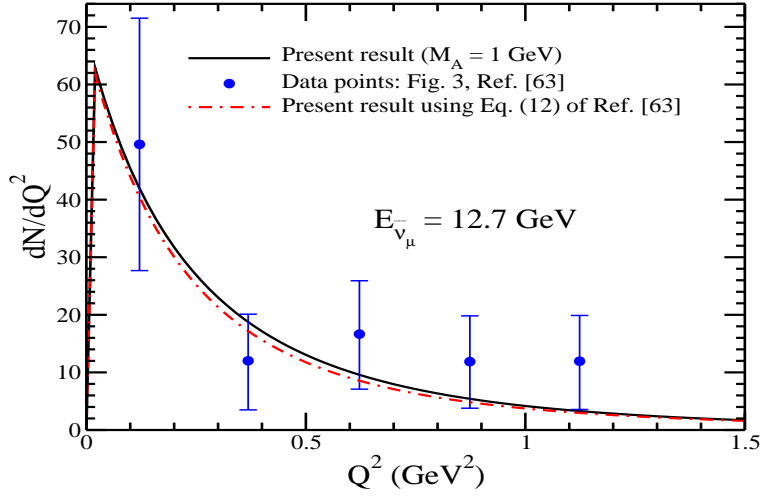}
  \end{center}
\caption{Comparison of present results for the $Q^2$ distribution with the results given in Fig. 3 of 
Ref.~\cite{Ammosov:1986jn}. Solid line represents the present results using $M_{A}$ = 1 GeV, dashed-dotted line 
represents the present results obtained using Eq.~(12) of Ref.~\cite{Ammosov:1986jn} and the data points are taken from 
Ref.~\cite{Ammosov:1986jn}.}\label{Ammosov}
 \end{figure}

  \begin{figure}
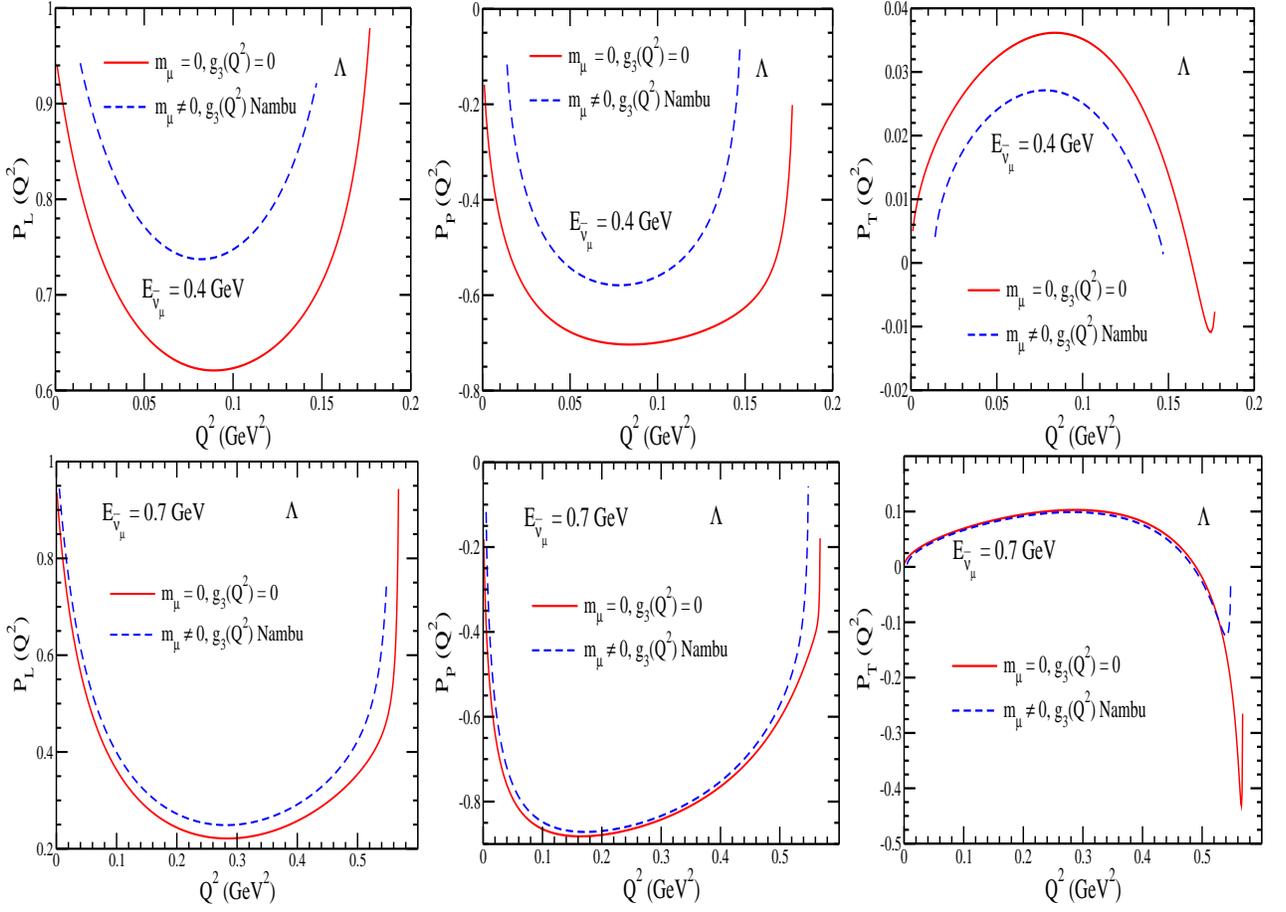

 \begin{center}
    \includegraphics[height=6cm,width=5.5cm]{Pl_Fp_variation_with_g2_1_lambda_enu_400MeV.eps} 
    \includegraphics[height=6cm,width=5.5cm]{Pp_Fp_variation_with_g2_1_lambda_enu_400MeV.eps}
    \includegraphics[height=6cm,width=5.5cm]{Pt_Fp_variation_with_g2_1_lambda_enu_400MeV.eps} \\
    \includegraphics[height=6cm,width=5.5cm]{Pl_Fp_variation_with_g2_1_lambda_enu_700MeV.eps} 
    \includegraphics[height=6cm,width=5.5cm]{Pp_Fp_variation_with_g2_1_lambda_enu_700MeV.eps}
    \includegraphics[height=6cm,width=5.5cm]{Pt_Fp_variation_with_g2_1_lambda_enu_700MeV.eps}   
       \end{center}
  \caption{$P_L (Q^2)$~(left panel), $P_P (Q^2)$~(middle panel) and $P_T (Q^2)$~(right panel) for the process 
  $ \bar{\nu}_\mu + p \rightarrow \mu^+ + \Lambda$ at $E_{\bar{\nu}_\mu} = 0.4$ GeV~(upper panel) and $0.7$ GeV~(lower 
  panel) with $m_\mu = 0,~g_3(Q^2)$ = 0~(solid line) and $m_\mu \neq 0$ and $g_3(Q^2)$ given by Nambu~(dashed line).}
  \label{pol-1}
 \end{figure}
       \begin{figure}
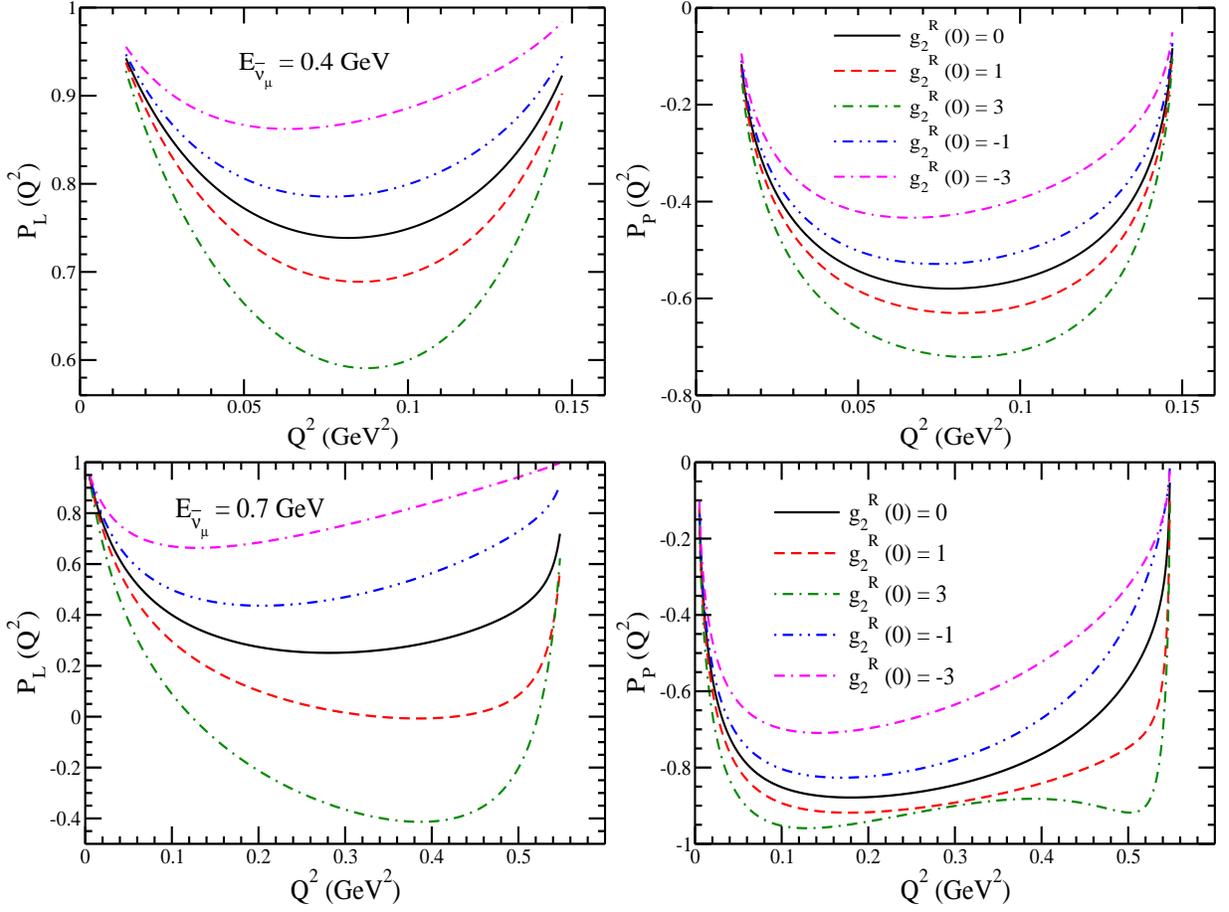

 \begin{center}
    \includegraphics[height=6cm,width=8cm]{Pl_real_g2_variation_lambda_enu_400MeV.eps} 
    \includegraphics[height=6cm,width=8cm]{Pp_real_g2_variation_lambda_enu_400MeV.eps} \\
    \includegraphics[height=6cm,width=8cm]{Pl_real_g2_variation_lambda_enu_700MeV.eps}    
    \includegraphics[height=6cm,width=8cm]{Pp_real_g2_variation_lambda_enu_700MeV.eps}
       \end{center}
  \caption{$P_L (Q^2)$~(left panel) and $P_P (Q^2)$~(right panel) for the process $ \bar{\nu}_\mu + p \rightarrow \mu^+ 
  + \Lambda$ at $E_{\bar{\nu}_\mu} = 0.4$ GeV~(upper panel) and 0.7 GeV~(lower panel) with $g_2^R(0) = $ 0~(solid 
  line), 1~(dashed line), 3~(dash-dotted line), -1~(dash-double-dotted line) and -3~(double-dash-dotted line).}
  \label{pol-2}
 \end{figure} 
 \begin{figure}
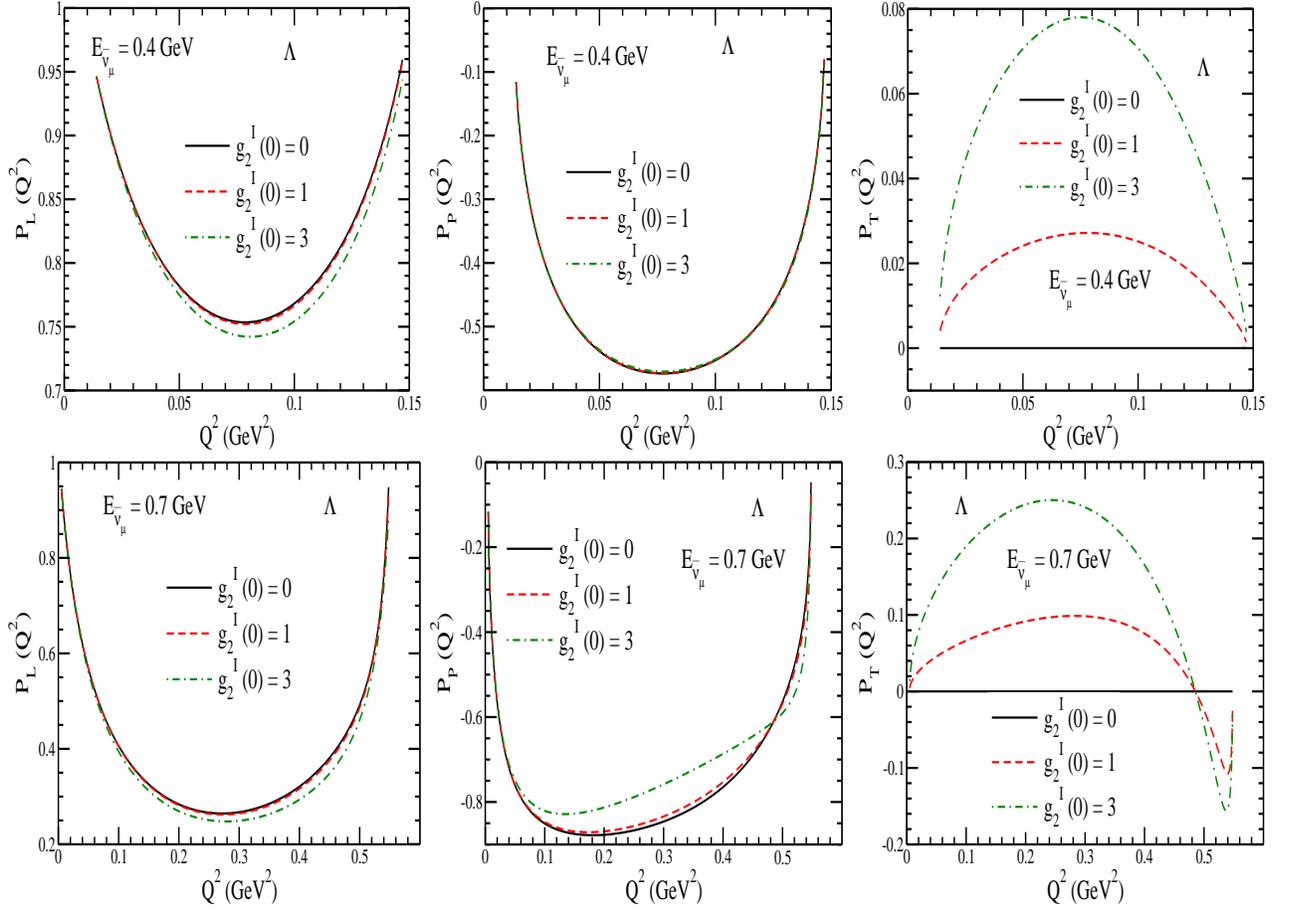

 \begin{center}
    \includegraphics[height=6cm,width=5.5cm]{Pl_enu_400MeV_g2_variation_with_Fp_lambda.eps} 
    \includegraphics[height=6cm,width=5.5cm]{Pp_enu_400MeV_g2_variation_with_Fp_lambda.eps}
    \includegraphics[height=6cm,width=5.5cm]{Pt_enu_400MeV_g2_variation_with_Fp_lambda.eps} \\
    \includegraphics[height=6cm,width=5.5cm]{Pl_g2_variation_with_Fp_enu_700MeV_lambda.eps} 
    \includegraphics[height=6cm,width=5.5cm]{Pp_g2_variation_with_Fp_enu_700MeV_lambda.eps}
    \includegraphics[height=6cm,width=5.5cm]{Pt_g2_variation_with_Fp_enu_700MeV_lambda.eps}    
       \end{center}
  \caption{$P_L (Q^2)$~(left panel), $P_P (Q^2)$~(middle panel) and $P_T (Q^2)$~(right panel) for the process 
  $ \bar{\nu}_\mu + p \rightarrow \mu^+ + \Lambda$ at $E_{\bar{\nu}_\mu} = 0.4$ GeV~(upper panel) and 0.7 GeV~(lower 
  panel) with $g_2^I(0) = $ 0~(solid line), 1~(dashed line) and 3~(dash-dotted line).}
  \label{pol-3}
 \end{figure}
 
  In Fig.~\ref{free_hyperon_dq2}, the results are presented for the differential cross section ($d\sigma/dQ^2$) as a 
 function of $Q^2$ for the $\Lambda$ and $\Sigma^-$ produced in the final state at the different antineutrino energies 
 viz. $E_{\bar{\nu}_\mu} = 0.5$ GeV, 0.75 GeV and 1 GeV at the two values of $M_A$ viz. $M_A = 1.026$ GeV and 1.2 GeV. 
 One may notice that the $Q^2$-distribution is not much sensitive to the choice of $M_A$.
 
 In Fig.~\ref{Ammosov}, we have presented the comparison of present results for the $Q^2$ distribution with the 
 results given in Fig. 3 of Ref.~\cite{Ammosov:1986jn}. Solid line represents the present results using $M_{A}$ = 1 
 GeV, dashed-dotted line represents the present results obtained using Eq.~(12) of Ref.~\cite{Ammosov:1986jn} and the 
 data points are taken from Ref.~\cite{Ammosov:1986jn}. We have multiplied our results with an arbitrary factor of 7 
 in order to compare our result with the experimental data points. 
 
 In order to compare with the experimental results of CERN experiment~\cite{Erriquez:1978pg}, we have performed the 
  numerical calculations for the flux averaged cross section $\langle \sigma \rangle$, longitudinal $\langle P_L 
  \rangle$ and perpendicular $\langle P_P \rangle$ polarization components relevant for the antineutrino flux of SPS 
  antineutrino beam of Gargamelle experiment at CERN~\cite{Armenise:1979zg} and present our results in 
  Table-\ref{mema_ff}. The results are compared with the available experimental results from CERN~\cite{Erriquez:1977tr,
  Eichten:1972bb,Erriquez:1978pg} experiment and the theoretical results quoted by Erriquez et 
  al.~\cite{Erriquez:1978pg}.

    \begin{table}[htbp]
\begin{tabular}{|c|c|c|c|}\hline\hline
& $\langle P_L \rangle$ & $\langle P_P \rangle $\footnote{One may note that, for present work we have considered the 
sign convention for perpendicular polarization which is opposite to that of used by Erriquez et 
al.~\cite{Erriquez:1978pg}.} &  $\langle \sigma \rangle$ $\times \;(10^{-40}$ cm$^2)$\\\hline
{\bf Experiments}&&&\\
Erriquez et al.~\cite{Erriquez:1978pg}   & -0.06$\pm$ 0.44 & 1.05 $\pm$ 0.30   & ~~~ 2.07 $\pm$ 0.75~~~~\\
Erriquez et al.~\cite{Erriquez:1977tr}   & -- &  -- &   1.40 $\pm$ 0.41(Propane)\\ 
Eichten et al.~\cite{Eichten:1972bb} &  --    &   --   & 1.3 $\pm ^{0.9}_{0.7}$(Freon) \\ \hline
{\bf Theory}&&&\\
Present work($M_A = 0.84$ GeV)& 0.10&--0.75&2.00\\
($M_A = 1.026$ GeV)&0.05 &  --0.85  & 2.15 \\
($M_A = 1.2$ GeV)&0.03&--0.89&2.31\\
Erriquez et al.~\cite{Erriquez:1978pg}($M_A = 0.84$ GeV)& 0.14 &0.73 &  2.07\\\hline\hline
\end{tabular}
\caption{Flux averaged cross section $\langle \sigma \rangle$(using Eq.~\ref{flux_sigma}), longitudinal $\langle P_L 
\rangle$ and perpendicular $\langle P_P \rangle$ components of polarization(using Eq.~\ref{flux_pol}) are given for the
process $\bar \nu_\mu p \to \mu^+ \Lambda$.}
  \label{mema_ff}
\end{table}
   
 Experimentally, one may get information about the polarization of hyperons through the structure of the angular 
 distribution of the pions, which are produced by the hyperon decay via. $Y \to N \pi$. The observation of the 
 components of the polarization provide an alternative method to determine the axial dipole mass, $M_A$, nature of the 
 second class current (whether with or without TRI) and the pseudoscalar form factor independent of the total and the 
 differential scattering cross sections. Moreover, the experimental observation of the transverse component of 
 polarization can be used to study the physics of T-violation. In Fig.~\ref{pol-1}, we have made an attempt to explore 
 the possibility of determining the pseudoscalar form factor $g_3^{NY}(Q^2)$ in $|\Delta S|=1$ sector and studied the 
 sensitivity of the $Q^2$-dependence on the polarization components $P_L(Q^2)$, $P_P(Q^2)$ and $P_T(Q^2)$ using the 
 expression of Nambu~\cite{Nambu:1960xd} in Eq.~(\ref{Nambu}) for the process $\bar\nu_\mu p \to \mu^+\Lambda$ at 
 $E_{\bar \nu_{_\mu}}$=0.4 and 0.7 GeV. We see that at $E_{\bar \nu_{_\mu}}=0.4$ GeV, $P_L(Q^2)$, $P_P(Q^2)$ and 
 $P_T(Q^2)$ are sensitive to $g_3^{NY}(Q^2)$, but with the increase in energy the difference in the results obtained 
 with and without $g_3^{NY}(Q^2)$ are almost the same. It seems, therefore, possible in principle, to determine the 
 pseudoscalar form factor in the $\Lambda$ polarization measurements at lower antineutrino energies. The total cross 
 section $\sigma$ and the differential cross section $d \sigma/dQ^2$ are not found to be very sensitive to the values 
 of $g_3^{NY} (Q^2)$ and are not shown here~\cite{Akbar:2016awk}.
 
 For the reaction ${\bar{\nu}_\mu + p \rightarrow \mu^+ + {\Lambda}}$, we have also studied the dependence of the 
 polarization components on the second class currents with T-invariance and showed the results for $P_L (Q^2)$ and 
 $P_P (Q^2)$ as a function of $Q^2$ in Fig.~\ref{pol-2}. These results are presented for the polarization components 
 using the second class current form factor in the presence of T invariance i.e. using the real values of $g_2^{np} (0) 
 = g_2^{R} (0) =$ 0, $\pm$1 and $\pm$3 and $M_2 = M_A$ in Eq.~(\ref{g2}) at the different values of $E_{\bar{\nu}_{\mu}}
 =$ 0.4 and 0.7~GeV. We find that $P_L (Q^2)$ shows large variations as we change $|g_2^{R} (0)|$ form 0 to 3 at high 
 antineutrino energies, $E_{\bar{\nu}_\mu}$ (say 0.7 GeV) in comparison to the lower energies (say 0.4 GeV). For 
 example, in the peak region of $Q^2$, the difference is 80$\%$ at $E_{\nu_{\mu}}$ = 0.7 GeV and it is 20$\%$ at 
 $E_{\nu_{\mu}}$ = 0.4 GeV as $g_2^R (0)$ is changed from 0 to 3. In the case of $P_P (Q^2)$ also, the $Q^2$ dependence 
 is quite strong and similar to $P_L (Q^2)$. 
 
 In Fig.~\ref{pol-3}, the results are presented for $P_L (Q^2)$, $P_P (Q^2)$ and $P_T (Q^2)$ as a function of $Q^2$ in 
 the presence of the second class current without T-invariance, using the imaginary values of the induced tensor form 
 factor i.e. $g_2^{np} (0) = i~g_2^{I} (0)$, where $g_2^I (0) =$ 0, 1 and 3, at the different values of 
 $E_{\bar{\nu}_{\mu}} =$ 0.4 and 0.7~GeV. We see that  while $P_L(Q^2)$ is less sensitive to $g_2^I (0)$ at 
 $E_{\bar{\nu}_{\mu}}$ in the range $0.4-0.7$ GeV. $P_P(Q^2)$ is almost insensitive to $g_2^{I}(0)$ at the lower 
 $E_{\bar{\nu}_{\mu}}$, say at $E_{\bar{\nu}_{\mu}} =$ 0.4 GeV. However, at higher antineutrino energies, say at 
 $E_{\bar{\nu}_{\mu}} =$ 0.7 GeV, $P_P(Q^2)$ is sensitive to $g_2^{I}(0)$. Moreover, $P_T(Q^2)$ is sensitive to 
 $g_2^I(0)$ at all antineutrino energies. $P_T(Q^2)$ shows 8$\%$ and 25$\%$ variations at $Q^2 = $ 0.08, and 0.25 
 GeV$^2$ at $E_{\bar{\nu}_{\mu}} =$ 0.4 and 0.7 GeV, respectively, when $g_2^{I} (0)$ is varied from 0 to 3.
   \begin{figure}
 \begin{center}
    \includegraphics[height=7.5cm,width=11cm]{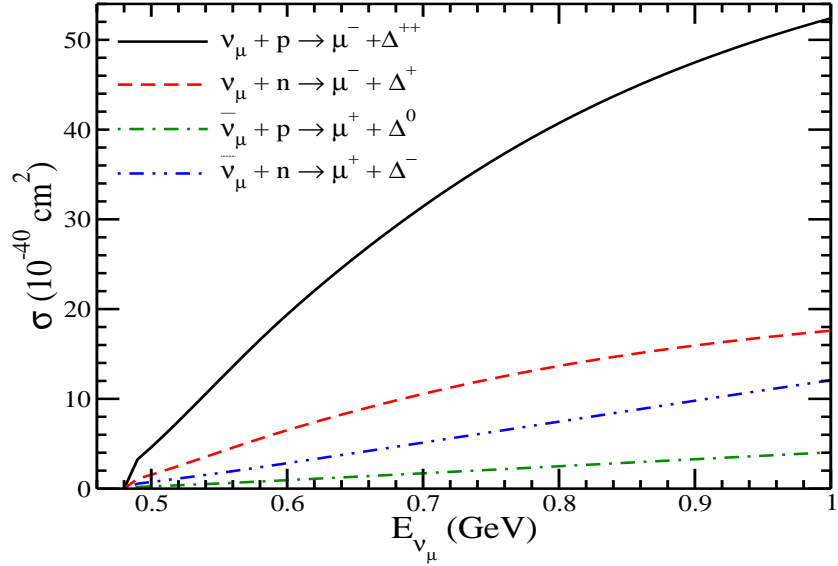} 
    \end{center}
  \caption{$\sigma$ vs. $E_{{\nu}_\mu}$ for the Delta production cross sections. Solid line represents the result 
  for $\Delta^{++}$ production cross section, dashed line represents the result for $\Delta^+$ cross section, 
  dash-dotted line represents the result for $\Delta^0$ production and the dash-double-dotted line represents the 
  result for the $\Delta^-$ production cross section.}
   \label{free_delta}
 \end{figure}
        \begin{figure}
 \begin{center}
    \includegraphics[height=6cm,width=5.5cm]{dsigma_dq2_delta_free_500MeV.eps} 
    \includegraphics[height=6cm,width=5.5cm]{dsigma_dq2_delta_free_750MeV.eps} 
    \includegraphics[height=6cm,width=5.5cm]{dsigma_dq2_delta_free_1GeV.eps}     
    \end{center}
  \caption{$d \sigma/dQ^2$ vs. $Q^2$ for the $\Delta$ production cross sections at the different antineutrino energies 
  viz. $E_{{\nu}_\mu} = 0.5$ GeV~(left panel), 0.75 GeV~(middle panel) and 1 GeV~(right panel). Lines and points 
  have the same meaning as in Fig.~\ref{free_delta}.}
   \label{free_delta_dq2}
 \end{figure}
       \begin{figure}
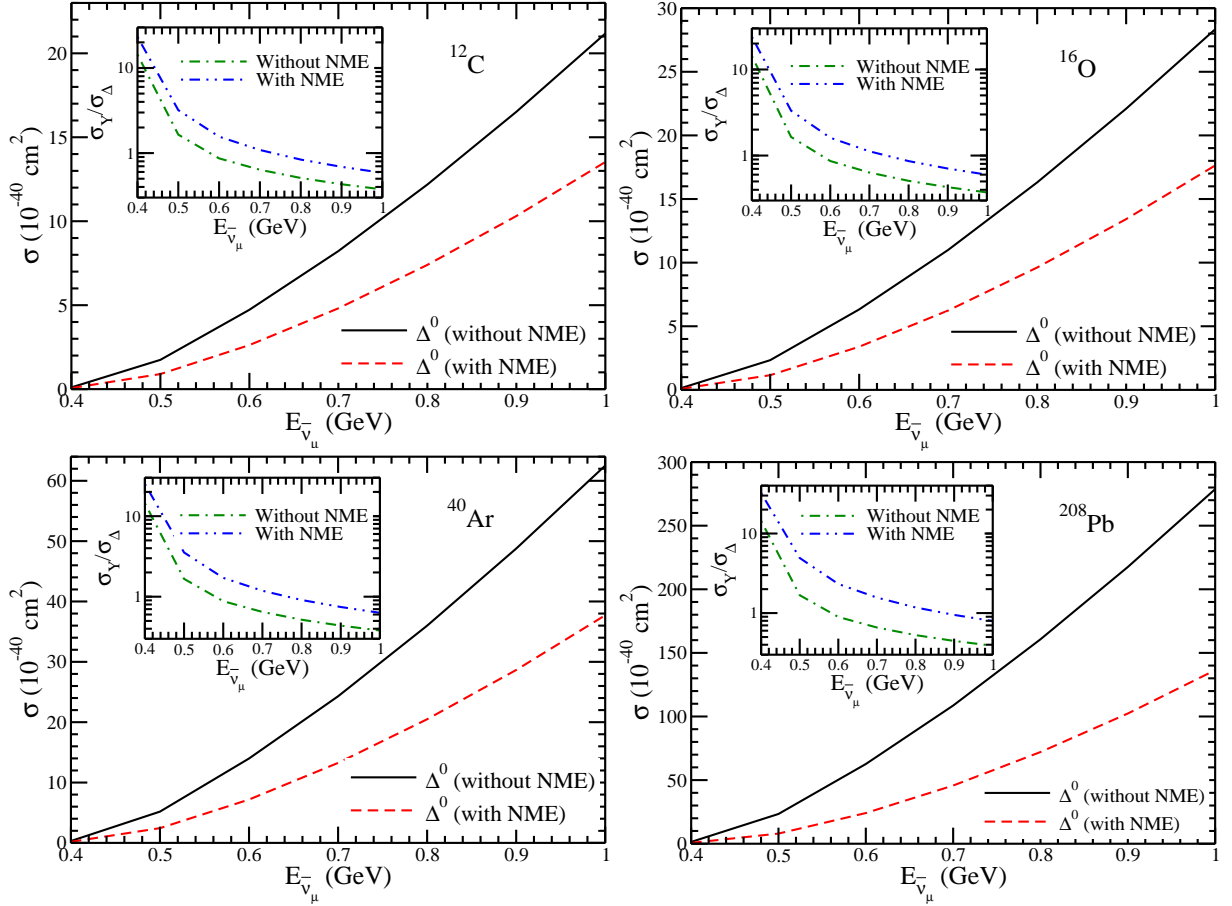

 \begin{center}
    \includegraphics[height=6cm,width=8cm]{total_sigma_delta0_NME_carbon.eps} 
    \includegraphics[height=6cm,width=8cm]{total_sigma_delta0_NME_oxygen.eps}\\
    \includegraphics[height=6cm,width=8cm]{total_sigma_delta0_NME_argon.eps}
    \includegraphics[height=6cm,width=8cm]{total_sigma_delta0_NME_lead.eps}
       \end{center}
  \caption{$\sigma$ vs. $E_{\bar{\nu}_\mu}$ for the $\Delta^0$ production cross sections without~(solid line) and 
  with~(dashed line) NME for the various nuclear targets. Ratio, $\sigma_Y/\sigma_{\Delta^0}$ vs. $E_{\bar{\nu}_\mu}$ 
  without~(dash-dotted line) and with~(dash-double-dotted) NME for the various nuclear targets are presented in the 
  inset.}
   \label{sigma_nucleus}
 \end{figure}
     \begin{figure}
 \begin{center}
    \includegraphics[height=7.5cm,width=11cm]{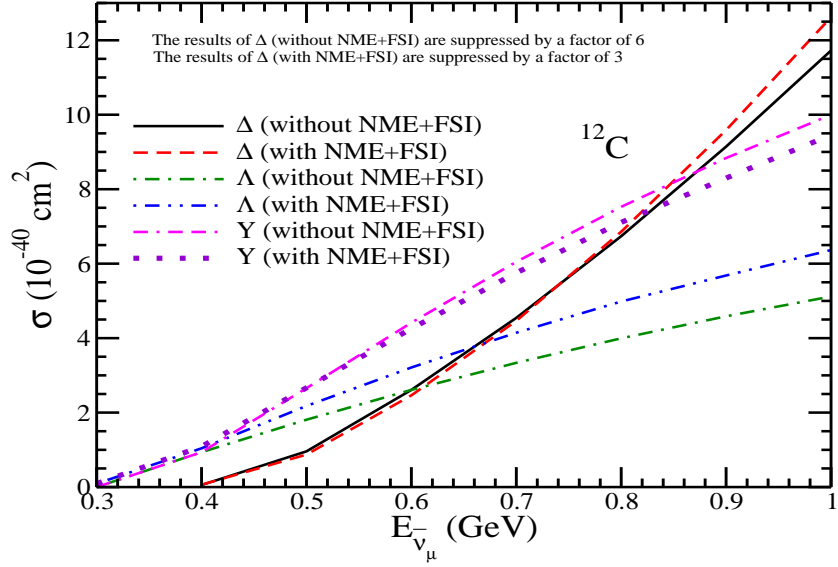} 
    \end{center}
  \caption{Results for the charged current $\pi^-$ production in $^{12}$C with and without NME+FSI.
  The results are presented for the pion production from the $\Delta$, $\Lambda$ and total hyperon $Y(=\Lambda + 
  \Sigma)$ with and without NME+FSI. Notice that the results of $\Delta$ without NME+FSI are suppressed by a factor of
  6 and the results with NME+FSI are suppressed by a factor of 3 to bring them on the same scale.}
   \label{c12-ME}
 \end{figure}
   \begin{figure}
 \begin{center}
    \includegraphics[height=7.5cm,width=11cm]{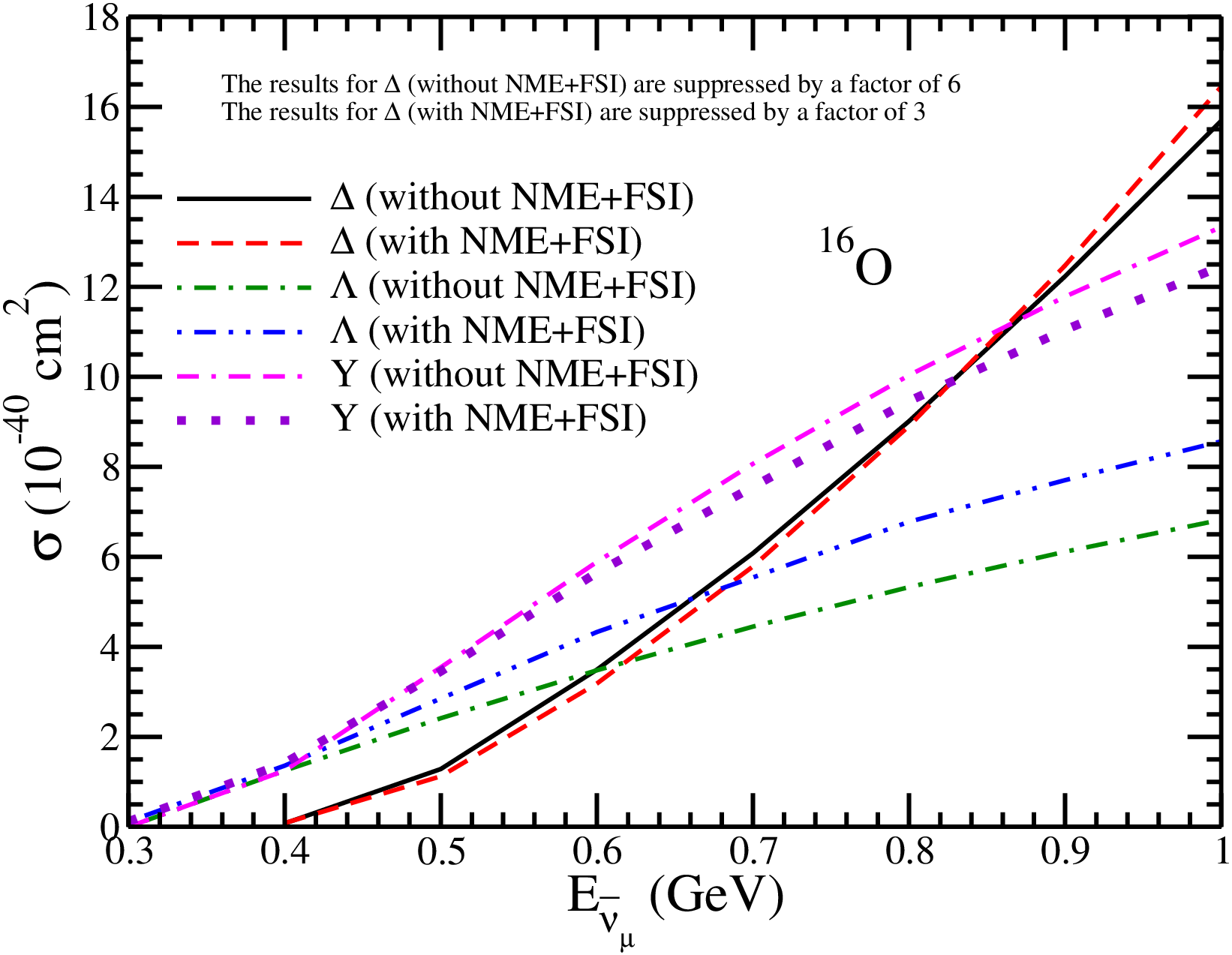}
       \end{center}
  \caption{Results for $\pi^-$ production in $^{16}$O. Lines and points have the same meaning as in Fig.\ref{c12-ME}.}
   \label{o16-ME}
 \end{figure}
  \begin{figure}
 \begin{center}
    \includegraphics[height=7.5cm,width=11cm]{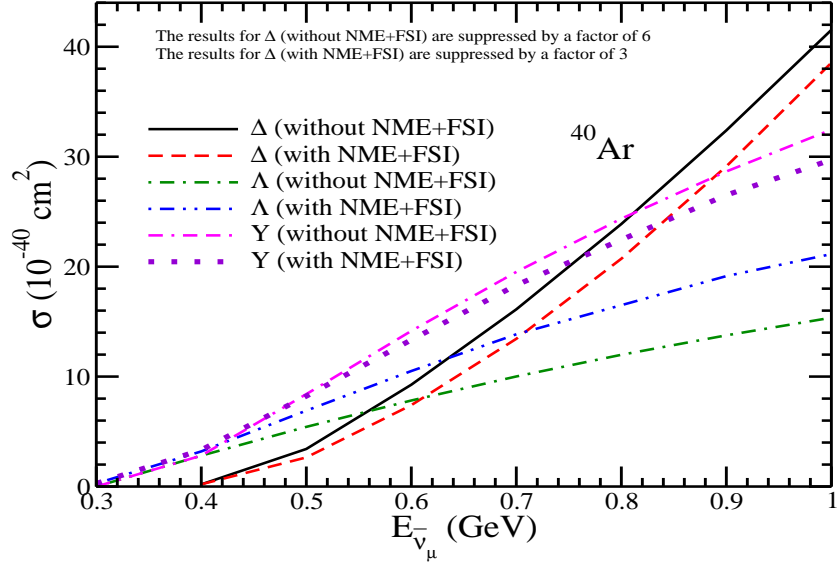} 
      \end{center}
  \caption{Results for $\pi^-$ production in $^{40}$Ar. Lines and points have the same meaning as in Fig.\ref{c12-ME}.}
   \label{ar40-ME}
 \end{figure}
  \begin{figure}
 \begin{center}
    \includegraphics[height=7.5cm,width=11cm]{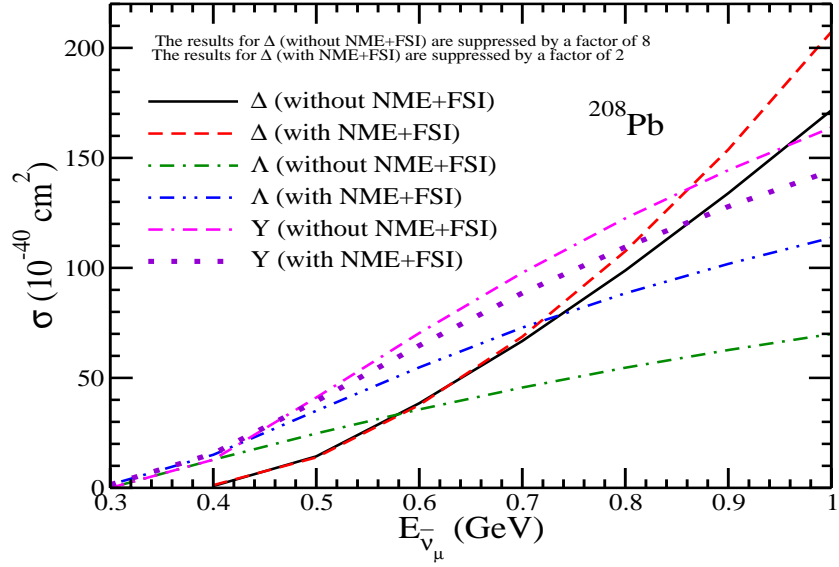}
     \end{center}
  \caption{Results for $\pi^-$ production in $^{208}$Pb. Lines and points have the same meaning as in Fig.\ref{c12-ME}. 
  Notice that the results of $\Delta$ without NME+FSI are suppressed by a factor of 8 and the results with NME+FSI are 
  suppressed by a factor of 2 to bring them on the same scale.}
   \label{pb208-ME}
 \end{figure}
 \begin{figure}   
 \begin{center}
    \includegraphics[height=7.5cm,width=11cm]{total_sigma_carbon_pi0_hyperon+delta_corrected.eps}
    \end{center}
  \caption{Results for the charged current $\pi^o$ production in $^{12}$C with and without NME+FSI. The results are 
  presented for the pion production from $\Delta$, $\Lambda$ and total hyperon $Y=\Lambda + \Sigma$ with and without 
  NME+FSI. Notice that the results of $\Delta$ without NME+FSI are suppressed by a factor of 3 and the results with 
  NME+FSI are suppressed by a factor of 2.}
   \label{c12-ME-NC}
 \end{figure}
  \begin{figure}
 \begin{center}
    \includegraphics[height=7.5cm,width=11cm]{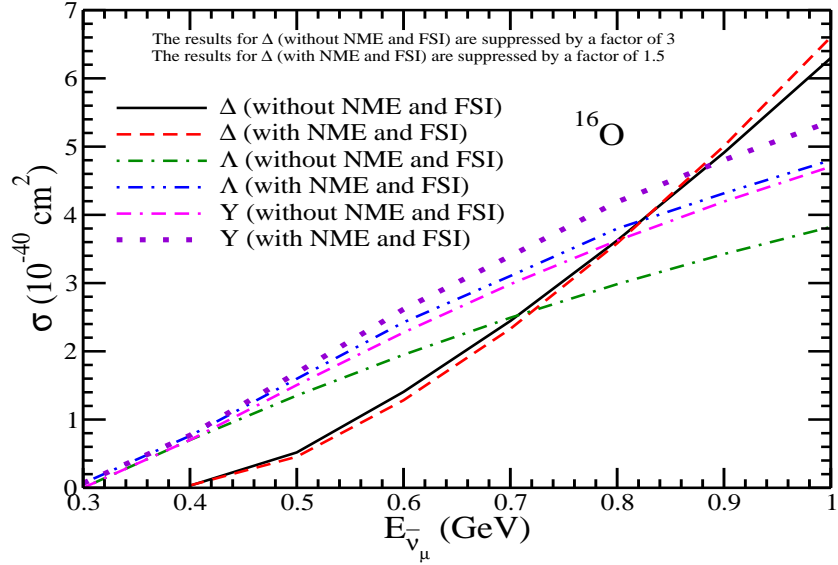} 
      \end{center}
  \caption{Results for $\pi^o$ production in $^{16}$O. Lines and points have the same meaning as in 
  Fig.~\ref{c12-ME-NC}. Notice that the results of $\Delta$ without NME+FSI are suppressed by a factor of 3 and the 
  results with NME+FSI are suppressed by a factor of 1.5.}
   \label{o16-ME-NC}
 \end{figure}
   \begin{figure}
 \begin{center}
    \includegraphics[height=7.5cm,width=11cm]{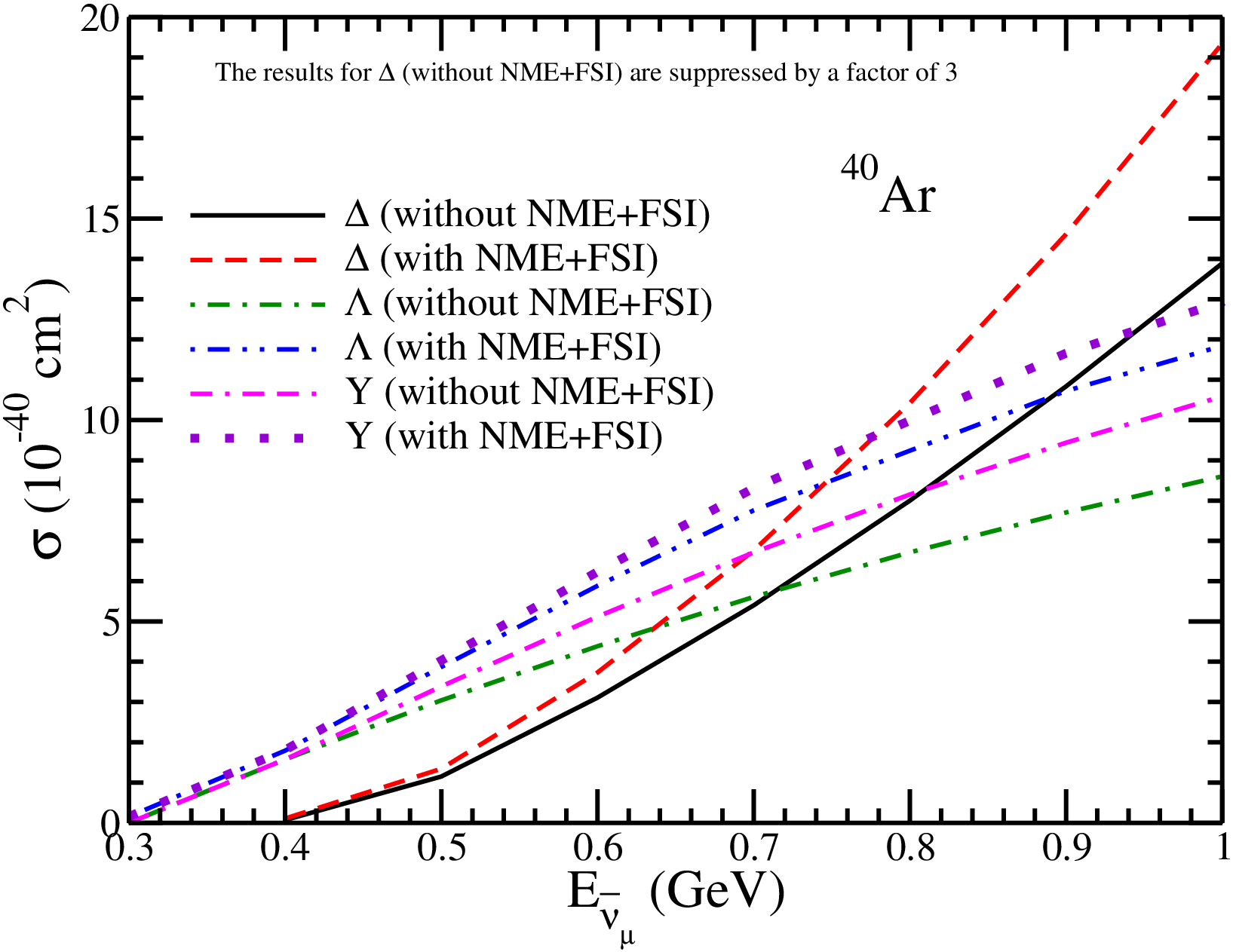} 
       \end{center}
  \caption{Results for $\pi^o$ production in $^{40}$Ar. Lines and points have the same meaning as in 
  Fig.~\ref{c12-ME-NC}. Notice that the results of $\Delta$ without NME+FSI are suppressed by a factor of 3.}
   \label{ar40-ME-NC}
 \end{figure}
   \begin{figure}
 \begin{center}
    \includegraphics[height=7.5cm,width=11cm]{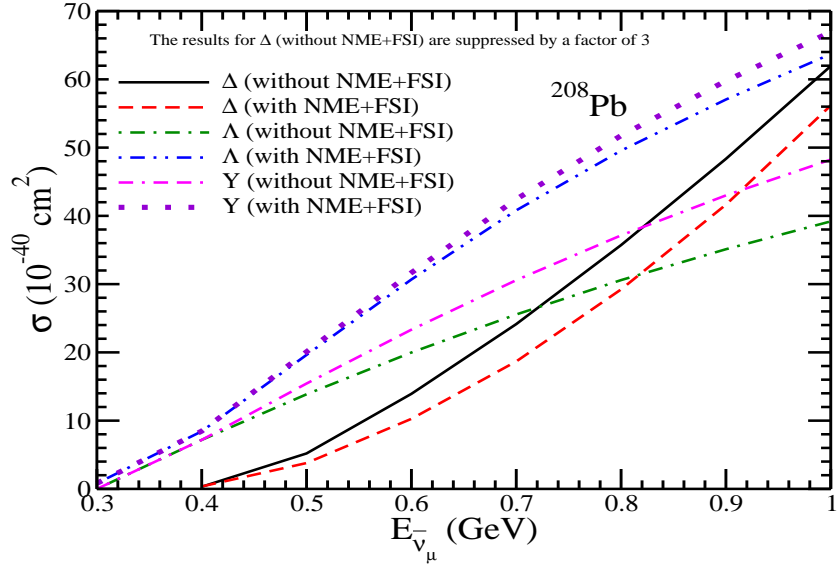}
       \end{center}
  \caption{Results for $\pi^o$ production in $^{208}$Pb. Lines and points have the same meaning as in 
  Fig.\ref{c12-ME-NC}. Notice that the results of $\Delta$ without NME+FSI are suppressed by a factor of 3.}
   \label{pb208-ME-NC}
 \end{figure}
     \begin{figure}
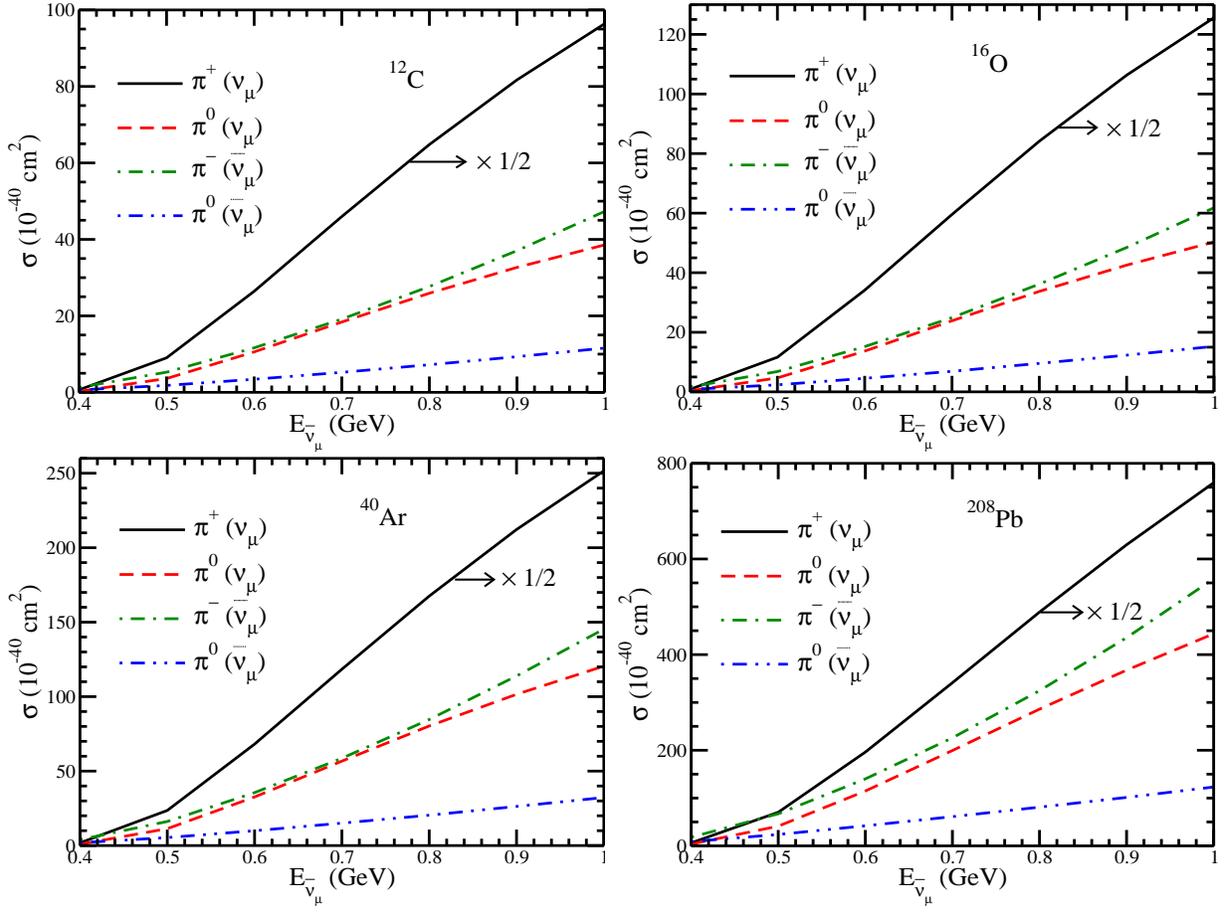

 \begin{center}
    \includegraphics[height=6cm,width=8cm]{total_sigma_pion_carbon.eps} 
    \includegraphics[height=6cm,width=8cm]{total_sigma_pion_oxygen.eps}\\
    \includegraphics[height=6cm,width=8cm]{total_sigma_pion_argon.eps}
    \includegraphics[height=6cm,width=8cm]{total_sigma_pion_lead.eps}
       \end{center}
  \caption{Results for the $\nu_\mu$ induced $\pi^+$ and $\pi^o$ production cross sections and ${\bar\nu}_\mu$ induced 
  $\pi^-$ and $\pi^o$ production cross sections. For $\nu_\mu$ scattering the contribution to the pions is coming from 
  the $\Delta$ only, while for ${\bar\nu}_\mu$ scattering the contribution to the pions is coming from the $\Delta$ as 
  well as the hyperons. The results are presented for $^{12}$C, $^{16}$O, $^{40}$Ar and $^{208}$Pb with NME+FSI. Notice 
  that $\pi^+$ production cross section has been reduced by half to bring it on the same scale.}
   \label{c-o-ar-pb}
 \end{figure}
 
 \subsubsection{${\Delta}$ production}
   The results for the $\Delta$ production cross sections are presented in Fig.~\ref{free_delta} for $\nu_\mu$ and 
   $\bar \nu_{_\mu}$ induced processes off the free nucleon target. For $\nu_\mu$ induced reaction, the leptonic 
   current in Eq.~(\ref{l}) will read as $l_\mu = \bar{u} (k^\prime) \gamma_\mu (1-\gamma_5) u(k)$. In the case of 
   $\Delta$ production, the cross sections are inhibited by the threshold effect at lower energies, and there is almost 
   no cross section until $E_{\bar{\nu}_\mu}$ = 0.4 GeV. In Fig.~\ref{free_delta_dq2}, the results for $d\sigma/dQ^2$ 
   are presented for the $\Delta$s produced in the final state at the different (anti)neutrino energies viz. 
   $E_{\nu_\mu, {\bar \nu_{_\mu}}}$ = 0.5, 0.75 and 1 GeV. The total cross section and its energy dependence as well as 
   the $Q^2$ dependence have been discussed elsewhere in literature~\cite{Athar:2007wd, prd1, Alam:2013cra, 
   Alam:2014bya, ruso}.
   
  \subsection{Hyperon and Delta production from nuclei}
  In Fig.~\ref{sigma_nucleus}, we have presented the results for $\sigma$ vs $E_{\bar{\nu}_\mu}$, for the $\Delta^0$ 
  produced off the proton bound in various nuclear targets like $^{12}$C, $^{16}$O, $^{40}$Ar and $^{208}$Pb with and 
  without the NMEs. It may be noticed that the NMEs due to the modification of the $\Delta$ properties in nuclei reduce 
  the cross section. In the case of lighter nuclei like $^{12}$C and $^{16}$O, this reduction is about $\sim$ 35$\%$ at 
  $E_{\bar \nu_{_\mu}}$=1 GeV. The reduction in the cross section increases with the increase in the nuclear mass 
  number and decreases with the increase in energy. For example, it becomes $\sim 40\%$ and 50$\%$ in $^{40}$Ar and 
  $^{208}$Pb, respectively at $E_{\bar \nu_{_\mu}} =1$ GeV. We find that the NMEs due to Pauli blocking and Fermi 
  motion effects, in the case of hyperons in the final state, are negligibly small and therefore we have not discussed 
  these effects and the results of the cross sections are almost the same as for the free hyperon 
  case~(Fig.~\ref{free_hyperon}). Moreover, when the hyperon-nucleon interaction i.e. the FSI effect in the hyperon 
  production, is taken into account the overall change in the hyperon production cross section is very small. These 
  results are used to obtain the ratio of total hyperon to $\Delta$ production cross sections i.e. $\frac{\sigma_Y}
  {\sigma_\Delta}$ which have been shown in the inset of these figures. It may be noticed that due to the threshold 
  effect initially the hyperon production cross section dominates and with the increase in energy the ratio reduces. 
  Due to the substantial reduction in the cross section for the $\Delta$ production, the ratio increases when NME is 
  taken into account in comparison to the free case. Moreover, this ratio is larger in heavy nuclei like $^{208}$Pb as 
  NME increases with the nucleon number.
  
   \subsection{Pion production}
   In this section the results are presented for the $\pi^-$ and $\pi^o$ productions respectively in the nuclei like 
   $^{12}$C, $^{16}$O, $^{40}$Ar and $^{208}$Pb. We give a preview of our main results for $\pi^-$ and $\pi^o$ 
   productions before they are presented in detail in Figs. Figs.~\ref{c12-ME}-\ref{pb208-ME} and 
   Figs.~\ref{c12-ME-NC}-\ref{pb208-ME-NC} for each case. These results are shown for the cross sections obtained 
   without and with NME+FSI effect for the pion production arising due to the $\Lambda$ production, total hyperon(Y) 
   production and the $\Delta$ production. In the case of hyperon production, NMEs in the production process as well as 
   the FSI due to hyperon-nucleon interactions have been taken into account. Moreover, we do not include the FSI of 
   pions within the nuclear medium which are produced as a result of hyperon decays. This is because the decay width of 
   pionic decay modes of hyperons is highly suppressed in the nuclear medium. Due to which these hyperons live long 
   enough to pass through the nucleus and decay outside the nuclear medium. Thus, the produced pions are less affected 
   by the strong interaction of nuclear field. This is not the case with the pion produced through strong decays of 
   $\Delta$, as they are further suppressed by the strong absorption of pions in the nuclear medium. Therefore, in the 
   low energy region the Cabibbo suppression in the case of pion production through hyperons get compensated by the 
   threshold suppression as well as by the strong pion absorption effects in the case of the pions produced through the 
   Delta excitation. On the other hand, FSI due to $\Sigma-N$ and $\Lambda-N$ interactions in various channels tend to 
   increase the $\Lambda$ production cross section and decrease the $\Sigma^{-}$ production cross section, which is 
   mainly a threshold effect. The quantitative increase~(decrease) in $\Lambda(\Sigma)$ yield due to FSI increases with 
   the increase in the nucleon number. The interaction of hyperons with the nucleons bound inside the nucleus, 
   separately affect $\Sigma^-$ and $\Sigma^0$ productions and the relation $ \sigma \left( \bar \nu_\mu + p \rightarrow
   \mu^+ + \Sigma^0   \right) = \frac12  \sigma \left(  \bar \nu_\mu + n \rightarrow \mu^+ + \Sigma^- \right)$ which 
   holds for the free case, does not hold for the case of nuclear targets. We must point out that although $\Sigma^+$ 
   is not produced from a free nucleon but can be produced through the final state interactions like $\Lambda p 
   \rightarrow \Sigma^+ n$ and $\Sigma^0 p \rightarrow \Sigma^+ n$, albeit the contributions would be small.  
  
  Using the results of $\sigma$, we have obtained the results for the ratio of hyperon to Delta production cross 
  sections, with and without NME+FSI, for $\pi^-$ as well as $\pi^o$ productions for all the nuclear targets 
  considered here by defining 
  \begin{equation}\label{r_without}
 \left. {\it R}_N={\frac{\sigma(Y \rightarrow N\pi)}{\sigma(\Delta \rightarrow N\pi)}}\right 
 \rvert_{\text{without NME+FSI effects}}
 \end{equation}
and
  \begin{equation}\label{r_with}
  \left. {\it R}_A=\frac{\sigma(Y \rightarrow N\pi)}{\sigma(\Delta \rightarrow N\pi)}\right 
  \rvert_{\text{with NME+FSI effects}}.
 \end{equation}
 This ratio directly tells us the enhancement of the ratio $R_A$ due to NME+FSI with the increase in the mass number of 
 the nuclear targets as the pions getting produced through the $\Delta$-resonant channel undergo a suppression due to 
 NME+FSI effect, while the pions getting produced from the hyperons~(all the interactions taken together i.e. $\Lambda$ 
 as well as $\Sigma$ contributions) have comparatively small NME+FSI effect.
  
  In Fig.~\ref{c12-ME} and Fig.~\ref{o16-ME}, we have presented the results for the total scattering cross section 
  $\sigma$ vs $E_{{\bar\nu}_\mu}$, for ${\bar\nu}_\mu$ scattering off the nucleon in $^{12}$C and $^{16}$O nuclear 
  targets giving rise to $\pi^-$. The results are presented for the pion production from $\Delta$, $\Lambda$ and $Y$ 
  with and without NME and FSI. In the case of hyperon production for $^{12}$C, the effect of FSI due to $Y-N$ 
  interaction increases the $\Lambda$ production cross section from the free case by about $23-24\%$ for 
  $E_{{\bar\nu}_\mu}=0.6 - 1$ GeV, while the change in the total hyperon production cross section results in a decrease 
  in the cross section due to the FSI effect which is about $3 - 5\%$ at these energies. We find that in the case of 
  pions produced through $\Delta$ excitations, NME+FSI lead to a reduction of around 50$\%$ in the $\pi^-$ production 
  for the antineutrino energies 0.6 $< E_{\bar\nu_\mu} <$ 1GeV. This results in the change in the ratio of 
  ${\it R}_N$~(Eq.~(\ref{r_without})) from 0.28 and 0.14 respectively at $ E_{\bar\nu_\mu}$=0.6 and 1GeV to 
  ${\it R}_A$~(Eq.~(\ref{r_with})) $\rightarrow$ 0.58 and 0.25 at these energies. In the case of $^{16}$O nuclear 
  target the observations are similar to what has been discussed above in the case of $^{12}$C nuclear target.
 
 In Fig.~\ref{ar40-ME}, we have presented the results for $\sigma$ vs $E_{{\bar\nu}_\mu}$, for ${\bar\nu}_\mu$ 
 scattering off $^{40}$Ar nuclear target. In the case of $\Lambda$ production, the effect of FSI leads to an increase 
 in the cross section by about $34-38\%$ for $E_{{\bar\nu}_\mu}=0.6 - 1$~GeV, however, the overall change in the 
 $\pi^-$ production from the hyperons results in a net reduction in the cross section from the free case, which is 
 about $6 - 8\%$ at these energies. In the case of pions produced through $\Delta$ excitations, NME+FSI leads to a 
 reduction of around $55 - 60\%$ in the $\pi^-$ production for the antineutrino energies 0.6 $\le E_{\bar\nu_\mu}\le$ 
 1 GeV, and the reduction is less at higher energies. This results in the change in the ratio of ${\it R}_N$ from 
 0.25 and 0.13 respectively at $ E_{\bar\nu_\mu} =0.6$ and 1 GeV to ${\it R}_A$, 0.6 and 0.26 at the corresponding 
 energies.  
 
 In the case of heavy nuclear target like $^{208}$Pb, the change in the cross section due to NME+FSI is 
 quite large and the results for $\sigma$ vs $E_{{\bar\nu}_\mu}$, for ${\bar\nu}_\mu$ scattering off the nucleon in 
 $^{208}$Pb nuclear target are shown in Fig.\ref{pb208-ME}. For example, the reduction in the cross section due to 
 NME+FSI when a $\Delta$ is produced as the resonant state, is about 75$\%$ at $ E_{\bar\nu_\mu}=$ 0.6 GeV and 70$\%$ 
 at $ E_{\bar\nu_\mu}=$ 1 GeV from the cross sections calculated without the medium effect. The enhancement in the 
 $\Lambda$ production cross section is about $55 - 60\%$ at these energies. While the overall change in the 
 $\pi^-$ production from the hyperons results in a net reduction which is about $8 - 12\%$. This results in the change 
 in the ratio of ${\it R}_N$ from 0.23 and 0.12 respectively at $ E_{\bar\nu_\mu}=0.6$ and 1 GeV to 
 ${\it R}_A~\rightarrow$ 0.86 and 0.35, respectively.    
 
 In Figs.~\ref{c12-ME-NC}, \ref{o16-ME-NC}, \ref{ar40-ME-NC} and \ref{pb208-ME-NC}, we have presented the results for 
 the total scattering cross section $\sigma$ vs $E_{{\bar\nu}_\mu}$, for ${\bar\nu}_\mu$ scattering off nucleon in 
 $^{12}$C, $^{16}$O, $^{40}$Ar and $^{208}$Pb nuclear targets giving rise to $\pi^o$. These results are presented for 
 the pion production from $\Delta$, $\Lambda$ and $Y$ with and without NME+FSI. In the case of $\pi^o$ arising due to 
 hyperon decay, the contribution comes from the $\Lambda$ and $\Sigma^0$ decay, while there is no contribution from 
 $\Sigma^-$. Due to the FSI effect there is substantial increase in the $\Lambda$ production cross section and 
 reduction in the $\Sigma^0$ production cross section from the free case, which leads to an overall increase in the 
 $\pi^o$ production. Therefore, unlike the $\pi^-$ production where there is overall reduction, in the case of $\pi^o$ 
 production there is an increase in the cross section which is about $13 - 14\%$ in $^{12}$C and $^{16}$O, $22 - 23\%$ 
 in $^{40}$Ar and $26 - 38\%$ in $^{208}$Pb for $E_{\bar\nu_\mu}$ = 0.6 to 1 GeV. The different Clebsch-Gordan 
 coefficients for $\Delta$~(in Eqs.~(\ref{chan_numu_pi+})--(\ref{channels_numubar_pi-})) and the branching ratios for 
 the hyperons~(in Eqs.~(\ref{channels_numubar_pi-_lam}), (\ref{channels_numubar_pi-_sig0}), 
 (\ref{channels_numubar_pi-_sig-})) give a different ratios of ${\it R}_N$ and ${\it R}_A$. This results in the change 
 in the ratio of ${\it R}_N$ from 0.58 and 0.26 respectively at $ E_{\bar\nu_\mu} = $ 0.6 and 1 GeV to ${\it R}_A~~
 \rightarrow$ 1.3 and 0.5 in $^{12}$C and $^{16}$O, from 0.55 and 0.25 respectively at $ E_{\bar\nu_\mu}=0.6$ and 1 GeV 
 to 1.68 and 0.66 in $^{40}$Ar, and from 0.56 and 0.26 respectively at $ E_{\bar\nu_\mu}=0.6$ and 1 GeV to 3 and 1.2 in 
 $^{208}$Pb. Thus, in the case of $\pi^o$ production, there is significant increase in $Y \rightarrow N \pi$ to 
 $\Delta \rightarrow N \pi$ ratio when NME+FSI are taken into account specially in the case of heavier nuclear targets.
 
  \begin{figure}
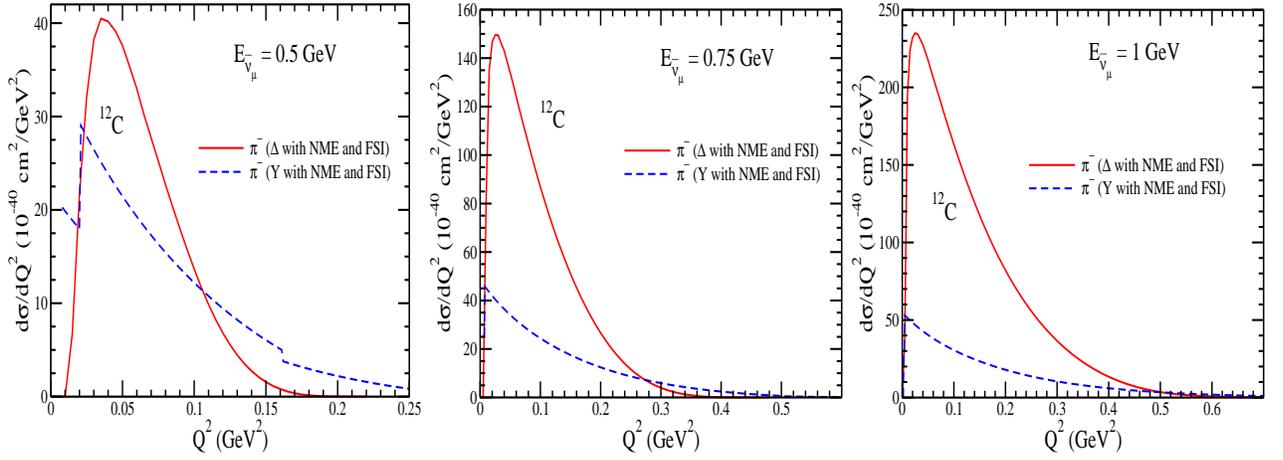

 \begin{center}
\includegraphics[height=6cm,width=5.5cm]{dsigma_dq2_enu_500MeV_delta+hyperon_FSI_carbon.eps}
\includegraphics[height=6cm,width=5.5cm]{dsigma_dq2_enu_750MeV_delta+hyperon_FSI_carbon.eps}
\includegraphics[height=6cm,width=5.5cm]{dsigma_dq2_enu_1GeV_delta+hyperon_FSI_carbon.eps}
\end{center}
  \caption{$\frac{d\sigma}{dQ^2}$ vs $Q^2$ at $E_{\bar{\nu}_\mu} = 0.5$ GeV~(left panel), $0.75$ GeV~(central panel) 
  and $1$ GeV~(right panel) for the ${\bar\nu}_\mu$ induced process in $^{12}$C nuclear target with NME+FSI. The 
  results are shown for the $\pi^-$ contribution from the $\Delta$ (solid line) and from the hyperons (dashed line).}
  \label{q2-C}
 \end{figure}
 \begin{figure}
 \begin{center}
\includegraphics[height=6cm,width=5.5cm]{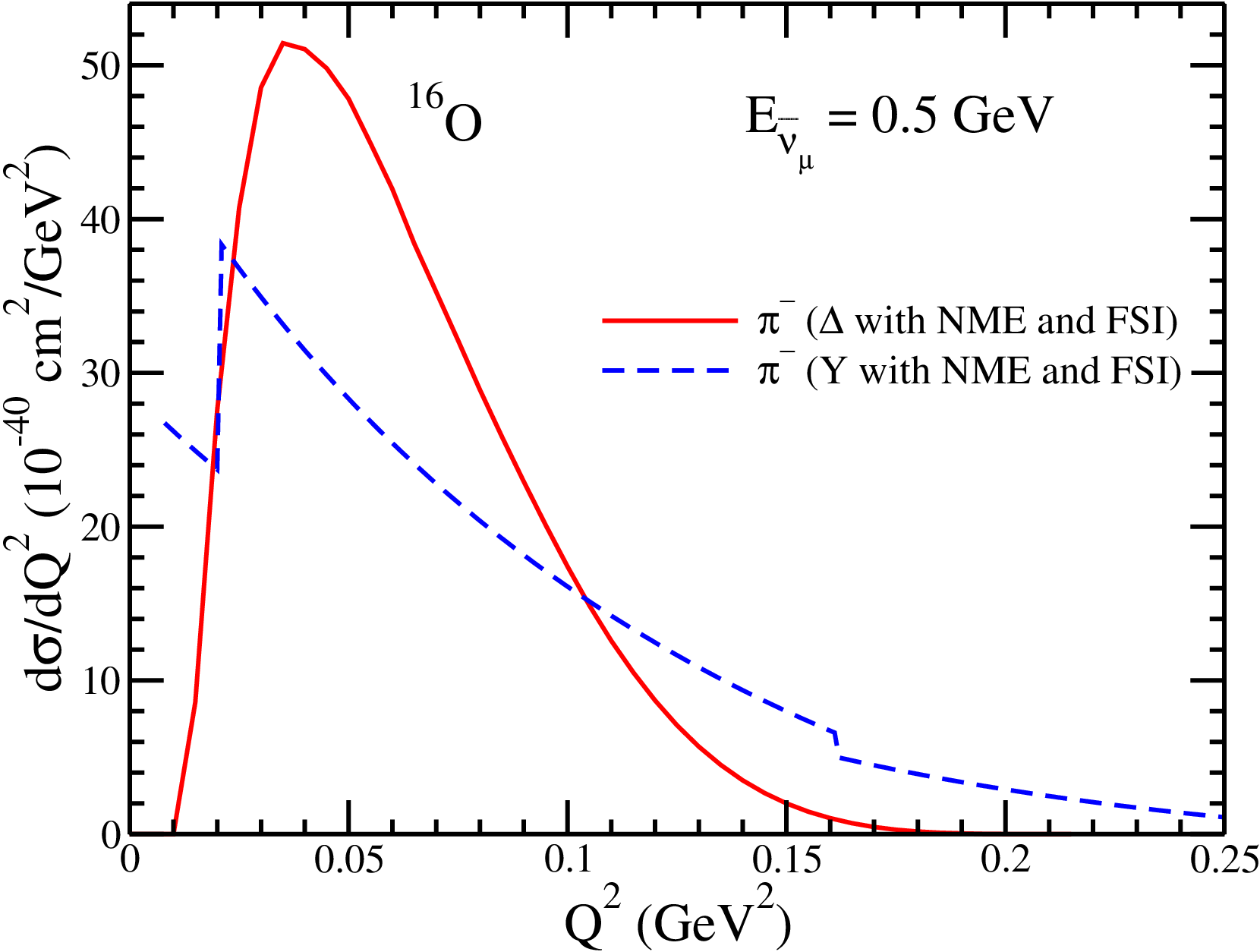}
\includegraphics[height=6cm,width=5.5cm]{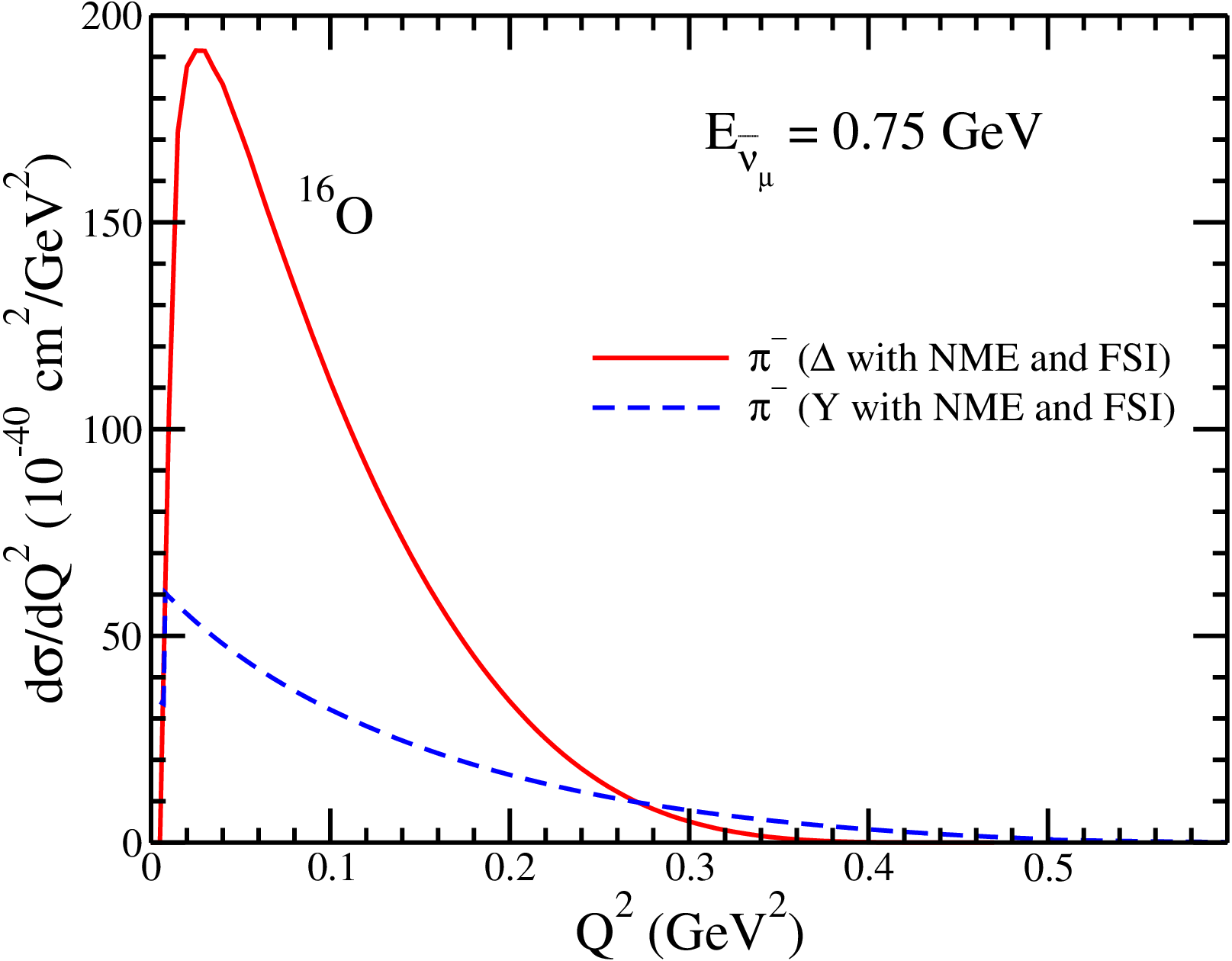}
\includegraphics[height=6cm,width=5.5cm]{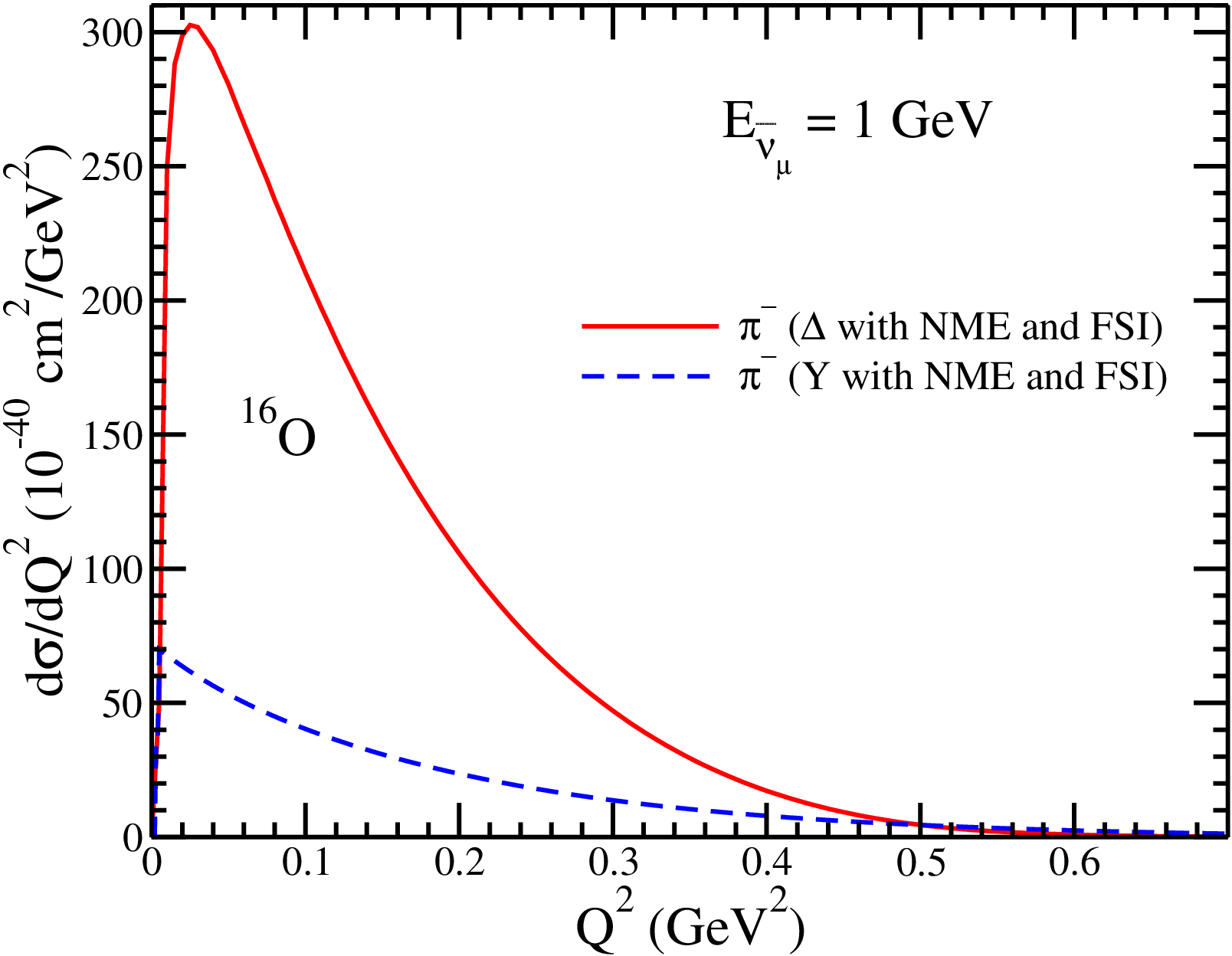}
  \end{center}
  \caption{$\frac{d\sigma}{dQ^2}$ vs $Q^2$  at $E_{\bar{\nu}_\mu} = 0.5$ GeV~(left panel), $0.75$GeV~(central panel) 
  and $1$GeV~(right panel) in $^{16}$O. Lines and points have the same meaning as in Fig.\ref{q2-C}.}
  \label{q2-O}
 \end{figure}
  \begin{figure}
 \begin{center}
\includegraphics[height=6cm,width=5.5cm]{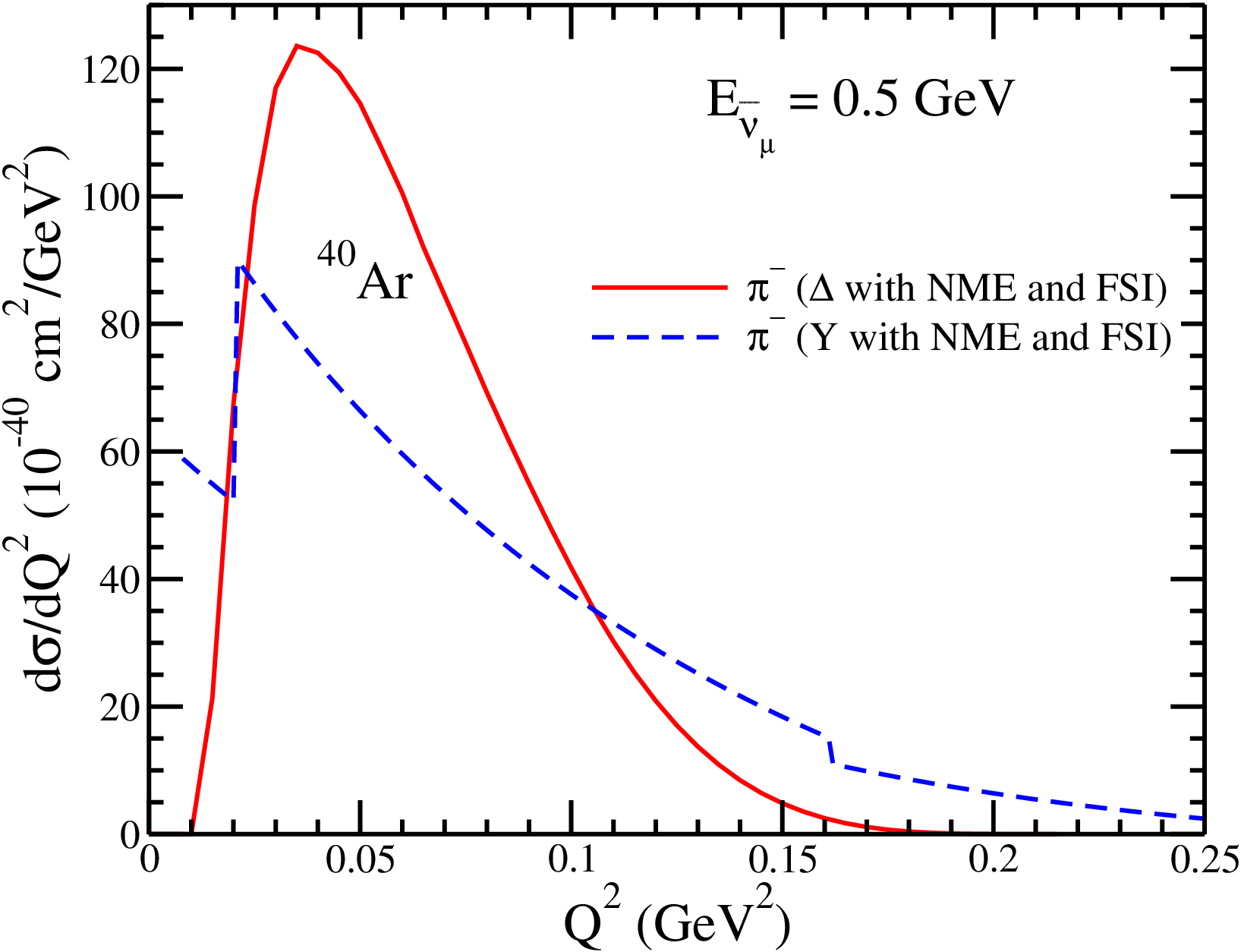}
\includegraphics[height=6cm,width=5.5cm]{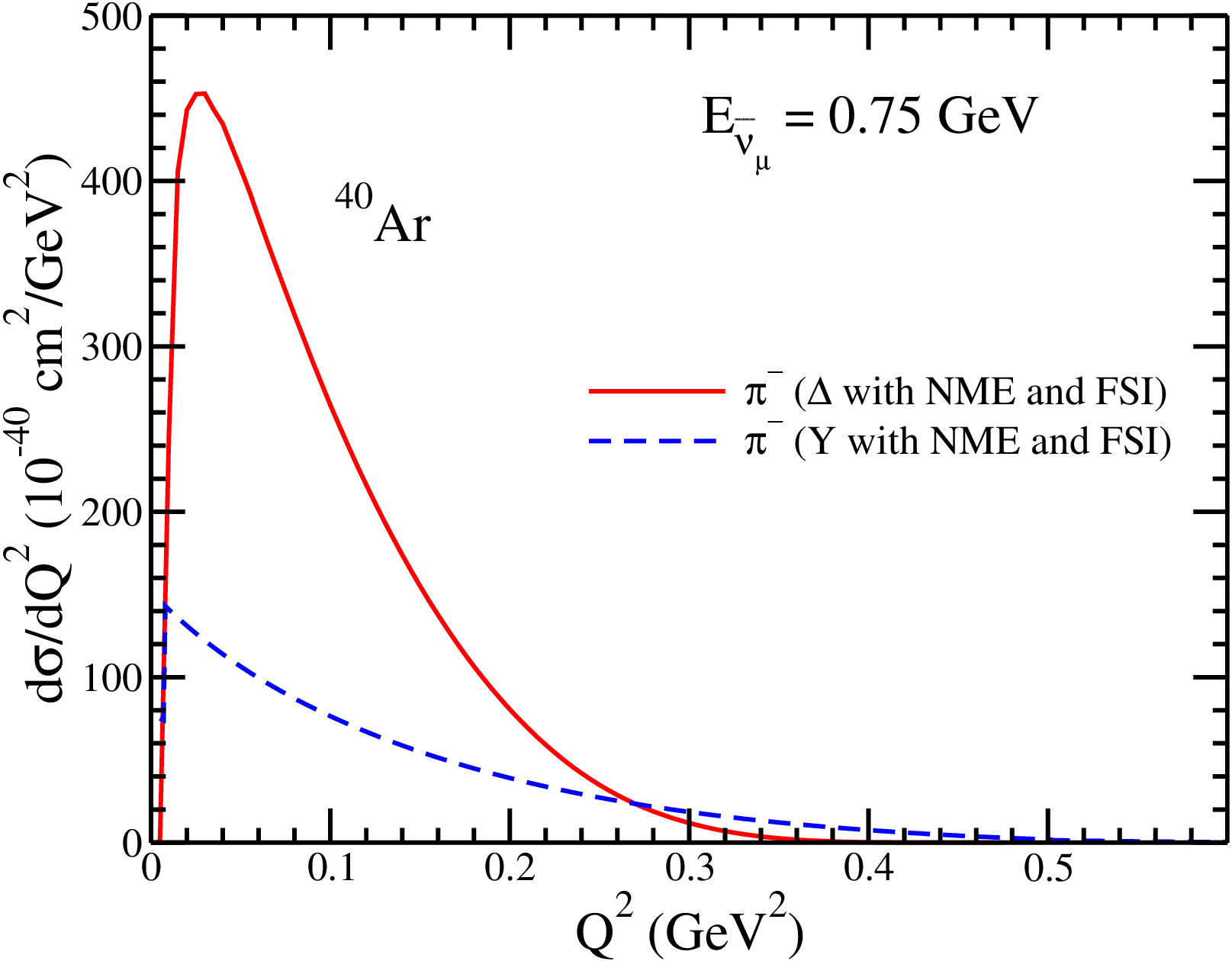}
\includegraphics[height=6cm,width=5.5cm]{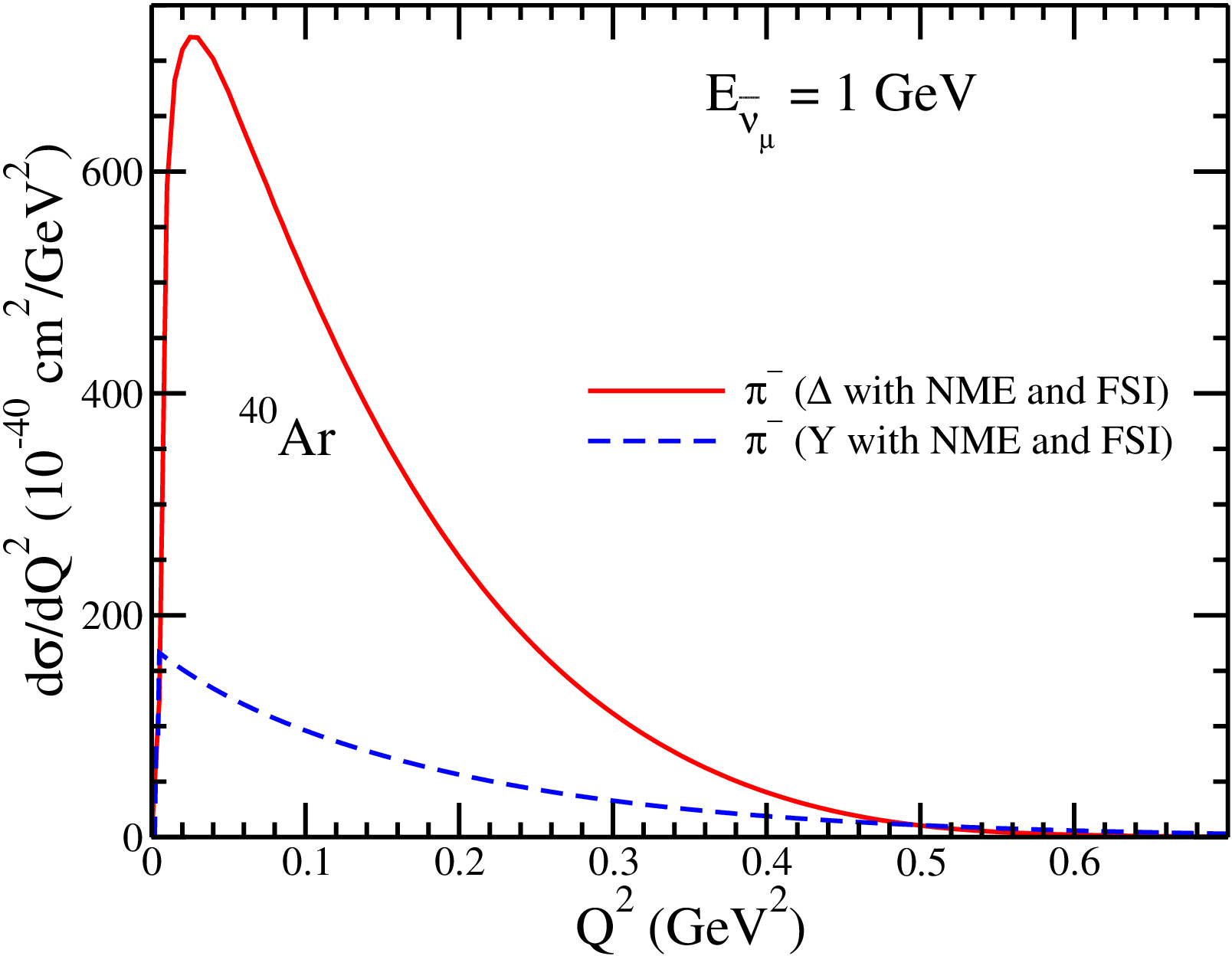}
  \end{center}
  \caption{$\frac{d\sigma}{dQ^2}$ vs $Q^2$  at $E_{\bar{\nu}_\mu} = 0.5$ GeV~(left panel), $0.75$GeV~(central panel) 
  and $1$GeV~(right panel) in $^{40}$Ar. Lines and points have the same meaning as in Fig.\ref{q2-C}.}
  \label{q2-Ar}
 \end{figure}
   \begin{figure}
 \begin{center}
\includegraphics[height=6cm,width=5.5cm]{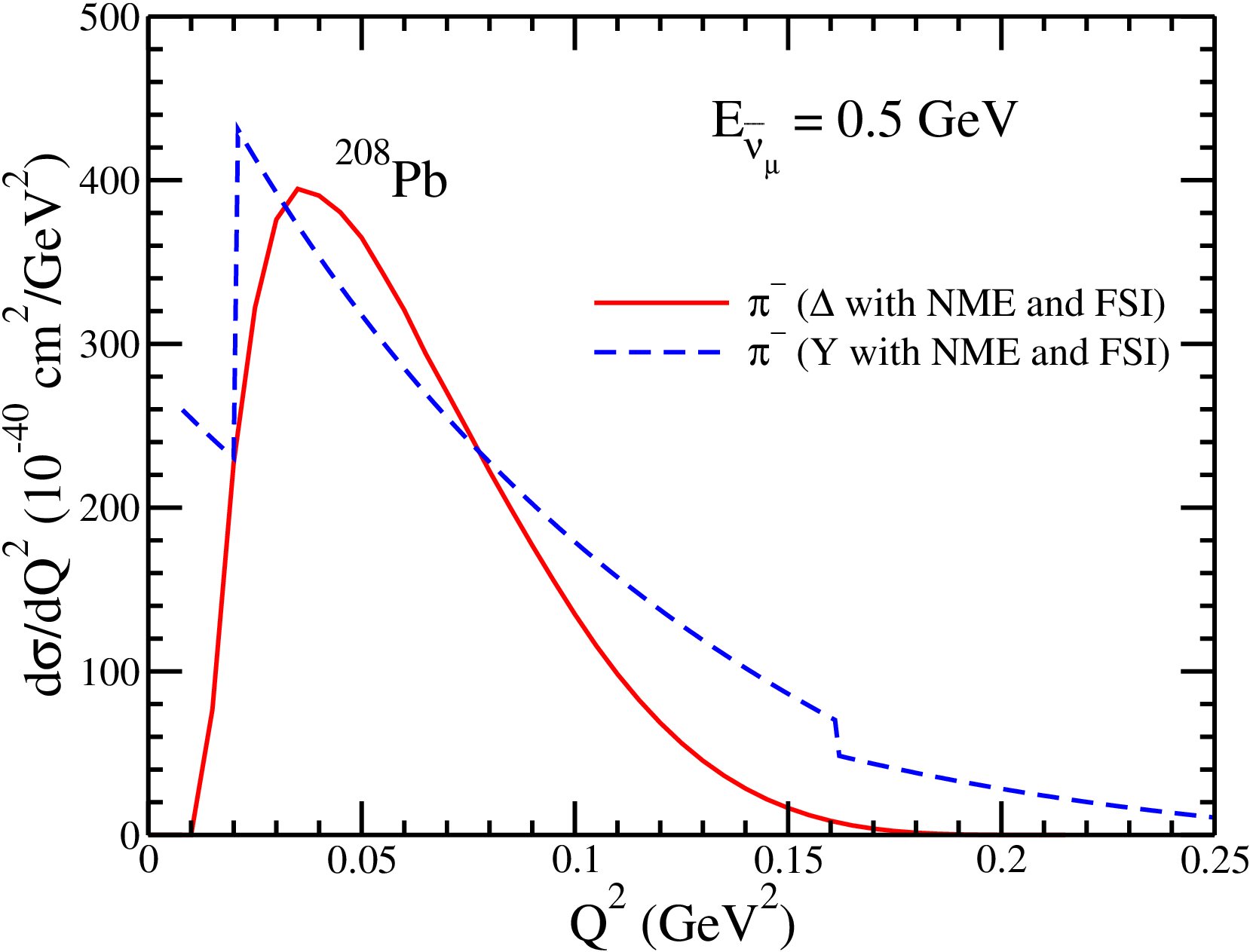}
\includegraphics[height=6cm,width=5.5cm]{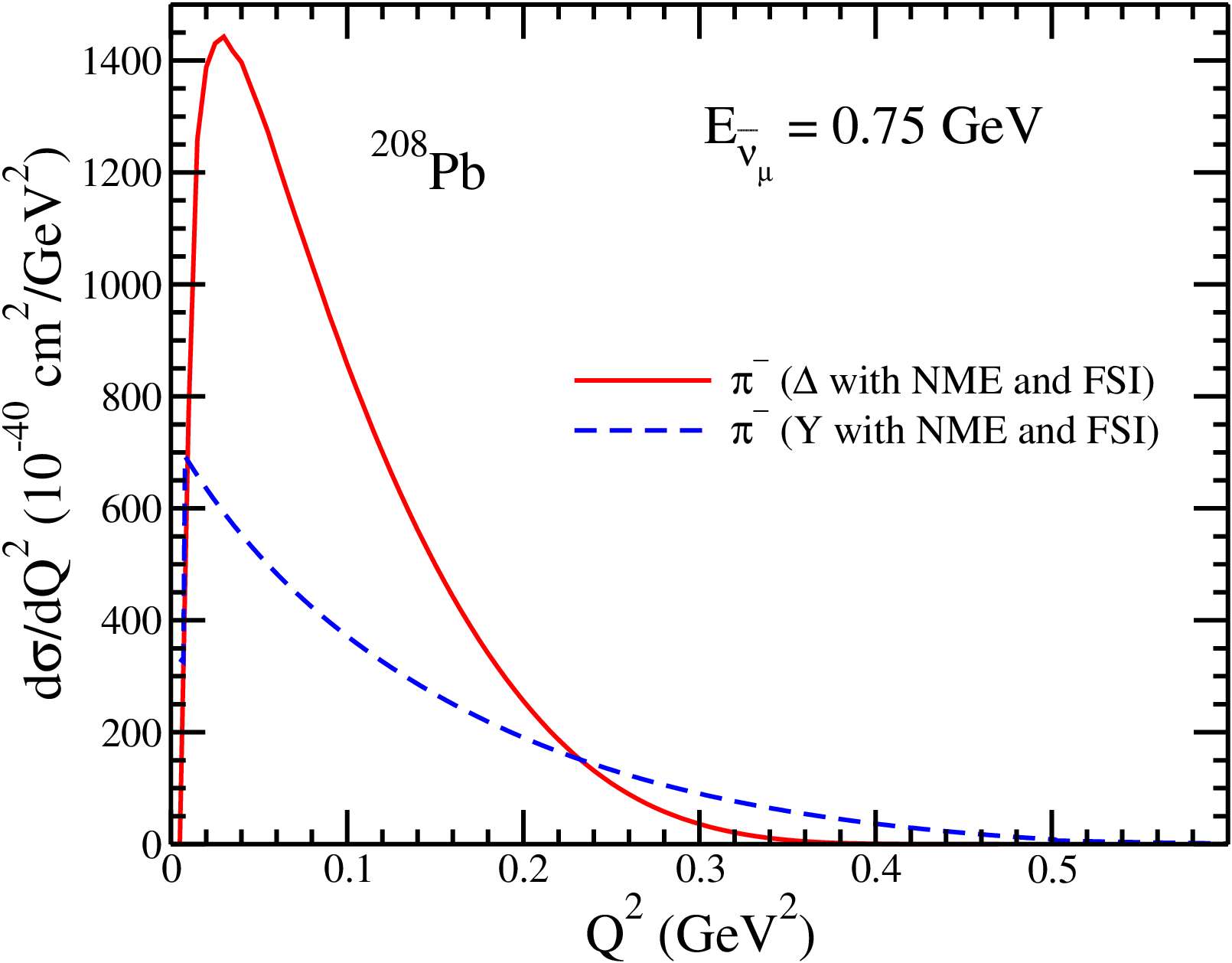}
\includegraphics[height=6cm,width=5.5cm]{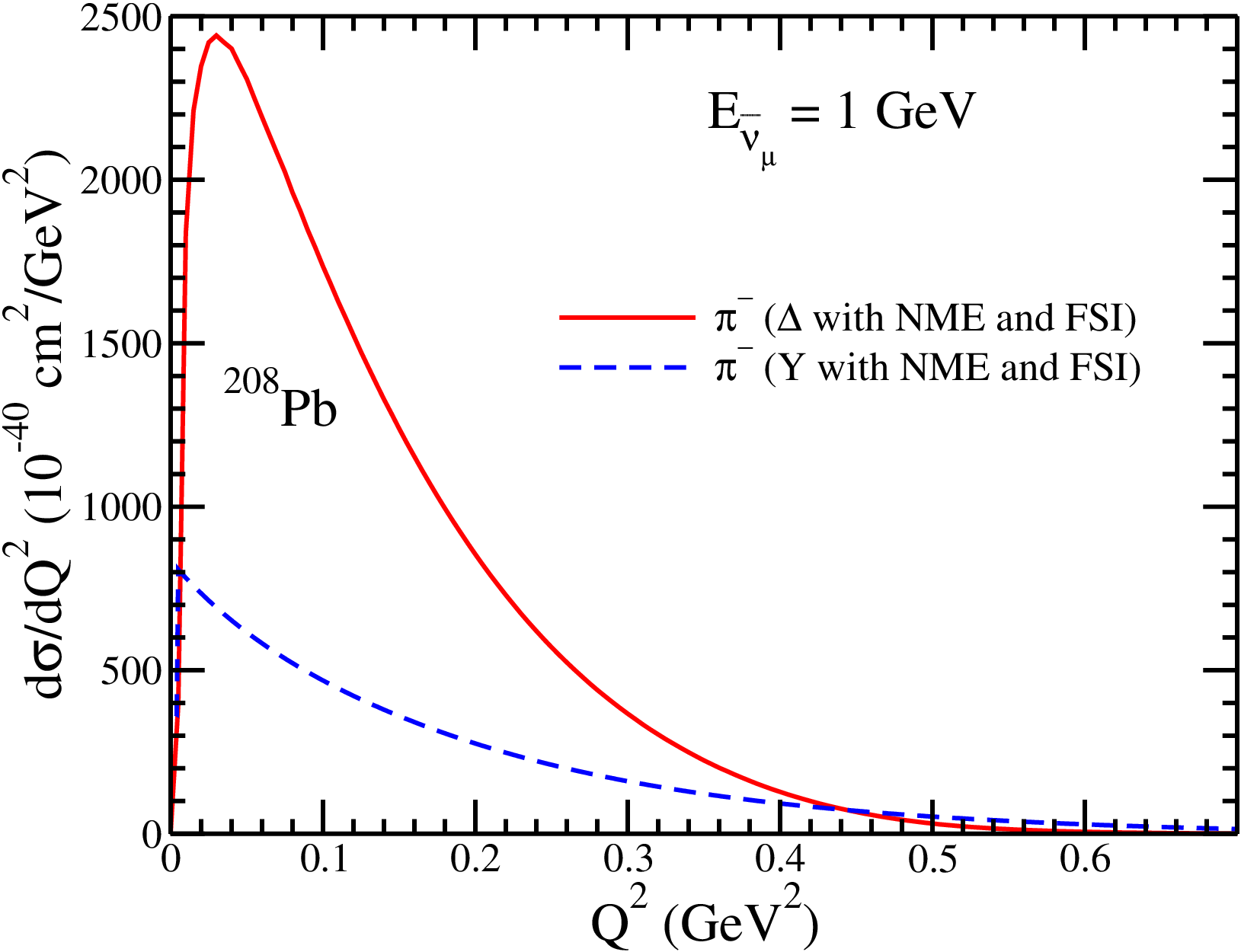}
  \end{center}
  \caption{$\frac{d\sigma}{dQ^2}$ vs $Q^2$  at $E_{\bar{\nu}_\mu} = 0.5$ GeV~(left panel), $0.75$GeV~(central panel) 
  and $1$GeV~(right panel) in $^{208}$Pb. Lines and points have the same meaning as in Fig.\ref{q2-C}.}
  \label{q2-Pb}
 \end{figure}
 
 In Fig.~\ref{c-o-ar-pb}, we have presented the results for the $\nu_\mu$ induced $\pi^+$ and $\pi^o$ productions and 
 ${\bar\nu}_\mu$ induced $\pi^-$ and $\pi^o$ productions. For $\nu_\mu$ induced reactions, the leptonic current in 
 Eq.~(\ref{l}) will read as $l_\mu = \bar{u} (k^\prime) \gamma^\mu (1 - \gamma_5) u (k)$ and the expression of 
 $ \rho_{N}(r)$ in Eq.~(\ref{rho}) will become $\rho_{N}(r)=~\rho_{p}(r)~+~\frac{1}{9}\rho_{n}(r) $. These results are 
 shown for $^{12}$C, $^{16}$O, $^{40}$Ar and $^{208}$Pb with NME+FSI. For $\nu_\mu$ scattering, the contribution to 
 the pions is coming from $\Delta$ only, while for ${\bar\nu}_\mu$ scattering the contribution to the pions is coming 
 from the $\Delta$ as well the hyperons. Though in the case of the pions produced through the hyperon 
  production, there is an overall suppression by a factor of sin$^{2}\theta_{c}$ but these are kinematically favored as 
  the $\Lambda$ production starts around $E_{\bar{\nu}_\mu}$ = 250 MeV, while $\Sigma^-$ and $\Sigma^0$ production 
  start around $E_{\bar{\nu}_\mu}$ = 325 MeV, and there is overall no NME effect on the total hyperon production and no 
  FSI effect on the outgoing pions, whereas the reduction is quite significant for the pions arising from the 
  $\Delta$s.    
   
  In Figs.~\ref{q2-C}, \ref{q2-O}, \ref{q2-Ar} and \ref{q2-Pb}, we have presented the results for the $Q^2$ 
  distribution i.e. $\frac{d\sigma}{dQ^2}$ vs $Q^2$ in $^{12}$C, $^{16}$O, $^{40}$Ar and $^{208}$Pb nuclear targets 
  with NME+FSI at the different incident antineutrino energies viz. $E_{\bar{\nu}_\mu} = 0.5$~GeV, $0.75$~GeV and 
  $1$~GeV. These results are presented for the $\pi^-$ contribution from the $\Delta$s and the hyperons. It may be 
  observed that at low $E_{\bar{\nu}_\mu}$, $\pi^-$ has significant contribution from the hyperons, like at 
  $E_{\bar{\nu}_\mu} = 0.5$ GeV, in the peak region of $Q^2$, hyperons contribute $\sim$40$\%$ in $^{12}$C, $^{16}$O 
  and $^{40}$Ar and 50$\%$ in $^{208}$Pb of the total $\pi^-$ production, while with the increase in energy the 
  contribution from the hyperons decreases, for example, at $E_{\bar{\nu}_\mu} = 1$ GeV hyperons contribute 16$\%$ in 
  $^{12}C$, $^{16}$O and $^{40}$Ar and 24$\%$ in $^{208}$Pb. The peak region of $Q^2$ for the hyperons shifts towards 
  the lower $Q^2$ than the $\Delta$s. In the case of $\pi^o$(not shown here), the results are similar except that the 
  contributions from the hyperons dominate at lower energies in all the nuclear targets in comparison to the $\Delta$ 
  contributions and the dominance increases with the increase in the nuclear mass number.
  
 \section{Summary and conclusions}\label{conclusions}
 In this work, we have presented a review of the theoretical and experimental work done on the quasielatic production 
 of hyperons induced by antineutrinos which was started more than 50 years ago soon after the $V-A$ theory of weak 
 interactions was extended to the strangeness sector by Cabibbo~\cite{Cabibbo:1964zza} using SU(3) symmetry properties 
 of the weak hadronic currents. The experimental results on the total cross sections and their $Q^2$ dependence 
 available from the older experiments at CERN~\cite{Erriquez:1977tr, Eichten:1972bb, Erriquez:1978pg}, 
 FNAL~\cite{Ammosov:1986jn, Ammosov:1986xv}, SKAT~\cite{Brunner:1989kw} and BNL~\cite{Fanourakis:1980si} and the 
 results on Lambda hyperon polarizations from CERN~\cite{Erriquez:1978pg} are compared with the most recent theoretical 
 results~\cite{Akbar:2016awk, Fatima:2018gjy}. 
 
 In view of the future experiments proposed with the antineutrinos at the accelerator and atmospheric antineutrino 
 experiments in the medium energy region of few GeV on the nuclear targets like $^{12}$C, $^{16}$O, $^{40}$Ar and 
 $^{208}$Pb. We have also presented a summary of the recent theoretical works on the total cross sections, polarization 
 components and their $Q^2$ dependence corresponding to many energies relevant for these experiments. These results may 
 be useful in determining the axial vector transition form factor in the strangeness sector specially for the 
 pseudoscalar form factor and the form factor corresponding to the second class currents with and without T-invariance 
 and test the validity of various symmetry properties of the weak hadronic currents. We have also studied the 
 contribution of hyperons produced in these reactions towards the $\bar{\nu}$ induced pion production cross sections in 
 the neutrino oscillation experiments being done at T2K, MINER$\nu$A, DUNE, SUPER-K and HYPER-K.
 \vspace{3mm}
 
 We summarize our results in the following:

 \begin{itemize}
  \item [(A)] In the case of the nucleon targets:
 
 \begin{itemize}
 \item [(i)] The hyperon production is generally Cabibbo suppressed as compared to the $\Delta$ production but in the
 low energy region of $E_{{\bar\nu}_\mu} <  0.6$ GeV it could be comparable to the $\Delta$ 
 production due to the threshold effects.
 
 \item [(ii)]  
 \begin{enumerate}
  \item [(a)] The $Q^2$ distributions are sensitive to the pseudoscalar form factor at lower antineutrino energies.
 
  \item [(b)] The longitudinal $P_L(Q^2)$ as well as the perpendicular $P_P(Q^2)$ components of the hyperon 
  polarization are sensitive to the pseudoscalar form factor $g_3 (Q^2)$ specially at lower energies.
 \end{enumerate}

   \item [(iii)]  In the presence of the second class currents with T-invariance:
   \begin{enumerate}
    \item [(a)]The $Q^2$ distribution is not much sensitive to the presence of the second class current until its 
    coupling strength $g_2^R(0)$ becomes large i.e. $|g_2^R(0)|>1$.
  
  \item [(b)] The longitudinal component of the polarization ${P}_L (Q^2)$ is positive at lower antineutrino energies 
  and becomes negative at higher energies for the values of $g_2^R(0)$ taken to be positive and large i.e. $g_2^R(0)>
  1$. 
  
  \item [(c)] The perpendicular component of the polarization ${P}_P (Q^2)$ is negative for all the values of 
  $g_2^R(0)$ taken to be positive or negative in the energy range of the present interest.
    \end{enumerate}

  \item [(iv)] In the presence of the second class currents without T-invariance:
  \begin{enumerate}
   \item [(a)] The $Q^2$ distribution is not much sensitive to the second class current unless $g_2^I(0)>1$.
  
  \item [(b)]  The transverse component of the polarization ${P}_T(Q^2)$ is nonvanishing and it increases with 
  $g_2^I(0)$. It changes sign when the sign of $g_2^I(0)$ is reversed and the absolute value of ${P}_T(Q^2)$ increases 
  with the increase in the energy. 
  
  \item [(c)]  The longitudinal($P_L(Q^2)$) and the perpendicular($P_P(Q^2)$) components of the polarization are not 
  very sensitive to the choice of $g_2^I(0)$.
  \end{enumerate}
  \end{itemize}

  \item [(B)] In the case of the nuclear targets:
 \end{itemize}

  \begin{itemize}
 \item [(i)]  The effect of NME and FSI is to increase the production of $\Lambda$-hyperon and to decrease the 
 production of $\Sigma$-hyperons in the nuclear medium due to charge exchange processes like $\Sigma~ N ~\rightarrow~
 \Lambda~ N$ and $\Lambda~ N ~\rightarrow~ \Sigma~ N$, but the total hyperon production remains the same.
 
         \item [(ii)] In the case of the $\Delta$ production cross sections, the NME+FSI effect reduces the cross 
         section significantly. This reduction in the cross section increases with the increase in mass number, for 
         example, at $E_{\bar{\nu}_\mu} = 1$ GeV in the case of $^{12}$C, $^{16}$O and $^{40}$Ar, the reduction in 
         the cross section is in the range of $40 - 50\%$ which becomes $70\%$ in the case of $^{208}$Pb.
        
   \item [(iii)] The reduction due to NME+FSI effects in the case of pions obtained from $\Delta$ excitation is large 
 enough to compensate for Cabibbo suppression of pions produced through the hyperon decay specially in the low energy 
 region. Because of this, the pion production from the hyperons is comparable to the pion production from the $\Delta$ 
 excitation up to the antineutrino energies of about 0.5 GeV for $\pi^-$ production and 0.65 GeV for $\pi^o$ 
 production.
     
      \item [(iv)] The ratio of pions produced through $Y$ and $\Delta$ excitations in nuclei increases with the mass 
 number due to the final state interactions of the pions as the pions coming from the $\Delta$ decays are suppressed 
 due to FSI as compared to the pions coming from the hyperons. This ratio decreases with the increase in the 
 antineutrino energies.
 
      \item [(v)] We have also presented the numerical results for $d\sigma/dQ^2$ 
      in the various nuclear targets like $^{12}$C, $^{16}$O, $^{40}$Ar and $^{208}$Pb at the different antineutrino 
      energies.
     
 \end{itemize}
 
 Thus, to conclude, the contribution of hyperon production to the  $\pi^-$ and $\pi^o$ productions induced 
    by the antineutrinos on the nuclear targets is important specially in the sub-GeV energy region. 
    
\section*{Acknowledgment}   
M. S. A. and S. K. S. are thankful to Department of Science and Technology (DST), Government of India for providing 
financial assistance under Grant No. EMR/2016/002285.


\begin{thebibliography}{0}
\bibitem{Abe:2014tzr} 
  K.~Abe {\it et al.} [T2K Collaboration],
    PTEP {\bf 2015}, 043C01 (2015).
  
  

\bibitem{Paley:2013sta} 
  J.~M.~Paley [NO$\nu$A and LBNE Collaborations],
    PoS ICHEP {\bf 2012}, 393 (2013).
  
  

\bibitem{NOvA:2018gge} 
  M.~A.~Acero {\it et al.} [NO$\nu$A Collaboration],
    arXiv:1806.00096 [hep-ex].
 
  

\bibitem{Abe:2015zbg} 
  K.~Abe {\it et al.} [Hyper-Kamiokande Proto- Collaboration],
    PTEP {\bf 2015}, 053C02 (2015).
  
  

\bibitem{Acciarri:2016ooe} 
  R.~Acciarri {\it et al.} [DUNE Collaboration],
    arXiv:1601.02984 [physics.ins-det].
  
  

\bibitem{Galymov:2016nwz} 
  V.~Galymov [LAGUNA-LBNO Consortium],
    Nucl.\ Part.\ Phys.\ Proc.\  {\bf 273-275}, 1854 (2016).
  
  

\bibitem{Goodman:2015gmv} 
  M.~Goodman,
    Adv.\ High Energy Phys.\  {\bf 2015}, 256351 (2015).
  
  

\bibitem{Abe:2017ufe} 
  K.~Abe {\it et al.} [T2K Collaboration],
    Phys.\ Rev.\ D {\bf 96}, 052001 (2017).
  
  

\bibitem{Tsai:2017pta} 
  Y.~T.~Tsai [MicroBooNE Collaboration],
    arXiv:1705.07800 [hep-ex].
  
  

\bibitem{AguilarArevalo:2010bm} 
  A.~A.~Aguilar-Arevalo {\it et al.} [MiniBooNE Collaboration],
    Phys.\ Rev.\ D {\bf 83}, 052007 (2011).
  
  

\bibitem{Mariani:2011xm} 
  C.~Mariani [SciBooNE Collaboration],
    J.\ Phys.\ Conf.\ Ser.\  {\bf 408}, 012038 (2013).
  
  

\bibitem{Aguilar-Arevalo:2013nkf} 
  A.~A.~Aguilar-Arevalo {\it et al.} [MiniBooNE Collaboration],
    Phys.\ Rev.\ D {\bf 91}, 012004 (2015).
  

\bibitem{Andreopoulos:2009rq} 
  C.~Andreopoulos {\it et al.},
    Nucl.\ Instrum.\ Meth.\ A {\bf 614}, 87 (2010).
  
  

\bibitem{Hayato:2009zz} 
  Y.~Hayato,
    Acta Phys.\ Polon.\ B {\bf 40}, 2477 (2009).
  
  

\bibitem{Golan:2012wx} 
  T.~Golan, C.~Juszczak and J.~T.~Sobczyk,
    Phys.\ Rev.\ C {\bf 86}, 015505 (2012).
  
  

\bibitem{Buss:2011mx} 
  O.~Buss {\it et al.},
    Phys.\ Rept.\  {\bf 512}, 1 (2012).
  
  

\bibitem{Alvarez-Ruso:2017oui} 
  L.~Alvarez-Ruso {\it et al.},
    Prog.\ Part.\ Nucl.\ Phys.\  {\bf 100}, 1 (2018).
  

\bibitem{Katori:2016yel} 
  T.~Katori and M.~Martini,
    J.\ Phys.\ G {\bf 45}, 013001 (2018).
  
  

\bibitem{Alvarez-Ruso:2014bla} 
  L.~Alvarez-Ruso, Y.~Hayato and J.~Nieves,
    New J.\ Phys.\  {\bf 16}, 075015 (2014).
  
  

\bibitem{Formaggio:2013kya} 
  J.~A.~Formaggio and G.~P.~Zeller,
    Rev.\ Mod.\ Phys.\  {\bf 84}, 1307 (2012)
  
  

\bibitem{Morfin:2012kn} 
  J.~G.~Morfin, J.~Nieves and J.~T.~Sobczyk,
    Adv.\ High Energy Phys.\  {\bf 2012}, 934597 (2012).
  
   

\bibitem{Alam:2015gaa} 
  M.~Rafi Alam, M.~Sajjad Athar, S.~Chauhan and S.~K.~Singh,
    Int.\ J.\ Mod.\ Phys.\ E {\bf 25}, 1650010 (2016).
  
    

\bibitem{Akbar:2015yda} 
  F.~Akbar, M.~Rafi Alam, M.~Sajjad Athar, S.~Chauhan, S.~K.~Singh and F.~Zaidi,
    Int.\ J.\ Mod.\ Phys.\ E {\bf 24}, 1550079 (2015).
  
  

\bibitem{prd1} S. Ahmad, M. Sajjad Athar and S. K. Singh, Phys. Rev. {\bf D 74}, 073008 (2006).
   
   

\bibitem{a6} S. K. Singh, M. Sajjad Athar and S. Ahmad, Phys. Rev. Lett. {\bf 96}, 241801 (2006).
 
 

\bibitem{Athar:2007wd} 
  M.~Sajjad~Athar, S.~Ahmad and S.~K.~Singh,
    Phys.\ Rev.\ D {\bf 75}, 093003 (2007).
 

\bibitem{Hernandez:2007qq} 
  E.~Hernandez, J.~Nieves and M.~Valverde,
    Phys.\ Rev.\ D {\bf 76}, 033005 (2007).
  
  

\bibitem{Adamson:2016xxw} 
  P.~Adamson {\it et al.} [NO$\nu$A Collaboration],
    Phys.\ Rev.\ D {\bf 93}, 051104 (2016).
  

\bibitem{Holstein}
B.~R.~Holstein,
\textit{Weak interaction in nuclei} (Princeton University Press, 1990).
  
  

\bibitem{Oset:1989ey} 
  E.~Oset, P.~Fernandez de Cordoba, L.~L.~Salcedo, and R.~Brockmann,
    Phys.\ Rep.\  {\bf 188}, 79 (1990).
 

\bibitem{Cabibbo:2003cu} 
  N.~Cabibbo, E.~C.~Swallow and R.~Winston,
    Ann.\ Rev.\ Nucl.\ Part.\ Sci.\  {\bf 53}, 39 (2003).    

\bibitem{Gaillard:1984ny} 
 J.~M.~Gaillard and G.~Sauvage,
     Ann.\ Rev.\ Nucl.\ Part.\ Sci.\  {\bf 34}, 351 (1984).

\bibitem{Gazia}
A.~Garcia, P.~ Kielanowski,
Lecture notes in Physics, vol. {\bf 222}, Springer (1985), Edited by A.~Bohm. 
 

\bibitem{Henley:1969uz} 
  E.~M.~Henley,
  Ann.\ Rev.\ Nucl.\ Part.\ Sci.\  {\bf 19}, 367 (1969).
  
   

\bibitem{Cannata:1970br} 
  F.~Cannata, R.~Leonardi and F.~Strocchi,
  Phys.\ Rev.\ D {\bf 1}, 191 (1970).
  
       

\bibitem{DeRujula:1970ek} 
  A.~De Rujula and E.~De Rafael,
  Phys.\ Lett.\  {\bf 32B}, 495 (1970).  
 
 

\bibitem{Adler:1963}
 S. L. Adler,
 Nuovo Cimento {\bf 30}, 1020 (1963), E {\bf 32}, 309 (1964).
    

\bibitem{Berman:1964zza} 
  S.~M.~Berman, M.~Veltman,
  Phys.\ Lett.\  {\bf 12}, 275 (1964).
  

\bibitem{Fujii2}
A. Fujii and Y. Yamaguchi,
Progress of Theoretical Physics {\bf 33}, 552 (1965).
  

\bibitem{Fujii1}   
   A. Fujii and Y. Yamaguchi,
 Nuovo Cimento {\bf 43}, 325 (1966).

\bibitem{Cabibbo:1964zza} 
  N.~Cabibbo,
  Phys.\ Lett.\  {\bf 12}, 137 (1964). 

\bibitem{Glashow:1965zz} 
  S.~L.~Glashow,
  Phys.\ Rev.\ Lett.\  {\bf 14}, 35 (1965). 
  
      

\bibitem{Okamura:1971pn} 
  H.~Okamura,
  Prog.\ Theor.\ Phys.\  {\bf 45}, 1707 (1971).
  
     

\bibitem{Ketley}  
     I. J. Ketley, 
  Nuovo Cimento {\bf 38}, 302 (1965). 
  
    

\bibitem{Egardt} 
    L. Egardt, 
  Nuovo Cimento {\bf 29}, 954 (1963).
  

\bibitem{Block:1965zol} 
  M.~M.~Block in {\it Symmetries in Elementary Particle Physics}, Edited by A. Zichichi, Academic Press (1965).
 
     

\bibitem{Block:NAL}
   M. Block, NAL Summer Study, {\bf 1} 215 (1968).

\bibitem{Cabibbo:1965zza} 
  N.~Cabibbo and F.~Chilton,
    Phys.\ Rev.\  {\bf 137}, B1628 (1965).
     
  

\bibitem{Singh:2006xp} 
  S.~K.~Singh and M.~J.~Vicente Vacas,
    Phys.\ Rev.\ D {\bf 74}, 053009 (2006). 
    

\bibitem{Alam:2014bya} 
   M.~Rafi~Alam, M.~Sajjad~Athar, S.~Chauhan and S.~K.~Singh,
     J.\ Phys.\ G {\bf 42}, 055107 (2015).
     
 

\bibitem{Cabibbo:1963yz} 
  N.~Cabibbo,
    Phys.\ Rev.\ Lett.\  {\bf 10}, 531 (1963).
  
  

\bibitem{Block:1964gj} 
  M.~M.~Block {\it et al.},
    Phys.\ Rev.\ Lett.\  {\bf 12}, 262 (1964).
  
  

\bibitem{Chilton:1964zza} 
  F.~Chilton,
    Nuovo Cim.\  {\bf 31}, 447 (1964).
    
   

\bibitem{sirlin}
   A. Sirlin, Nuovo Cim.\  {\bf 37}, 137 (1965).
   
   

\bibitem{Finjord:1975zy} 
  J.~Finjord and F.~Ravndal,
    Nucl.\ Phys.\ B {\bf 106}, 228 (1976).
    

\bibitem{Marshak}
R.~E.~Marshak, Riazuddin, C.~P.~Ryan,
\textit{Theory of Weak Interactions in Particle Physics} (Wiley-Interscience, 1969).

\bibitem{LlewellynSmith:1971uhs} 
  C.~H.~Llewellyn Smith,
  Phys.\ Rept.\  {\bf 3}, 261 (1972). 

\bibitem{Pais:1971er} 
  A.~Pais,
  Annals Phys.\  {\bf 63}, 361 (1971).

\bibitem{Erriquez:1977tr}
  O.~Erriquez et al.,
    Phys.\ Lett.\ B {\bf 70}, 383 (1977).
    

\bibitem{Eichten:1972bb} 
  T.~Eichten  et al.,
    Phys.\ Lett.\ B {\bf 40}, 593 (1972).

\bibitem{Erriquez:1978pg} 
  O.~Erriquez et al.,
    Nucl.\ Phys.\ B {\bf 140}, 123 (1978).
 

\bibitem{Fanourakis:1980si} 
  G.~Fanourakis et al.,
    Phys.\ Rev.\ D {\bf 21}, 562 (1980).
      

\bibitem{Ammosov:1986jn}
 V.~V.~Ammosov et al.,
  Z \ Phys.\ C {\bf 36}, 377 (1987).

\bibitem{Ammosov:1986xv} 
  V.~V.~Ammosov  et al.,
    JETP Lett.\  {\bf 43}, 716 (1986)
  [Pisma Zh.\ Eksp.\ Teor.\ Fiz.\  {\bf 43}, 554 (1986)].
  
  

\bibitem{Brunner:1989kw} 
  J.~Brunner  et al. [SKAT Collaboration],
    Z.\ Phys.\ C {\bf 45}, 551 (1990).
    

\bibitem{Kuzmin:2003ji} 
  K.~S.~Kuzmin, V.~V.~Lyubushkin and V.~A.~Naumov,
   Mod.\ Phys.\ Lett.\ A {\bf 19}, 2815 (2004), 
  [Phys.\ Part.\ Nucl.\  {\bf 35}, S133 (2004)].
  
  

\bibitem{Alam:2013cra} 
  M.~Rafi Alam, S.~Chauhan, M.~Sajjad Athar and S.~K.~Singh,
    Phys.\ Rev.\ D {\bf 88}, 077301 (2013).
  
  

\bibitem{JPARC}
  http://j-parc.jp/index-e.html
  
    

\bibitem{fermi}
  http://www.fnal.gov/
  
  

\bibitem{Kuzmin:2008zz} 
  K.~S.~Kuzmin and V.~A.~Naumov,
    Phys.\ Atom.\ Nucl.\  {\bf 72}, 1501 (2009).
  

\bibitem{Bilenky:2013fra} 
  S.~M.~Bilenky, E.~Christova,
    J.\ Phys.\ G {\bf 40}, 075004 (2013).
    

\bibitem{Bilenky:2013iua} 
  S.~M.~Bilenky, E.~Christova,
    Phys.\ Part.\ Nucl.\ Lett.\  {\bf 10}, 651 (2013).  

\bibitem{Akbar:2017qsf} 
  F.~Akbar, M.~Sajjad Athar, A.~Fatima and S.~K.~Singh,
    Eur.\ Phys.\ J.\ A {\bf 53}, 154 (2017).

\bibitem{Graczyk:2017rti} 
  K.~M.~Graczyk and B.~E.~Kowal,
    Phys.\ Rev.\ D {\bf 97}, no. 1, 013001 (2018).
 

\bibitem{Kuzmin:2004ke} 
  K.~S.~Kuzmin, V.~V.~Lyubushkin and V.~A.~Naumov,
  Mod.\ Phys.\ Lett.\ A {\bf 19}, 2919 (2004).       

\bibitem{Graczyk:2004vg} 
  K.~M.~Graczyk, 
  Nucl.\ Phys.\ Proc.\ Suppl.\  {\bf 139}, 150 (2005). 

\bibitem{Hagiwara:2004gs} 
  K.~Hagiwara, K.~Mawatari and H.~Yokoya,
  Nucl.\ Phys.\ Proc.\ Suppl.\  {\bf 139}, 140 (2005).  
  

\bibitem{Graczyk:2004uy} 
  K.~M.~Graczyk,
  Nucl.\ Phys.\ A {\bf 748}, 313 (2005).

\bibitem{Valverde:2006yi} 
  M.~Valverde, J.~E.~Amaro, J.~Nieves and C.~Maieron,
  Phys.\ Lett.\ B {\bf 642}, 218 (2006). 
  
  

\bibitem{Fatima:2018tzs} 
  A.~Fatima, M.~Sajjad Athar and S.~K.~Singh,
    arXiv:1806.08597 [hep-ph].
  
  

\bibitem{Fatima:2018gjy} 
  A.~Fatima, M.~Sajjad Athar and S.~K.~Singh,
    Eur.\ Phys.\ J.\ A {\bf 54}, 95 (2018).
  
  

\bibitem{Akbar:2016awk} 
  F.~Akbar, M.~Rafi Alam, M.~Sajjad Athar and S.~K.~Singh,
    Phys.\ Rev.\ D {\bf 94}, 114031 (2016).
  
  

\bibitem{Mintz:2004eu} 
  S.~L.~Mintz,
  J.\ Phys.\ G {\bf 30}, 565 (2004).
  
 

\bibitem{Mintz:2002cj} 
  S.~L.~Mintz, M.~A.~Barnett,
  Phys.\ Rev.\ D {\bf 66}, 117501 (2002).
  
  

\bibitem{Mintz:2001jc} 
  S.~L.~Mintz,
  Nucl.\ Phys.\ A {\bf 690}, 711 (2001).
  
  

\bibitem{Wu:2013kla} 
  J.~J.~Wu and B.~S.~Zou,
    Few Body Syst.\  {\bf 56}, no. 4-5, 165 (2015).
  
  

\bibitem{Drakoulakos:2004gn} 
  D.~Drakoulakos {\it et al.} [MINER$\nu$A Collaboration],
    hep-ex/0405002.
  
  

\bibitem{DUNE}
  http://www.dunescience.org/
  

\bibitem{Bradford:2006yz} 
  R.~Bradford et al.,
    Nucl.\ Phys.\ Proc.\ Suppl.\  {\bf 159}, 127 (2006). 
    
    

\bibitem{Weinberg:1958ut} 
  S.~Weinberg,
    Phys.\ Rev.\  {\bf 112}, 1375 (1958).
  

\bibitem{Nambu:1960xd}
Y.~Nambu,
Phys.\ Rev.\ Lett.\ {\bf 4}, 380 (1960).
    

\bibitem{Bilenky}
S.~M.~Bilenky,
\textit{Basics of Introduction to Feynman Diagrams and Electroweak Interactions Physics} (Editions Fronti\`{e}res, 
1994). 
 

\bibitem{paschos2} O. Lalakulich, E. A. Paschos and G. Piranishvili, Phys. Rev. {\bf D 74}, 014009 (2006).
  
  

\bibitem{AlvarezRuso:1998hi} 
  L.~Alvarez-Ruso, S.~K.~Singh and M.~J.~Vicente Vacas,
    Phys.\ Rev.\ C {\bf 59}, 3386 (1999).
  


\bibitem{ruso} S. K. Singh, M. J. Vicente Vacas and E. Oset,
 Phys. Lett. {\bf B 416}, 23 (1998)
  Erratum: [Phys.\ Lett.\ B {\bf 423}, 428 (1998)].


\bibitem{a4} E. Oset and L. L. Salcedo, Nucl. Phys. {\bf A 468}, 631
 (1987); C. Garcia Recio, E. Oset, L. L. Salcedo, D. Strottman and
 M. J. Lopez, Nucl. Phys. {\bf A 526}, 685 (1991).
 
   

\bibitem{vries} C. W. de Jager, H.de Vries and C.de Vries, At. Data Nucl. Data Tables {\bf 14}, 479 (1974).

\bibitem{a5} M. J. Vicente Vacas, M. Kh Khankhasaev and S. G. Mashnik, nucl-th/9412023, M. J. Vicente Vacas, Private 
Communication.

\bibitem{Armenise:1979zg} 
  N.~Armenise  et al.,
    Nucl.\ Phys.\ B {\bf 152}, 365 (1979).

\end{thebibliography}
\end{document}